%% file: EXO-13-009_temp.tex
\pdfoutput=1

\documentclass[11pt,twoside,a4paper,cmspaper,final,collab]{cms-tdr}
\def\svnVersion{258627}\def\cmsCernNo{2013-037}\def\cmsCernDate{\today}\def\cmsMessage{Published in the Journal of High Energy Physics as \href{http://dx.doi.org/10.1007/JHEP08(2014)174}{\doi{10.1007/JHEP08(2014)174.}}}

\begin{document}\cmsNoteHeader{EXO-13-009}

\hyphenation{had-ron-i-za-tion}
\hyphenation{cal-or-i-me-ter}
\hyphenation{de-vices}

\RCS$Revision: 249604 $
\RCS$HeadURL: svn+ssh://svn.cern.ch/reps/tdr2/papers/EXO-13-009/trunk/EXO-13-009.tex $
\RCS$Id: EXO-13-009.tex 249604 2014-07-03 16:18:58Z alverson $
\newlength\cmsFigWidth
\setlength\cmsFigWidth{0.48\textwidth}
\providecommand{\NA}{---}
\newcommand{\THISLUMI} {19.7\xspace}
\newcommand{\MINMWWMASS} {600\xspace}
\newcommand{\MAXMWWMASS} {2500\xspace}
\newcommand{\MINMWWMASSCONCL} {0.8\xspace}
\newcommand{\MAXMWWMASSCONCL} {2.5\xspace}
\newcommand{\MINMWWMASSFIT} {700\xspace}
\newcommand{\SIGNALSHAPEMASS} {1500\xspace}
\newcommand{\SFTTBARELEHP} {\ensuremath{0.96\pm0.03}\xspace}
\newcommand{\SFTTBARMUHP} {\ensuremath{0.97\pm0.02}\xspace}
\newcommand{\SFTTBARELELP} {\ensuremath{1.39\pm0.08}\xspace}
\newcommand{\SFTTBARMULP} {\ensuremath{1.31\pm0.05}\xspace}
\newcommand{\SFWTAGHP} {\ensuremath{0.89\pm0.08}\xspace}
\newcommand{\SFWTAGLP} {\ensuremath{1.28\pm0.30}\xspace}
\newcommand{\WMASSDATA} {\ensuremath{84.7\pm0.4}\xspace}
\newcommand{\WMASSMC} {\ensuremath{83.4\pm0.3}\xspace}
\newcommand{\WRESDATA} {\ensuremath{7.9\pm0.6}\xspace}
\newcommand{\WRESMC} {\ensuremath{7.2\pm0.4}\xspace}
\newcommand{\WJETNORMUNCERT} {below 10\%\xspace}
\newcommand{\TTBARNORMUNCERT} {below 5\%\xspace}
\newcommand{\VVNORMUNCERT} {20\%\xspace}
\newcommand{\VTAGUNCERTHP} {9\%\xspace}
\newcommand{\VTAGUNCERTLP} {24\%\xspace}
\newcommand{\LUMIUNCERT} {2.6\%\xspace}
\newcommand{\MEANUNCERT} {1.5\%\xspace}
\newcommand{\WIDTHUNCERT} {4.5\%\xspace}
\newcommand{\RANGELIMITS} {from 70\unit{fb} to 3\unit{fb}\xspace}
\newcommand{\pp}{\Pp\Pp}%
\newcommand{\Wo}{\PW\xspace}%
\newcommand{\Wp}{\PWp\xspace}%
\newcommand{\Wm}{\PWm\xspace}%
\newcommand{\Zo}{\cPZ\xspace}%
\newcommand{\Vo}{\ensuremath{\mathrm{V}}\xspace}%
\newcommand{\MN}{\Pgm\Pgm\xspace}%
\newcommand{\EN}{\Pe\Pgn\xspace}%
\newcommand{\LN}{\ensuremath{\ell\Pgn}\xspace}%
\newcommand{\MW}{\ensuremath{m_\Wo}\xspace}%
\newcommand{\MZ}{\ensuremath{m_\Zo}\xspace}%
\providecommand{\MT}{\ensuremath{M_\mathrm{T}}\xspace}%
\newcommand{\MLL}{\ensuremath{m_{\ell\ell}}\xspace}%
\newcommand{\nunubar}{\Pgn\Pagn\xspace}%
\newcommand{\qqbar}{\Pq\Paq\xspace}%
\newcommand{\qqbarpr}{\ensuremath{\Pq\Paq^({}'^){}}\xspace}
\newcommand{\HZZ}{\ensuremath{\PXXG\to\Zo\Zo}}%
\newcommand{\HZZllqq}{\ensuremath{\HZZ\to\qqbar\,\LL}}%
\newcommand{\Hllqq}{\ensuremath{\PXXG\to\LL\qqbar}}%
\newcommand{\MPl}{\ensuremath{{M_{\text{Pl}}}}\xspace}%
\newcommand{\RedMPl}{\ensuremath{\overline{M}_{\text{Pl}}}\xspace}%
\newcommand{\Grav}{\ensuremath{\PXXG_{\mathrm{bulk}}}}%
\newcommand{\GZZ}{\ensuremath{\PXXG\to\Zo\Zo}}%
\newcommand{\GZZllqq}{\ensuremath{\GZZ\to\qqbar\,\LL}}%
\newcommand{\Gllqq}{\ensuremath{\Grav\to\LL\qqbar}}%
\newcommand{\XWW}{\ensuremath{X\to\Wo\Wo}}%
\newcommand{\XZZ}{\ensuremath{X\to\Zo\Zo}}%
\newcommand{\XVV}{\ensuremath{X\to \Vo\Vo}}%
\newcommand{\Xllqq}{\ensuremath{X\to\LL\qqbar}}%
\newcommand{\XZZllqq}{\ensuremath{\XZZ\to\LL\qqbar}}%
\newcommand{\Mg}{\ensuremath{m_{\PXXG}}}%
\newcommand{\mG}{\ensuremath{\text{M}_{\text{G}}}}%
\newcommand{\mX}{\ensuremath{\text{M}_{\text{X}}}}%
\newcommand{\wX}{\ensuremath{\Gamma_{\text{X}}}}%
\newcommand{\mZ}{\ensuremath{m_{\Zo}}}%
\newcommand{\mW}{\ensuremath{m_{\Wo}}}%
\newcommand{\mH}{\ensuremath{m_{\PXXG}}}%
\newcommand{\mZZ}{\ensuremath{m_{\Zo\Zo}}\xspace}%
\newcommand{\mWV}{\ensuremath{m_{\Wo \Vo}}\xspace}%
\newcommand{\mWW}{\ensuremath{m_{\Wo \Wo}}\xspace}%
\newcommand{\mVV}{\ensuremath{m_{\Vo\Vo}}\xspace}%
\newcommand{\mVW}{\ensuremath{m_{\Wo \Vo}}\xspace}%
\newcommand{\mll}{\ensuremath{m_{\ell\ell}}\xspace}%
\newcommand{\mLL}{\ensuremath{m_{\ell\ell}}\xspace}%
\newcommand{\mjj}{\ensuremath{m_\mathrm{jj}}}%
\newcommand{\mJ}{\ensuremath{m_{\text{jet}}}}%
\newcommand{\nsubj}{\ensuremath{\tau_{21}}}%
\newcommand{\ktilde}{\ensuremath{k/\overline{M}_\mathrm{Pl}}\xspace}%
\newcommand{\JHUGEN} {{\textsc{jhugen}}\xspace}
\newcommand{\BulkG}{\ensuremath{\PXXG_{\text{bulk}}}\xspace}\mathchardef\mhyphen="2D
\newcommand{\lnujet}{\ensuremath{\ell \nu}+V-jet\xspace}
\newcommand{\lljet}{\ensuremath{\ell \ell}+V-jet\xspace}
\newcommand\T{\relax}
\cmsNoteHeader{EXO-13-009} 
\title{Search for massive resonances decaying into pairs of boosted bosons in semi-leptonic final states at $\sqrt{s} = 8\TeV$}

\date{\today}

\abstract{
  A search for new resonances
  decaying to WW, ZZ, or WZ is presented. Final states are considered in which
  one of the vector bosons decays leptonically and the other hadronically.
  Results
  are based on data corresponding to an integrated luminosity of
  19.7\fbinv recorded in proton-proton collisions at $\sqrt{s}=8$\TeV
  with the CMS detector at the CERN LHC.
  Techniques aiming at identifying jet substructures
  are used to analyze signal events in which the
  hadronization products from the decay of highly boosted W or Z
  bosons are contained within a single reconstructed jet.
 Upper limits on the production of generic WW, ZZ, or WZ resonances are set
  as a function of the resonance mass and width. We increase the sensitivity of the analysis by statistically
  combining the results of this search with a complementary study of the all-hadronic final state.
  Upper limits at 95\%
  confidence level are set on the bulk graviton production cross
  section in the range from 700 to 10\unit{fb} for resonance masses between 600 and 2500\GeV,
  respectively. These limits on the bulk graviton model are the most stringent
  to date in the diboson final state.
 }

\hypersetup{%
pdfauthor={CMS Collaboration},%
pdftitle={Search for massive resonances decaying into pairs of boosted bosons in semi-leptonic final states at sqrt(s) = 8 TeV},%
pdfsubject={CMS},%
pdfkeywords={LHC, CMS, new physics, diboson resonances, boosted topology, jet substructure}}

\maketitle 

\section{Introduction}

The standard model (SM) of particle physics
has been very successful
in describing the high-energy physics phenomena investigated so far. One of the predictions of the SM is the existence
of a scalar particle, known as the Higgs boson, associated with the
spontaneous breaking of the electroweak (EW) symmetry and responsible
for the masses of the SM particles
\cite{Englert:1964et,Higgs:1964ia,Higgs:1964pj,Guralnik:1964eu,Higgs:1966ev,Kibble:1967sv}.
The recent discovery by the ATLAS and CMS Collaborations of a particle
compatible with the SM predictions for the Higgs boson
provides further verification of the SM
\cite{Chatrchyan201230,CMSHiggsDiscoveryLong,ATLASHiggsDiscovery}.
In view of large loop corrections to the Higgs boson mass, the question arises whether
the measured Higgs boson mass is the result of fine-tuned constants of nature
within the SM or whether new physics at the \TeVns{} scale stabilizes the Higgs field vacuum.
This question can be reformulated in terms of
the large difference between the mass of the Higgs boson
and the Planck scale \MPl, where the
gravitational force is expected to have the same strength as the other
fundamental forces ($\MPl \sim 10^{16}$\TeV ).

In many theoretical extensions of the SM, the spontaneous breaking of the EW
symmetry is associated with new strong dynamics appearing at the \TeVns{}
scale.  For instance, the origin of the new dynamics may be due to new
interactions~\cite{Weinberg:1975gm,Weinberg:1979bn,Susskind:1978ms} or
a composite Higgs
boson~\cite{Kaplan:1983fs,Contino:2006nn,Giudice:2007fh}.
These extensions of the SM predict
the existence
of new resonances coupling to pairs of massive vector bosons (VV,
where $\mathrm{V}=\PW\text{ or }\cPZ$). Results from previous direct searches at CMS
\cite{Chatrchyan:2012kk,Chatrchyan:2012baa,Chatrchyan:2012rva,CMS:EXO11095}
and
ATLAS~\cite{Collaboration:2012iua,Aad:2012vs,Aad:2012nev,Aad:2013wxa},
and from indirect bounds from the EW sector and from flavor physics~\cite{Delaunay:2010dw,Agashe:2003zs}
generally place lower limits on the masses of these VV resonances above
the \TeVns{} scale.

Models extending the number of spatial dimensions are of particular
interest in the attempt to explain the apparently large difference
between the EW and the gravitational scale.  Some of these models predict the
existence of a so-called tower of Kaluza--Klein (KK) excitations of a spin-2
boson, the KK graviton. The WW and ZZ channels are some of the
possible decay modes of the Randall--Sundrum (RS)
graviton~\cite{RS1Graviton1} in warped extra dimension models. The
original RS model (here denoted as RS1) can be extended to
the \textit{bulk graviton} (\Grav) model, which addresses the flavor structure of
the SM through localization of fermions in the warped extra
dimension~\cite{RSGravAgashe1,RSGravFitz1,RSGravAntipin1}.  In this
scenario, coupling of the graviton to light fermions is highly
suppressed and the decays into photons are negligible. On the other
hand, the production of gravitons from gluon fusion and their decays
into a pair of massive gauge bosons can be sizable at hadron colliders.
The model has two free parameters: the mass of the first mode of the
KK bulk graviton, \mG, and the ratio \ktilde, where $k$ is the unknown
curvature scale of the extra dimension, and $\RedMPl \equiv \MPl / \sqrt{8\pi}$ is
the reduced Planck mass.  Previous direct searches set limits on the
cross section times branching fraction for the production of \Grav~as a function of
\mG~\cite{Aad:2012nev,Chatrchyan:2012baa}.
It should be noted that a revised version of the theoretical
calculations has been recently released, superseding the previous
one~\cite{RSGravAgashe1,Oliveira:2014kla}.
With the new calculation, which predicts production cross sections four times 
smaller, previous limits may have to be revised. For example, the lower limit on the 
graviton mass quoted in Ref.~\cite{Chatrchyan:2012baa} is affected, though the experimental 
bounds on the graviton production cross section times branching fraction as a function of mass remain valid. This paper supersedes 
results from Ref.~\cite{Chatrchyan:2012baa} for graviton masses above 600\GeV, 
while the limit on the production cross section for graviton masses below 600\GeV from Ref.~\cite{Chatrchyan:2012baa} 
remains the most stringent CMS result for the final state considered in the reference.

\begin{figure}[t!hb]
\centering
\includegraphics[width=0.45\linewidth]{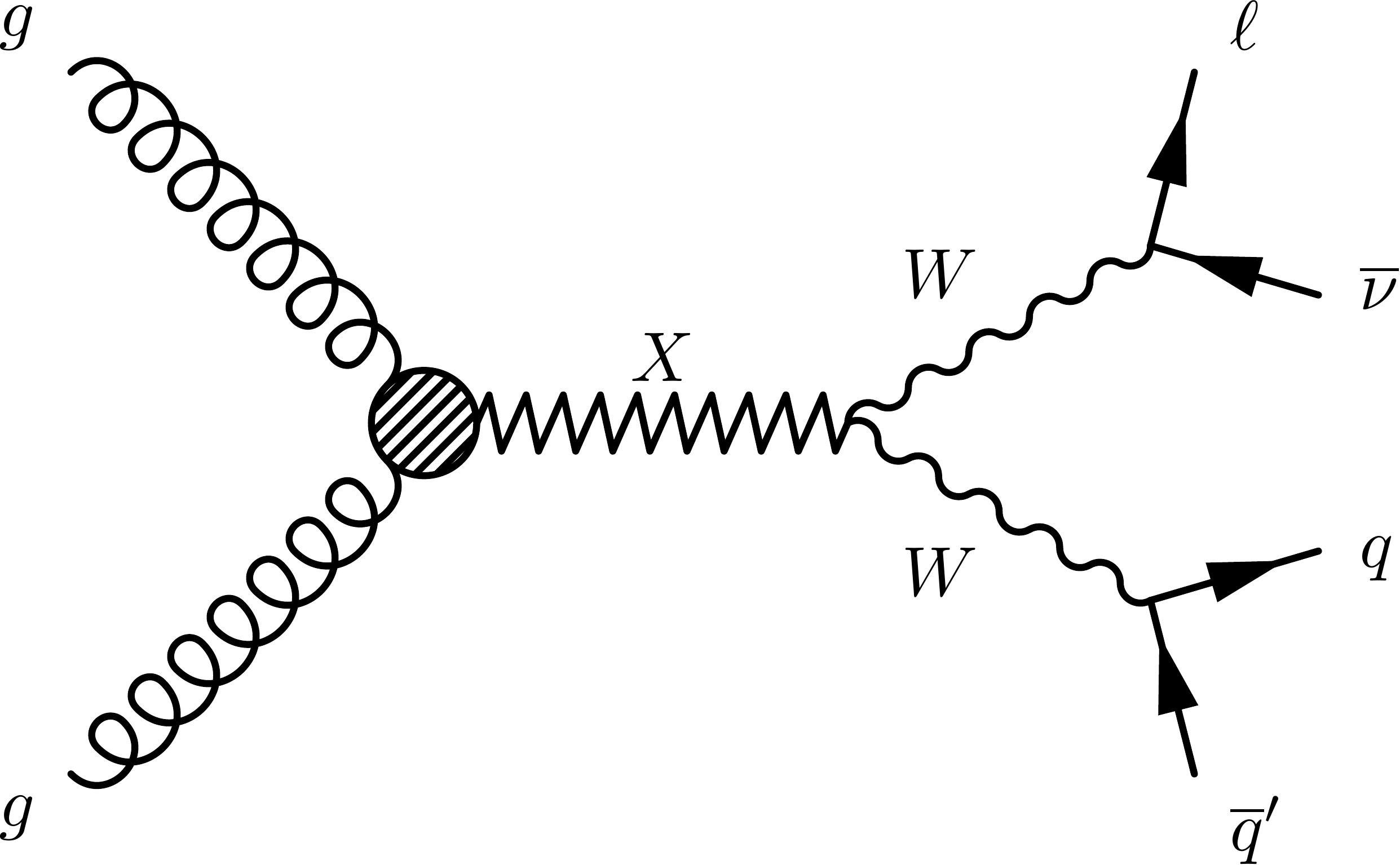}
\includegraphics[width=0.45\linewidth]{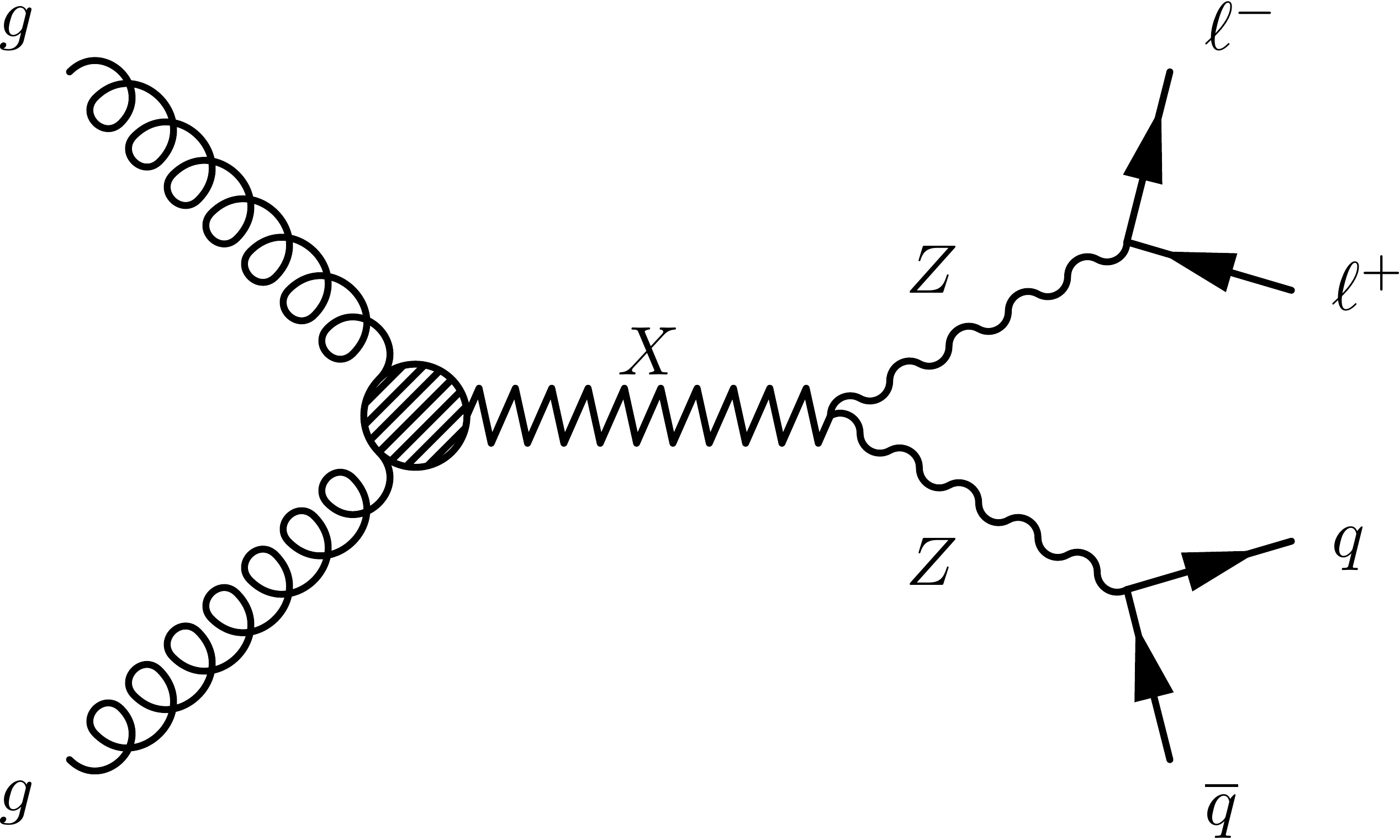}
\caption{Two Feynman diagrams for the production of a generic resonance X
 decaying to some of the final states considered in this study.}
\label{fig:diagram}
\end{figure}

We present a search for new resonances decaying to WW, ZZ, or WZ in which one
of the bosons decays leptonically and the other hadronically.
The analysis is based on the proton-proton
collision data at $\sqrt{s}=8$\TeV collected by the CMS experiment at
the CERN Large Hadron Collider (LHC) during 2012 and corresponding to
an integrated luminosity of \THISLUMI\fbinv.
The final states considered are either
$\ell \nu \qqbarpr$ or $\ell \ell \qqbarpr$,
resulting in events with a charged lepton, a neutrino and
a single reconstructed-jet (\lnujet channel) or
two charged leptons and a single reconstructed-jet (\lljet channel).
Figure~\ref{fig:diagram} shows two Feynman
diagrams relevant to the production and decays of a generic resonance X.
The search is limited to final states where $\ell=\mu$ or e; however the results include the case
in which $\PW  \to \tau\nu$ or $\cPZ  \to \tau\tau$ where
the tau decay is $\tau \to \ell\nu\nu$.
The gain in sensitivity from the decay channels including $\tau$ leptons
is limited, because of the small branching ratios involved.

For large values of the resonance mass, the two quarks originating from
the hadronically decaying W or Z bosons are highly collimated and are
typically reconstructed as a single massive jet (``V jet'').
Final states where two jets from a V decay are well resolved in the detector give a
negligible contribution to the sensitivity for the resonance masses considered in this search.
This analysis uses the additional information from jet substructure to perform jet
``V tagging'' and to further suppress the SM background, which mainly
originates from the SM production of V + jets and non-resonant VV
events~\cite{CMS:JME13006}. In the \lnujet channel \ttbar events also contribute to the background.
The signal is
characterized as a local enhancement in the WW, ZZ, or WZ invariant mass
distribution ($m_{\mathrm{VV}}$).  The invariant mass of the WW system
is determined by estimating
the neutrino transverse momentum with the measured
missing transverse energy (\ETmiss) in the event, while an estimate of the
neutrino longitudinal momentum is derived by imposing the
constraint of the W mass on the invariant mass of the $\ell\nu$
system.  The mass distributions for the dominant W+jets and Z+jets
backgrounds are determined from events with a reconstructed jet mass
not compatible with the W or Z hypothesis.  This analysis is optimized for WW
and ZZ resonances, but  because of the loose requirement on the V-jet
mass it is also sensitive to charged
resonances decaying to WZ.

The results of this analysis are combined with limits derived in a companion CMS search for resonances decaying to VV final states in the all-hadronic decay channel~\cite{CMS:EXO12024}. The all-hadronic analysis uses the same V-tagging techniques as presented here to separate the signal from the large multijet background.

In this paper, Section~\ref{sec:detector} briefly describes the CMS detector; Section~\ref{sec:samples} gives an overview of the simulations used in this analysis. Section~\ref{sec:reco} provides a detailed description of the reconstruction and event selection. In Section~\ref{sec:ttbarcontrol} we demonstrate the performance of the V tagging by studying a sample of events enriched in top quarks. Section~\ref{sec:sigbkg} describes the background estimation and the signal modeling. Systematic uncertainties are discussed in Section~\ref{sec:sysunc}. The results of the search for a bulk graviton and for generic resonances are presented in Section~\ref{sec:results}.
Appendix~\ref{sec:mod-indep-instr} contains detailed instructions
for applying the results presented here to new models with diboson resonances.

\section{The CMS detector}
\label{sec:detector}
The central feature of the CMS detector is a 3.8\unit{T} superconducting
solenoid with a 6\unit{m} internal diameter.  Within the field volume are
the silicon tracker, the crystal electromagnetic calorimeter (ECAL),
and the brass and scintillator hadron calorimeter (HCAL). The
calorimeters are supplemented by a steel/quartz-fiber Cherenkov
detector (HF) to extend the calorimetric coverage in the forward
direction.  The muon system is installed outside the solenoid and
embedded in the steel flux-return yoke.
The CMS experiment uses a right-handed coordinate system, with the origin at the nominal
collision point, the $x$-axis pointing to the center of the LHC ring,
the $y$-axis pointing up (perpendicular to the LHC plane), and the
$z$-axis along the anticlockwise beam direction. The polar angle
($\theta$) is measured from the positive $z$-axis and the azimuthal
angle ($\phi$) is measured from the positive $x$-axis in the
$x$--$y$ plane. The radius ($r$) denotes the distance from the
$z$-axis and the pseudorapidity ($\eta$) is defined as $\eta = -\ln[\tan(\theta / 2)]$.
The CMS tracker consists of 1440 silicon pixel and 15\,148 silicon strip detector modules.  The
ECAL consists of nearly 76\,000 lead tungstate crystals, which provide
coverage in pseudorapidity $\abs{\eta}< 1.479$ in the central barrel region and $1.479 <\abs{\eta} < 3.0$ in the two forward endcap regions.
The HCAL consists of a sampling calorimeter which utilizes alternating layers
of brass as an absorber and plastic scintillator as an active material, covering the pseudorapidity range $\abs{\eta}< 3$, which is extended to $\abs{\eta}< 5$ in combination with the HF.
Muons are measured in the pseudorapidity range $\abs{\eta}< 2.4$,
with detection planes
which employ
three technologies: drift tubes,
cathode strip chambers, and resistive plate chambers.
A detailed description of the CMS detector can be found in Ref.~\cite{Chatrchyan:2008aa}.

\section{Simulated samples}
\label{sec:samples}
The W+jets and Z+jets SM processes are simulated with
\MADGRAPH v5.1.3.30~\cite{Alwall:2011uj},
\ttbar and single top quark events are generated with
\POWHEG 1.0 r1380~\cite{Nason:2004rx,Frixione:2007vw,Alioli:2010xd,Alioli:2009je,Re:2010bp,Alioli:2011as}, while diboson (WW, WZ, and ZZ) processes are produced
with \PYTHIA v6.424 \cite{Sjostrand:2006za}. The parton showering and
hadronization are performed with \PYTHIA using the Z2*
tune~\cite{Chatrchyan:2011id}.
The CTEQ6L~\cite{Pumplin:2002vw} parton distribution functions (PDF)
are used in all generated samples, except for the \POWHEG \ttbar
sample, where the CT10 PDF set~\cite{Lai:2010vv}  is used. All generated
samples are processed through a
\GEANTfour-based~\cite{Agostinelli:2002hh} simulation of the CMS
detector. The simulated background samples are normalized using
inclusive cross sections calculated at next-to-leading order (NLO), or
next-to-next-to-leading order (NNLO) where available, calculated with
\textsc{mcfm} v6.6 \cite{MCFM:VJets,MCFM:VV,MCFM:TT,MCFM:SingleTop} and
\textsc{fewz} v3.1 \cite{FEWZ3}.

The bulk graviton model is used as a benchmark signal process, with the graviton forced to decay to the WW and ZZ final states. In this specific
model, the vector gauge bosons are  produced
with a longitudinal polarization ($\text{V}_{\text{L}}$)
in more than 99\% of the cases.
The graviton masses considered lie in the range \MINMWWMASS to 2500\GeV.
The events are generated with \JHUGEN v3.1.8~\cite{Gao:2010qx}, which
properly treats the spin correlations in the final state,
while the values for the bulk graviton cross sections and decay rates
are calculated at leading order with \textsc{CalcHEP} v3.4.1~\cite{Belyaev:2012qa}.
The total cross section of the process $\Pp\Pp\to\Grav$ at $\sqrt{s}=8$\TeV is 15.1\unit{fb},
for a graviton mass of 1\TeV and $\ktilde=0.5$. At the same resonance mass,
the branching fraction of $\Grav\to\PW\PW$ ($\Grav\to \cPZ\cPZ$) is 18.7\% (9.5\%).

Supplementary minimum bias interactions are added to the generated
events in order to match the additional particle production observed in data
from the large number of proton-proton interactions occurring per LHC bunch
crossing (pileup).
The simulated samples are corrected for observed
differences between data and simulation in the efficiencies of lepton
trigger, lepton identification/isolation, and selection criteria
identifying jets originating from hadronization of b quarks (b jets).

\section{Reconstruction and selection of events}
\label{sec:reco}

\subsection{Trigger and basic offline selection}
In the \lnujet channel, candidate signal events are selected online
with a trigger requiring either one muon or one electron, without
isolation requirements and with loose identification criteria.  The
transverse momentum (\pt) measured online must be higher than
40\GeV for the muons while the minimum transverse energy threshold is
80\GeV for the electrons. The trigger efficiencies for the single-muon
trigger vary between 82\% and 94\% depending on the value of the $\eta$ of
the muon.  The efficiency is above 98\% for the single-electron
trigger.

In the \lljet channel, events are selected online with a
trigger requiring either two muons or two electromagnetic
energy deposits, with loose identification criteria.
The trigger used for the electron
channel
rejects candidates if there is significant energy in the HCAL associated with the ECAL cluster.
The clusters are required to be loosely matched to the trajectories of
tracks with associated hits in the pixel detector.
No lepton isolation requirements are applied at the
trigger level. The \pt thresholds applied in the online selection of
the muons are 22\GeV for the highest-\pt muon and 8\GeV for the
second highest-\pt muon.
The transverse energy threshold for the ECAL clusters is set
at 33\GeV. The trigger efficiency of the double-muon
trigger varies between 80\% and 98\% depending on the value of $\eta$
of the leptons.  The efficiency of the double-electron trigger is
above 99\%.

Offline, all events are required to have at least one primary vertex
reconstructed within a 24\unit{cm} window along the beam axis, with a
transverse distance from the nominal pp interaction region of less
than 2\unit{cm}~\cite{CMS:TRK10005}. In the presence of more than one vertex
passing these requirements, the primary-event vertex is chosen to be the one
with the highest total $\pt^{2}$, summed over all the associated
tracks.

\subsection{Muon reconstruction and selection}

\textit{Tracker} muons are reconstructed using the inner tracker with an additional requirement of
a matching hit in the muon system~\cite{CMS:MUO10004}. Tracker muons must satisfy requirements
on the impact parameter of the track and on the number of hits in the silicon tracker.
Muons reconstructed with a fit using both the inner tracking system and the
muon spectrometer are defined as \textit{global}
muons~\cite{CMS:MUO10004}. Compared to tracker muons, global muons must pass additional
requirements on the number of hits in the muon detectors.
These quality selections ensure a precise measurement of the four-momentum and reject misreconstructed
muons. A large fraction of isolated high-\pt muons is usually identified as both tracker and global muons.
For large values of the mass of a ZZ resonance, the two
charged leptons originating from the high-\pt Z boson are highly
collimated because of the large Lorentz boost and  are characterized
by small values of their angular separation, $\Delta R =
\sqrt{\smash[b]{(\Delta\eta)^2 + (\Delta\phi)^2}}$.
While the global muon
reconstruction and identification are optimized for the case of
well-separated muons, inefficiencies in the global-muon reconstruction are observed when two muons from a
boosted Z are very close, typically causing the loss of one of them.
In order to recover the inefficiency in the muon identification,
the \lljet selection requires two tracker muons of which at least one
should be reconstructed and identified as a global muon.
Wherever possible, the kinematic quantities are calculated with the global fit.

An isolation requirement is applied in order to suppress the
background from multijet events where jet constituents are identified
as muons.  A cone of radius $\Delta R= 0.3$ is constructed around the
muon direction.  The isolation parameter is defined as the scalar sum
of the transverse momenta of all the additional reconstructed tracks
within the cone, divided by the muon \pt. The contribution from any
other muon candidate in the cone is excluded from the computation in
order to retain high signal efficiency when the two muons originate
from a boosted \Zo and are collimated to the point of entering in
each other's isolation cone.  Muon candidates with an isolation
parameter smaller than 0.1 are considered isolated and used in the
rest of the analysis. The efficiency of this muon selection has been
measured with a tag-and-probe method using \Zo bosons
\cite{CMS:FirstInclZ}, and it has a negligible dependence on the
number of reconstructed primary vertices in the event.
In the \lljet channel, events must have at least two muons
with $\abs{\eta} <2.4$ of which one should have $\pt > 40$\GeV
and the other $\pt > 20$\GeV.
In the \lnujet channel, we require exactly one
global muon with $\pt > 50$\GeV and $\abs{\eta} < 2.1$.

\subsection{Electron reconstruction and selection}

Electron candidates are reconstructed by matching energy deposits in
the ECAL with reconstructed tracks
\cite{Chatrchyan:2013dga}. In order to suppress multijet background,
electron candidates must pass stringent quality criteria tuned for
high-\pt objects and an isolation selection
~\cite{Chatrchyan:2012meb}.  The total scalar sum of the \pt of all
the tracks in a cone of radius $\Delta R = 0.3$ around the electron direction,
excluding tracks within an inner cone of $\Delta R = 0.04$ to remove the contribution
from the electron itself, must be less than 5\GeV.  A calorimetric isolation parameter is
 calculated by summing the energies of reconstructed deposits in both ECAL and HCAL, not associated with the electron
itself, within a cone of radius
$\Delta R = 0.3$ around the electron. The upper threshold for this
isolation parameter depends on the electron kinematic quantities and the average
amount of additional energy coming from pileup interactions.  When
evaluating the isolation parameter in the \lljet channel, the
contribution from any nearby electron candidate is excluded from the
calculation.  This is done in order to retain high signal efficiency
 when the two leptons from a \Zo decay are highly boosted and one of
them enters into the isolation cone of the other.
In the \lljet channel, we require at least two electrons with
$\pt > 40$\GeV and $\abs{\eta} < 2.5$. In the \lnujet channel, we
require exactly one electron with $\pt > 90$\GeV and $\abs{\eta} < 2.5$.
In both channels, the electrons must fall outside the overlap region between the ECAL
barrel and endcaps ($1.44 < \abs{\eta} <1.56$).

\subsection{Jets and missing transverse energy reconstruction}

Hadronic jets are clustered from the four-momenta of the particles reconstructed by
the CMS particle-flow (PF) algorithm~\cite{CMS-PAS-PFT-09-001,CMS-PAS-PFT-10-001},
using the \textsc{FastJet} software package~\cite{Cacciari:2011ma}.  The PF algorithm
reconstructs individual particles by combining information from all sub-detector
systems. The reconstructed PF constituents are assigned to one of the five candidate
categories (electrons, muons, photons, charged hadrons, and neutral hadrons).
In the jet clustering procedure charged PF particles not associated with the primary-event vertex are excluded.
Jets used for identifying the hadronically decaying W and \Zo bosons
are clustered using the Cambridge--Aachen algorithm \cite{Wobisch:1998wt} with a
distance parameter $R = 0.8$ (``CA8 jets'').
In order to identify b jets, the anti-\kt jet clustering algorithm is
used~\cite{Cacciari:2008gp} with a distance parameter $R = 0.5$ (``AK5
jets'') and the combined secondary vertex b-tagging algorithm
\cite{CMS:BTAG7TeV} is applied to the reconstructed AK5 jets.
The ratio of the b-tagging efficiency between data and simulation is
used as a scale factor to correct the simulated events.
A correction based on the projected area of the jet on the front face of
the calorimeter
is used to take into account the extra energy
clustered in jets due to neutral particles coming from pileup.
Jet energy corrections are derived from
simulation and from dijet and photon+jet events in data~\cite{CMS:JetCalibration}.
Additional quality criteria are applied to the jets in order to remove spurious
jet-like features originating from isolated noise patterns in the calorimeters or the
tracker. The efficiency of these jet quality requirements for signal events is above
99\%.  The CA8 (AK5) jets are required to be separated from any well-identified
electron or muon by $\Delta R>0.8$ (0.3). All jets must have $\pt>30\GeV$ and
$\abs{\eta}<2.4$ in order to be considered in the subsequent steps of the analysis.

The missing transverse energy \ETmiss is defined as the magnitude
of the vector sum of the transverse momenta of the reconstructed
PF objects.
The raw \ETmiss value is modified to account for corrections to the
energy scale of all the reconstructed AK5 jets in the event. More details on the
\ETmiss performance in CMS can be found in Refs.~\cite{CMS:METperformances,CMS-PAS-JME-12-002}.
The requirement $\ETmiss > 40~(80)\GeV$ is applied only for the
muon (electron) channel in the \lnujet analysis. The threshold is
higher in the electron channel to further suppress the larger background from
 multijet processes.

\subsection{\texorpdfstring{$\PW\to \ell \nu$ and $\cPZ\to \ell \ell$}{W to l nu and Z to ll} reconstruction and identification}
\label{subsec:Vlept}
In the \lnujet channel, identified electrons or muons are associated with the
$\PW \to \ell \nu$ candidate.
The transverse momentum of the undetected neutrino is assumed to be equal
to the \ETmiss.
The longitudinal momentum of the neutrino is obtained by solving a
second-order equation that sets the $\ell\nu$ invariant mass to be
equal to the known W-boson mass \cite{Beringer:1900zz}. In the case of two
real solutions, the smaller one is chosen; in the case of two complex
solutions, their real part is used.  The four-momentum of the
neutrino is used to build the four-momentum of the $\PW \to \ell \nu$
candidate. The same procedure is applied for $\PW \to \tau \nu$
candidates, where the $\tau$ decays to one electron or muon and two neutrinos. In this case,
the \ETmiss represents the transverse momentum of the three-neutrino system.

In the \lljet channel, the leptonic \Zo-boson candidate is reconstructed
by combining two oppositely charged lepton candidates of the same
flavor. The invariant mass of the dilepton system is required to be
between 70 and 110\GeV, consistent with the \Zo-boson mass. This
requirement is introduced to reduce significantly the Drell--Yan and top-quark backgrounds,
at the cost of a suppression of the small $\cPZ \to \tau\tau \to (\ell \nu \nu)\:(\ell \nu \nu) $ signal contribution.

\subsection{\texorpdfstring{$\PW\to \Pq\Paq'$ and $\cPZ\to \Pq\Paq$}{W to q anti-q' and Z to q anti-q} identification using jet substructure}
\label{subsec:Vhadr}

CA8 jets are used to reconstruct the W-jet and Z-jet candidates from
hadronic decays of boosted W and \Zo bosons, respectively.  In order to
discriminate against multijet backgrounds we exploit both the reconstructed
jet mass, which is required to be close to the W- or \Zo-boson mass, and
the two-prong jet substructure produced
by the particle cascades of two high-\pt quarks merging into one jet.

As the first step in exploring potential substructure,
the jet constituents are subjected to a jet grooming
algorithm, that improves the resolution on the jet
mass and reduces the effect of pileup
\cite{CMS:SMP12019}. The goal of jet grooming is to re-cluster the jet
constituents while applying additional requirements that eliminate soft,
large-angle quantum chromodynamic (QCD) radiation coming from sources other than the hard
interaction responsible for the V boson. Different jet grooming
algorithms have been explored at CMS and their performance on jets in
multijet processes has been studied in detail \cite{CMS:SMP12019}. In this
analysis, we use the \textit{jet pruning} algorithm
\cite{jetpruning1,Ellis:2009me}.
Jet pruning reclusters each jet starting
from all its original constituents using the CA algorithm,
discarding ``soft'' recombinations in each step of the iterative CA procedure. The combination of two input four-vectors $i$ and $j$ is
considered soft if either (i) $\pt^i$ or $\pt^j$ is small compared to the \pt of their combination, or
 (ii) the separation angle between $i$ and $j$ is large.
With $\tilde{\pt}$ the transverse momentum of the result of
the recombination of $i$ and $j$, the two possible tags of a soft recombination are expressed as (i)
$\min(\pt^{i},\pt^{j}) / \tilde{\pt} < 0.1$
and (ii) $\Delta R_{ij}>m^{\text{orig}}/\pt^{\text{orig}}$,
with $m^{\text{orig}}$ and $\pt^{\text{orig}}$ representing the
mass and \pt of the original un-pruned CA jet. Soft recombinations are
rejected, in which case the input four-vector with the smallest \pt is discarded and the input four-vector
with the highest \pt is retained for further recombinations.
A jet is considered as a W-jet candidate if its pruned mass, \mJ,
computed from the sum of the four-momenta of the constituents
surviving the pruning, falls in the range $65<\mJ<105\GeV$. Similarly,
a Z-jet candidate is required to have $70<\mJ<110\GeV$.

Further discrimination against jets from gluon and single-quark hadronization
is obtained from the quantity called \textit{N-subjettiness}~\cite{Thaler:2010tr}. The constituents of the jet
before the pruning procedure are re-clustered with the $\kt$
algorithm~\cite{Catani:1993hr,Ellis:1993tq}, until N joint objects (\textit{subjets}) remain in the iterative combination procedure of the $\kt$
algorithm. The N-subjettiness, $\tau_{N}$, is then defined as
\begin{equation}
\tau_N = \frac{1}{d_{0}} \sum_{k} p_{\mathrm{T},k} \min( \Delta R_{1,k}, \Delta R_{2,k},\ldots,\Delta R_{N,k}),
\end{equation}
where the index $k$ runs over the PF constituents of the jet and the distances
$\Delta R_{n,k}$ are calculated with respect to the axis of the
$n$th subjet. The normalization factor $d_{0}$ is
calculated as $d_{0}=\sum_{k} p_{\mathrm{T},k}R_{0}$, setting $R_{0}$ to the
jet radius of the original jet. The variable $\tau_{N}$ quantifies the
capability of clustering the jet constituents in exactly $N$ subjets,
with small values representing configurations more compatible
with the $N$-subjets hypothesis.
The ratio between 2-subjettiness and
1-subjettiness, $\nsubj=\tau_{2}/\tau_{1}$, is found to be a powerful
discriminant between jets originating from hadronic V decays and from gluon and single-quark hadronization.
We reject V-jet candidates with $\nsubj>0.75$. The remaining
events are further categorized according to their value of \nsubj~in
order to enhance the sensitivity of the analysis. Jets coming from
hadronic W or \Zo decays in signal events are characterized by lower
values of $\tau_{21}$ compared to the SM backgrounds.

\subsection{Final event selection and categorization}

After reconstructing the two vector bosons, we apply the final
selections used
for the search. In the \lnujet
(\lljet) channel, both the leptonic and the hadronic V-boson
candidates must have a \pt greater than 200 (80)\GeV.
The larger threshold for the \lnujet channel is related to the higher
trigger thresholds and the larger multijet background in this category of events.
In addition, there are specific topological selection criteria in the \lnujet
channel requiring that the two W bosons from the decay of a massive resonance are approximately back-to-back:
the $\Delta R$ distance between the lepton and the W-jet
is greater than $\pi/2$;
the azimuthal angular separation between the missing transverse energy vector and the W-jet
is greater than 2.0 radians; and the azimuthal angular separation between the $\PW \to \ell \nu$
and W-jet candidates
is greater than 2.0 radians.
To further reduce the level of the \ttbar background in the \lnujet channel, events are
rejected  if there is one or more b-tagged AK5 jet in the event, using a working
point of the b-tagging algorithm tuned to provide a misidentification rate of $\sim$1\%
and efficiency of $\sim$70\%. This veto preserves about 90\% of signal events.
The looser selections in the \lljet channel allow the extension of the range of probed
masses to lower values. The minimum value of $m_{\mathrm{VV}}$ is 700 (500)\GeV
for the \lnujet (\lljet) channel, respectively.

To enhance the analysis sensitivity, we distinguish two V-jet
categories:
\begin{itemize}
\item high-purity (HP) category: $\tau_{21}\leq0.5$;
\item low-purity (LP) category: $0.5<\tau_{21}<0.75$.
\end{itemize}

Although it is expected that the HP category dominates the total
sensitivity of the analysis, the LP is retained, since for large
masses of a new resonance it provides improved signal efficiency with
only moderate background contamination.
 The final categorization is
based on four classes of events, depending on their lepton flavor
(muon or electron) and V-jet purity (LP and HP).  In case several
distinct diboson resonance candidates are present in the same event,
only one is kept for further analysis. Diboson pairs in the HP
category are preferred to those in the LP category and, in case
multiple choices are still possible, the candidate with the V-jet with
the highest \pt is retained. After the final selection no events with multiple leptonic Z candidates remain.

\begin{figure}[htbp]
\centering
\includegraphics[width=0.45\linewidth]{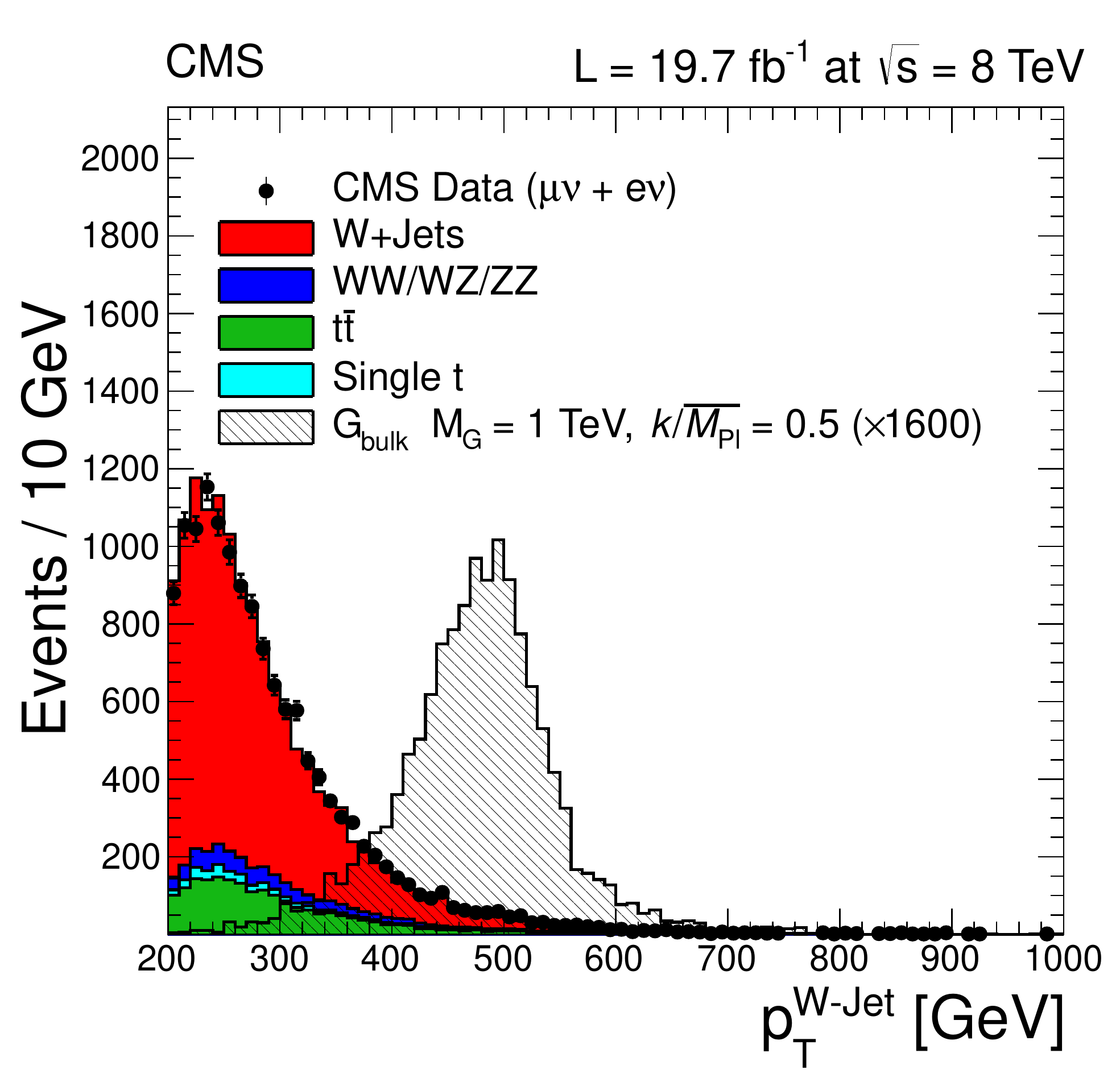}
\includegraphics[width=0.45\linewidth]{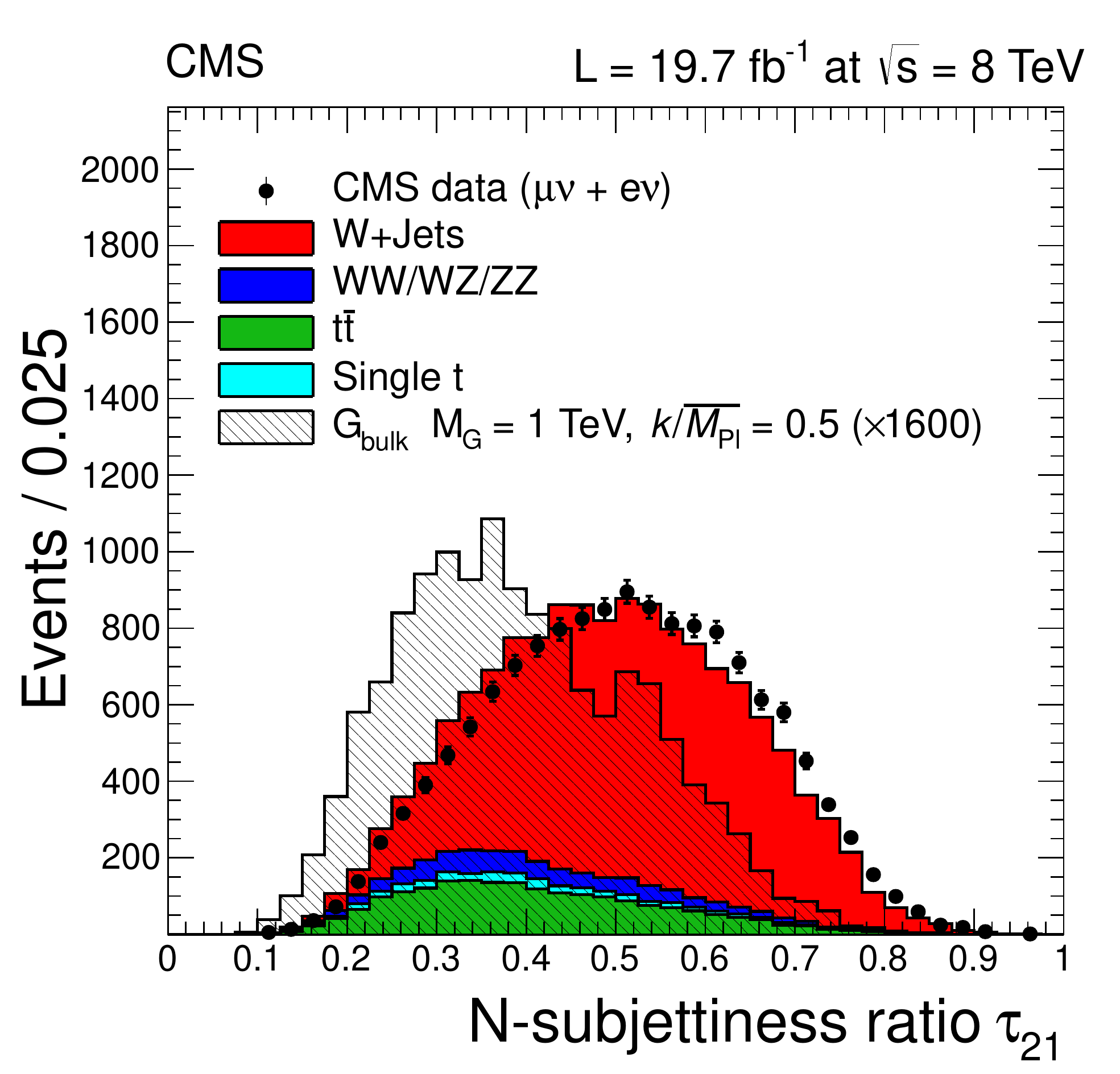}
\caption{Hadronic W \pt and N-subjettiness ratio $\tau_{21}$ distributions for the combined muon and electron channels and with
$65 < \mJ < 105$\GeV. The VV, \ttbar, and single-t backgrounds are taken from simulation and are normalized
to the integrated luminosity of the data sample. The W+jets background is rescaled such that the total number of background
events matches the number of events in data.
The signal is scaled by a factor of 1600 for better visualization.}
\label{fig:KinePlotsW}
\end{figure}

\begin{figure}[htbp]
\centering
\includegraphics[width=0.45\linewidth]{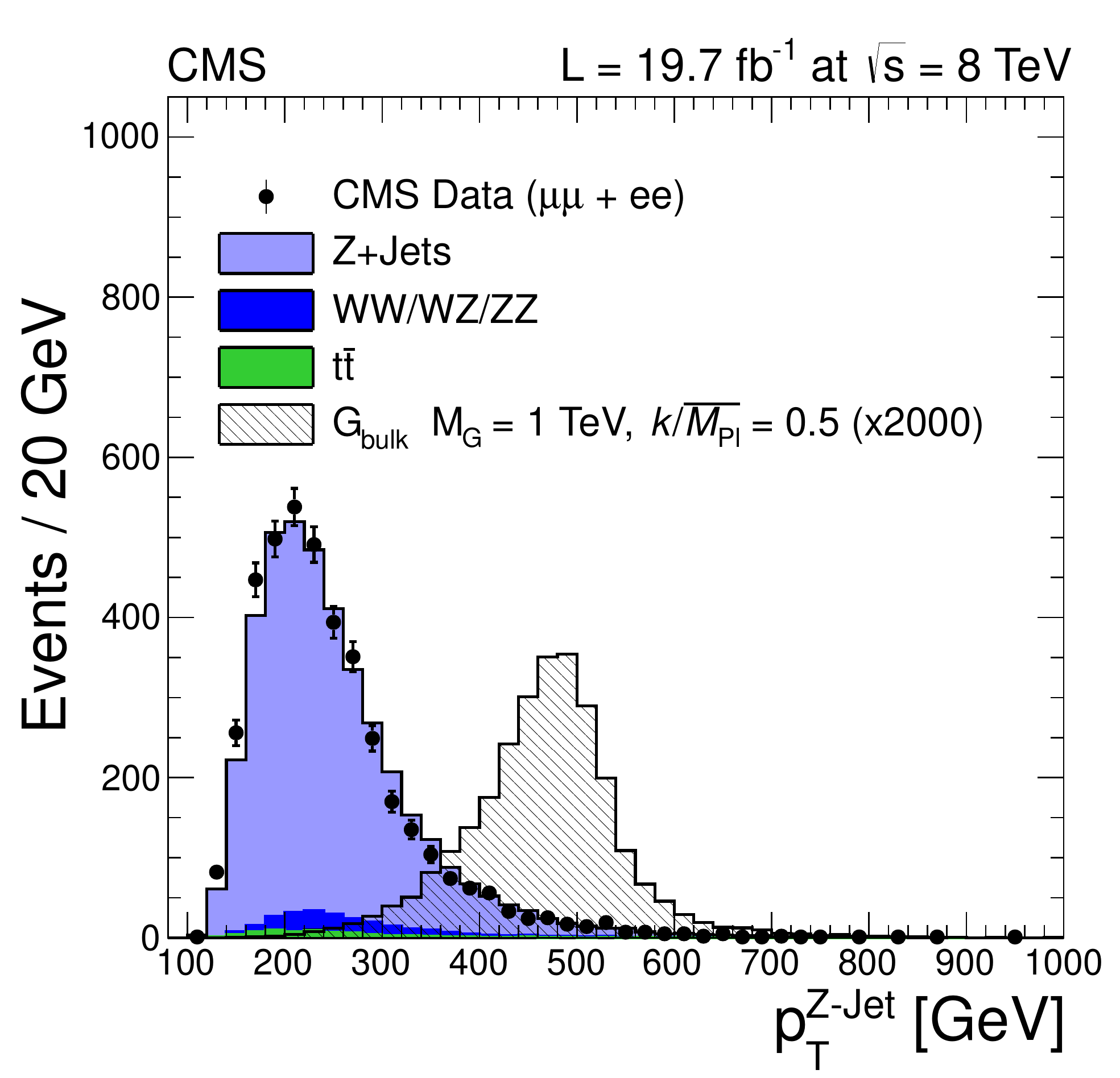}
\includegraphics[width=0.45\linewidth]{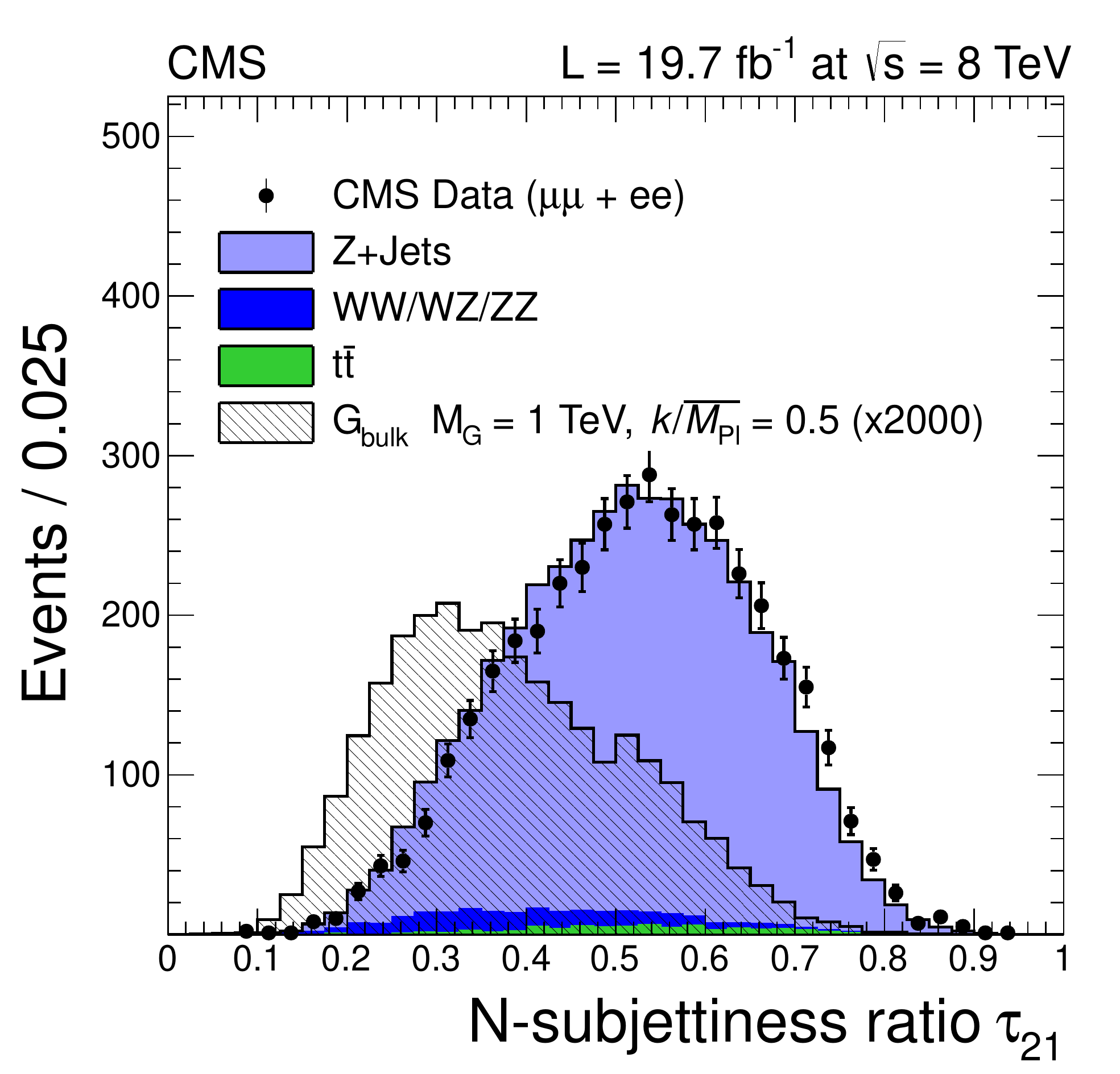}
\caption{Hadronic Z \pt and N-subjettiness ratio $\tau_{21}$ distributions for the combined muon and electron channels and with
$70 < \mJ < 110$\GeV. The VV and \ttbar backgrounds are taken from simulation and are normalized
to the integrated luminosity of the data sample. The Z+jets background is rescaled such that the total number
of background events matches the number of events in data. The signal is scaled by a factor of 2000
for better visualization.}
\label{fig:KinePlotsZ}
\end{figure}

The \pt and $\tau_{21}$ distributions for the hadronic W (Z) boson
candidate after the \lnujet (\lljet) selection are shown in
Fig.~\ref{fig:KinePlotsW} (Fig.~\ref{fig:KinePlotsZ}), after applying
a $65<m_{\text{jet}}<105$\GeV ($70<m_{\text{jet}}<110$\GeV)
requirement.  The $\tau_{21}$ distribution shows some disagreement
between data and simulation.
Previous studies suggest that part of this discrepancy
can be attributed to a mismodeling of the parton showering in simulation~\cite{CMS:JME13006}.
The analysis is designed to be robust against
differences between data and simulation independent of their specific sources,
as described in the next sections.

\section{W tagging in a top-quark enriched control sample}
 \label{sec:ttbarcontrol}

The data/simulation discrepancy observed in the key variable
\nsubj~(Figs.~\ref{fig:KinePlotsW} and \ref{fig:KinePlotsZ}) is of
particular concern as the mismodeling of the variable could
bias the signal efficiency estimated from the simulated samples.  It
is important to study the mismodeling in a signal-free sample with the
characteristics of the jets similar to those expected for a genuine
signal.  In this way one can extract correction factors to apply to
the signal efficiency suggested by the simulation and obtain a small systematic uncertainty related to this effect.
A sample of high-\pt W bosons decaying hadronically, and reconstructed as a single CA8 jet, can be isolated
in \ttbar and single top-quark events. The control sample is selected
by applying all analysis requirements but inverting the b-jet veto.
The data are compared with the predictions from simulation.
Discrepancies between data and simulation are corrected in the
analysis using the scale factors for top-quark background
normalization, V-tagging efficiency, and peak and resolution of the
V-jet mass distribution derived in this section.  Since the jet
substructure produced in simulation depends on the modeling of
the parton shower, \PYTHIA v6.426 is used for this part of the event
simulation. In this way the results of this study can be consistently
applied to the signal MC samples, that are also generated with the
same parton showering.

\begin{figure}[htbp]
\centering
\includegraphics[width=\cmsFigWidth]{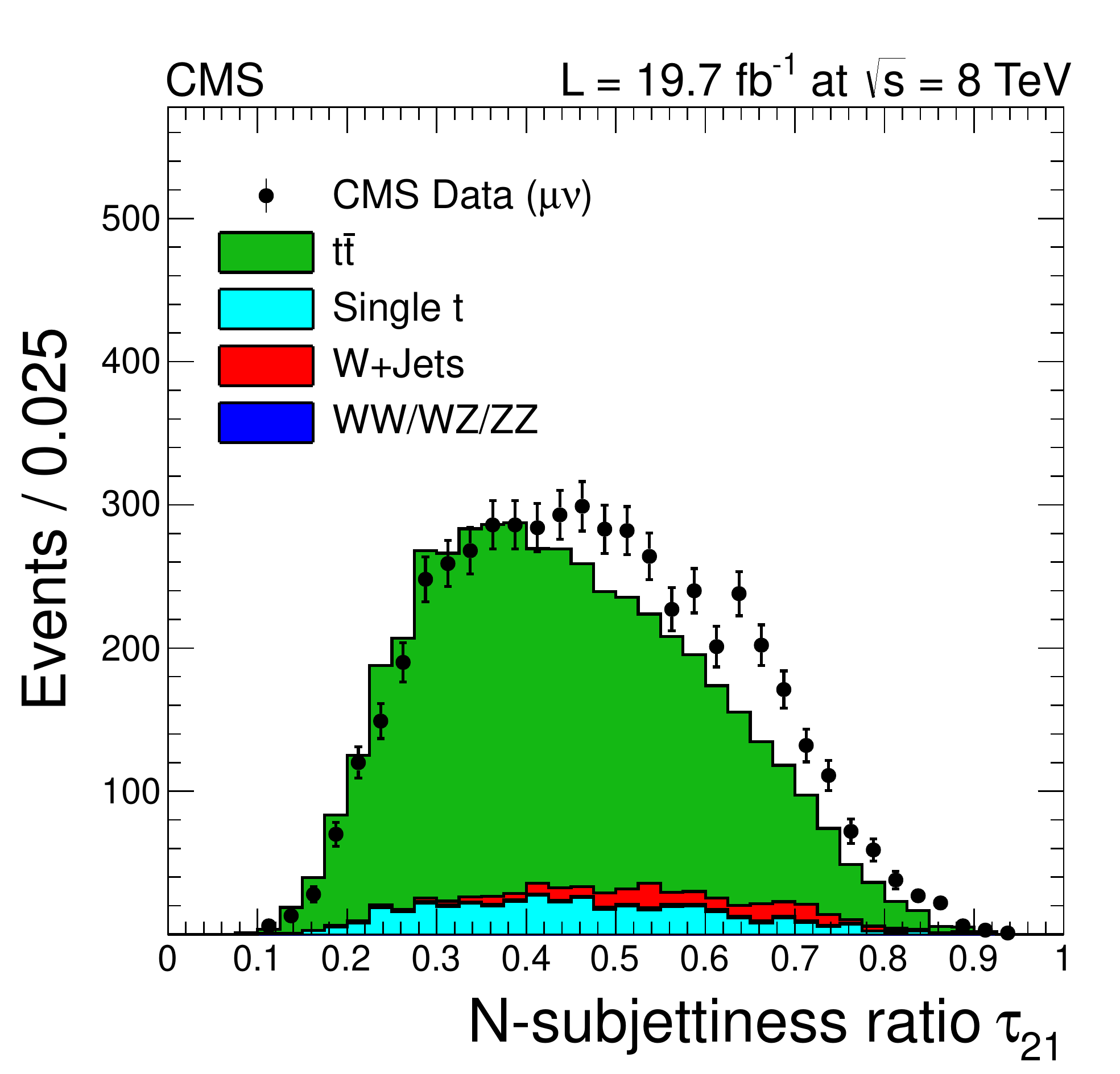}
\includegraphics[width=\cmsFigWidth]{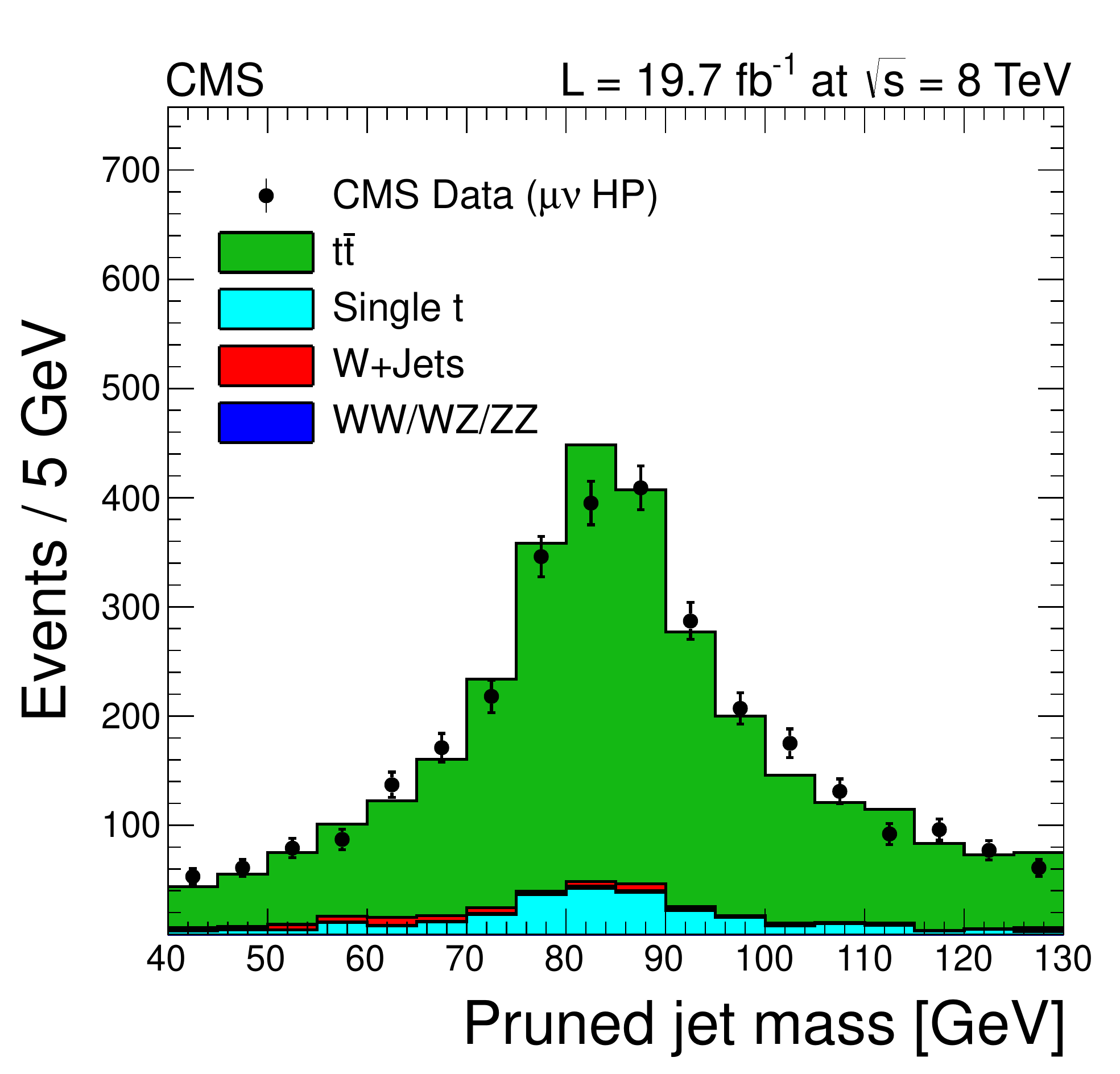}
\caption{
Distributions from the top-quark enriched control sample in the muon channel. 
Left: N-subjettiness ratio $\tau_{21}$, Right: \mJ~after requiring $\tau_{21}<0.5$. 
The distributions show some disagreement between data and simulation. 
The simulation is corrected for these discrepancies using the method based on data 
described in Section~\ref{sec:ttbarcontrol}. 
This approach ensures that the analysis is robust against differences 
between data and simulation, independent of their sources.
}
\label{fig:ttbarControlCut}
\end{figure}

The \nsubj~distribution in the top-quark enriched control sample is
shown in the left plot of Fig.~\ref{fig:ttbarControlCut}, while the
right plot shows the pruned jet mass distribution after applying the HP selection
of $\tau_{21} < 0.5$.  The pruned jet mass plot shows a clear peak for
events with an isolated W boson decaying to hadrons (W-signal component), as well
as a combinatorial component mainly due to events where the extra b jet
from the top-quark decay is in the proximity of the W.  From the
comparison between data and simulation, a normalization correction
factor for \ttbar and single top-quark background processes is evaluated in
the signal region ($65<m_{\text{jet}}<105$\GeV).  The measured
data-to-simulation scale factors are \SFTTBARMUHP (\SFTTBARELEHP) in
the muon (electron) channel for the high-purity category, and
\SFTTBARMULP (\SFTTBARELELP)
for the low-purity category.  These scale factors (including both the
W-signal and the combinatorial components) are used to correct the
normalization of the \ttbar and single top-quark simulated background
predictions in the signal region.

A simultaneous fit to the jet mass distributions, before and after
the \mJ~and \nsubj~requirements, is
performed to separate the W-signal from the combinatorial components
in the top-quark enriched sample, in both data and simulation.
The fit results are used to extract the
efficiencies for identifying an isolated hadronic W boson (W tagging
based on \mJ~and \nsubj~requirements).
Differences in the resulting W-tagging efficiencies will be driven by
the discrepancy between data and simulation in the \nsubj~distribution.  The
ratio of the efficiency in data and simulation yields W-tagging scale
factors that are used to correct the total signal efficiency
predicted by the simulation. The scale factor for W tagging is \SFWTAGHP
(\SFWTAGLP) for the high-purity (low-purity) category, combining the muon
and electron channels.

In addition, the W-jet
mass peak position and resolution
are extracted from the same fit and are measured to be
\WMASSMC and \WRESMC\GeV, respectively, in the simulation and
\WMASSDATA and \WRESDATA\GeV in the data, where the uncertainties
given are statistical only. The mass peak position is slightly shifted with
respect to the W-boson mass because of the presence of extra energy deposited in
the jet cone coming from pileup, underlying event, and initial-state radiation not completely
removed by the jet pruning procedure. For events with top quarks, additional energy
contributions come also from the possible presence of a b jet close to the W-jet candidate.

The same corrections are used also in the case where the V-jet is
assumed to come from a \Zo boson. The kinematic properties of W-jets and
Z-jets are very similar
and the agreement between data and simulation is
expected to be equally good.

\section{Modeling of background and signal}
\label{sec:sigbkg}

\subsection{Background estimation}
\label{sec:bkgd}

After the full selection, the dominant background comes from SM
V+jets events.  A procedure based on data has been developed in order
to estimate this background.  Other minor sources of background, such as
\ttbar, single top-quark, and $\Vo\Vo$ production, are estimated using the simulated samples
after applying correction factors based on control samples in data, as described in the
previous sections.
A signal-depleted control region is defined around the
\mJ~mass window described in Section~\ref{subsec:Vhadr}. For the
\lnujet channel, lower and upper sideband regions are defined in
the \mJ~ranges [40, 65] and [105, 130]\GeV, respectively. In
the \lljet channel, the sidebands are defined in the \mJ~ranges [50, 70] and [110, 130]\GeV.

\begin{figure}[htbp]
\centering
\includegraphics[width=\cmsFigWidth]{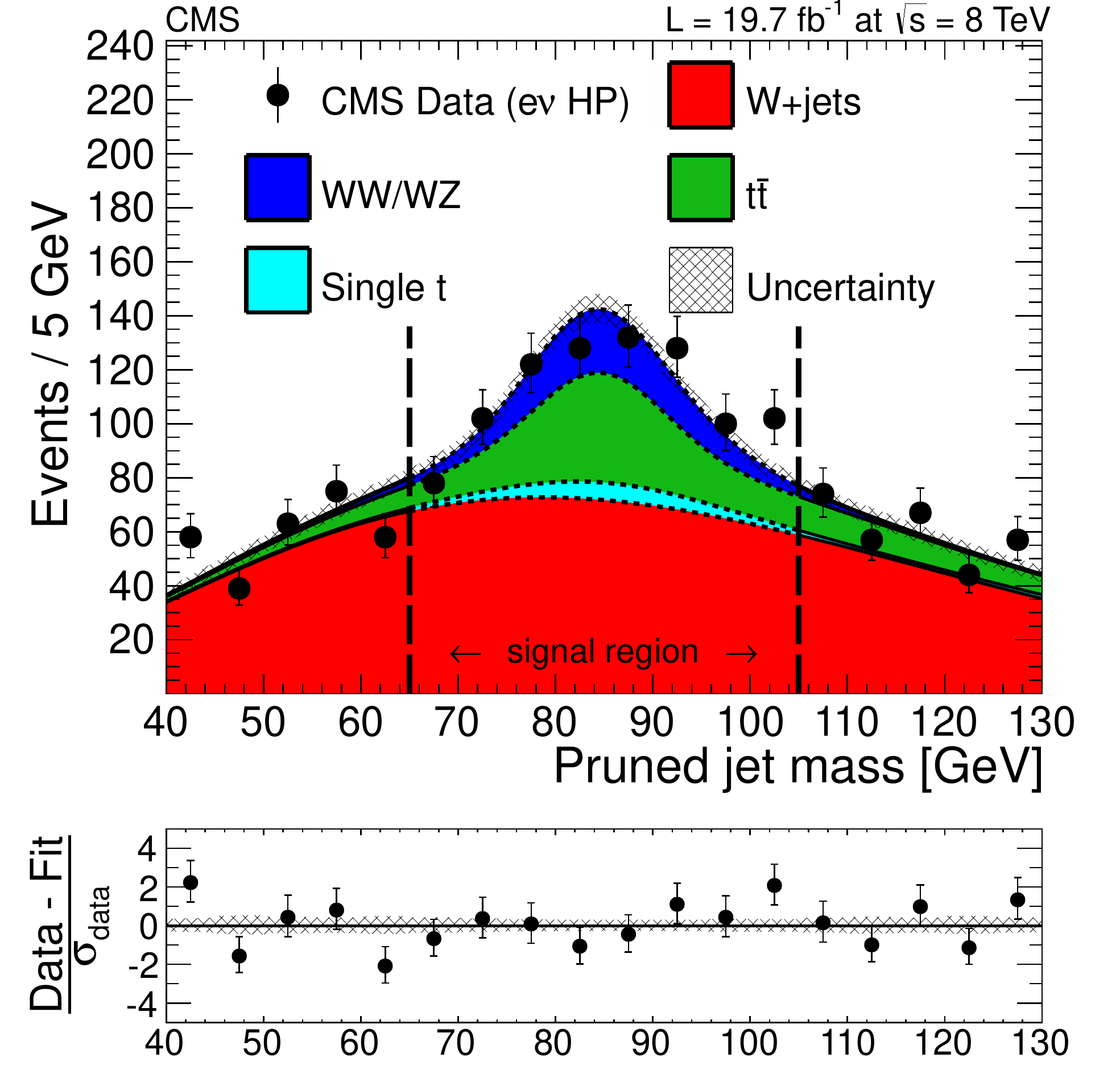}
\includegraphics[width=\cmsFigWidth]{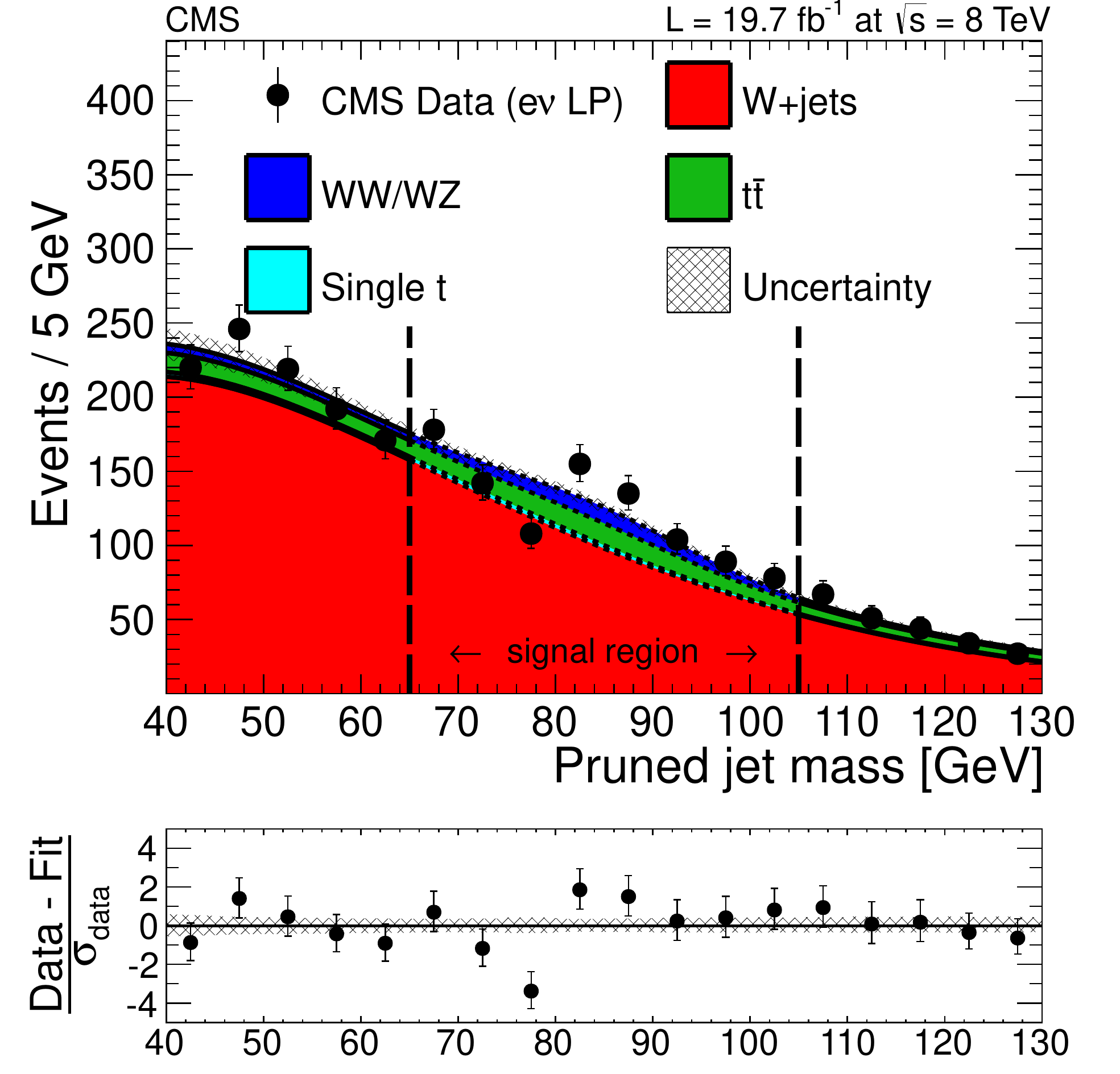}
\caption{Distributions of the pruned jet mass, \mJ, in the \lnujet analysis in the
electron channel. The left (right) panel shows the distribution for the HP (LP)
category. All selections are applied
except the final \mJ~signal window requirement.  Data are shown as black
markers.
The prediction of the non-resonant W+jets
background comes from a fit excluding the signal region (between the vertical dashed lines),
while the predictions for the minor backgrounds come from the simulation.
The MC resonant shapes are corrected using the differences
between data and simulation in the W peak position and width measured
in the \ttbar control region (see Section~\ref{sec:ttbarcontrol}).
At the bottom of each plot, the bin-by-bin fit residuals, (data-fit)/$\sigma_\text{data}$,
are shown together with the uncertainty band of the fit normalized by $\sigma_\text{data}$.}
\label{fig:WJetsNormalization}
\end{figure}

\begin{figure}[htbp]
\centering
\includegraphics[width=\cmsFigWidth]{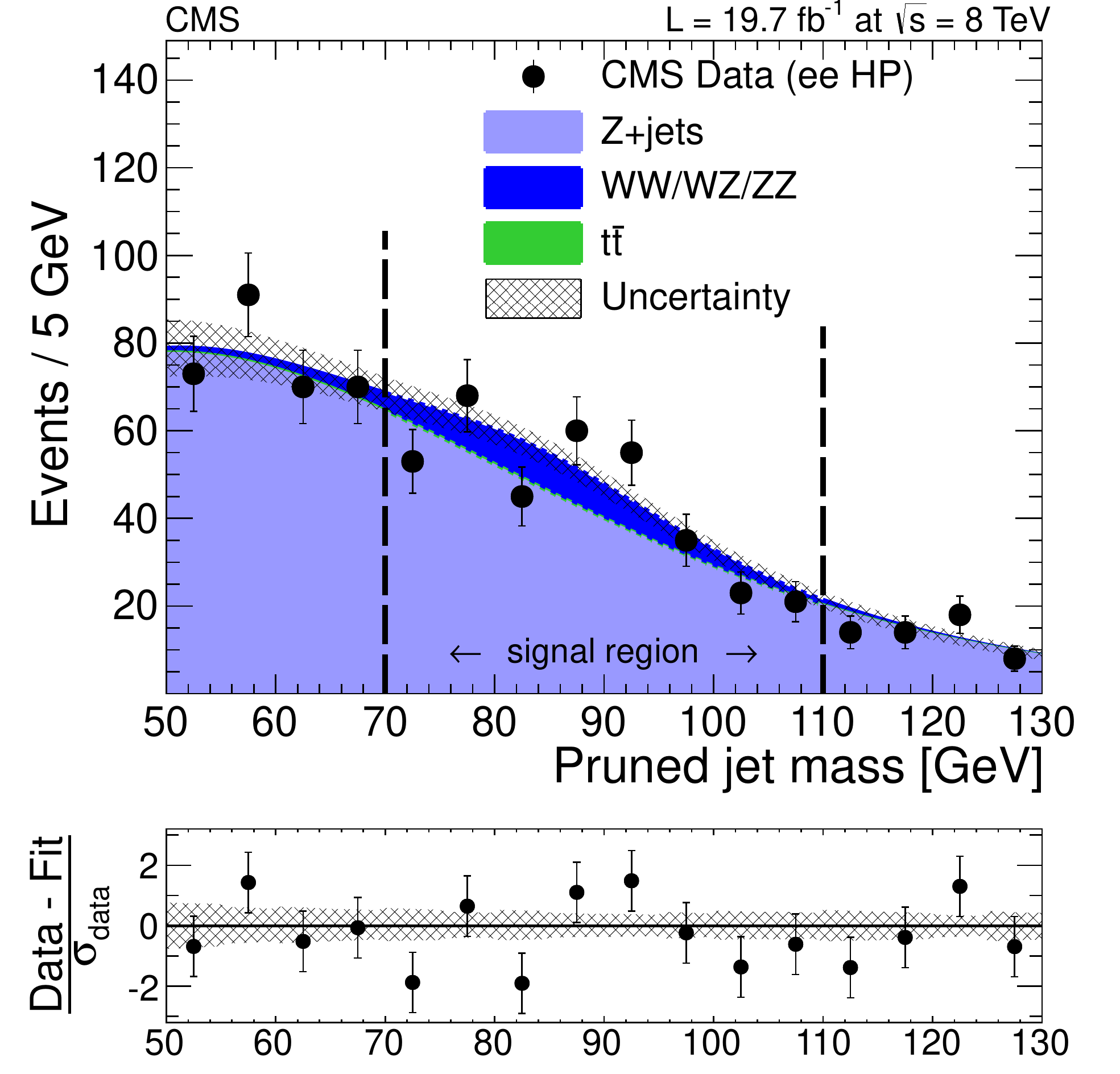}
\includegraphics[width=\cmsFigWidth]{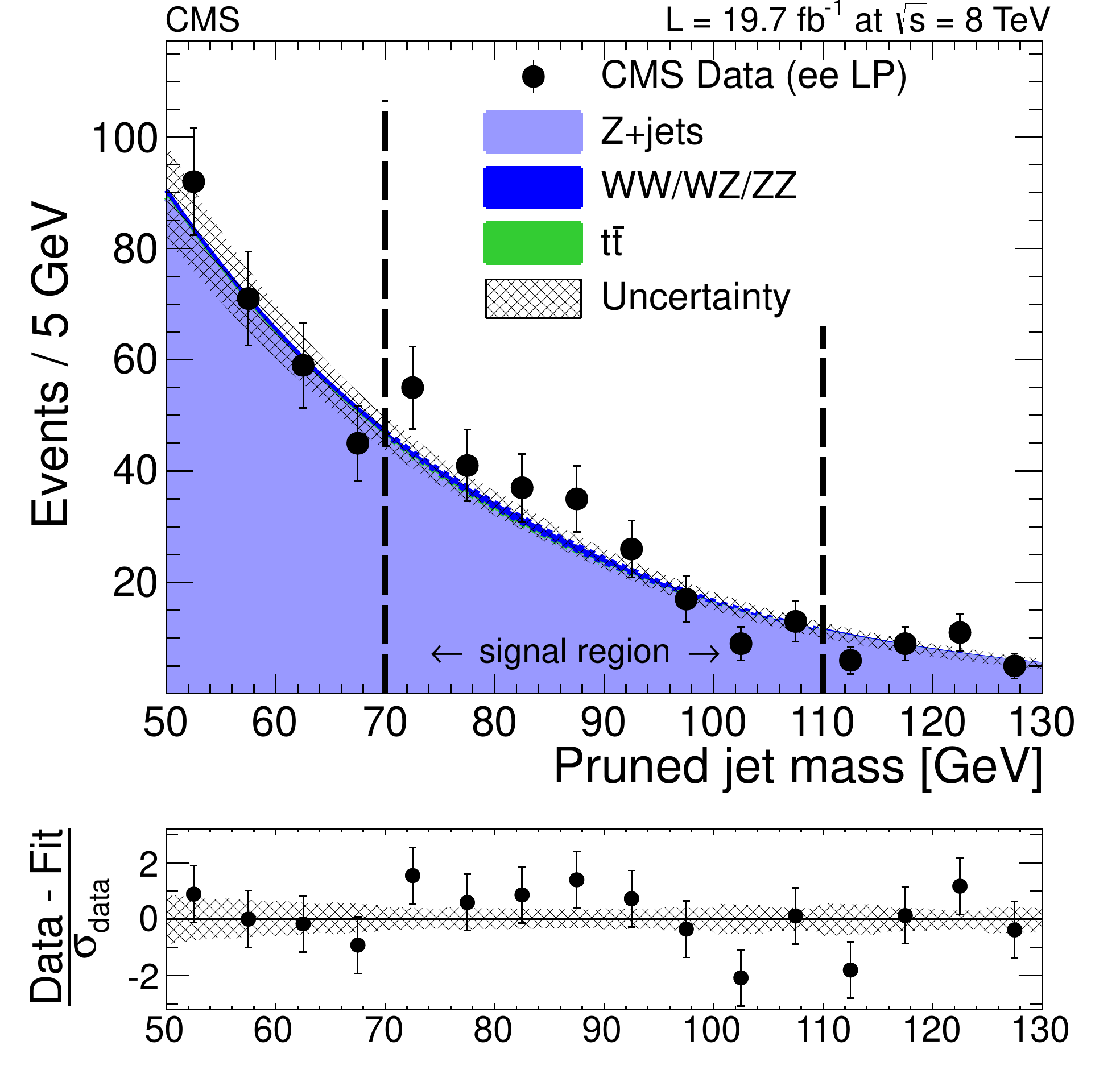}
\caption{Distributions of the pruned jet mass, \mJ, in the \lljet analysis in the
electron channel. The left (right) panel shows the distribution for the HP (LP) category.
All selections are applied except
the final \mJ~signal window requirement.
Data are shown as black markers.
The prediction of the non-resonant
Z+jets background comes from a fit excluding the signal region (between the vertical dashed lines), while
the predictions for the minor backgrounds come from the simulation.
The MC resonant shapes are corrected using the differences
between data and simulation in the W peak position and width measured
in the \ttbar control region (see Section~\ref{sec:ttbarcontrol}).
At the bottom of each plot, the bin-by-bin fit residuals, (data-fit)/$\sigma_\text{data}$,
are shown together with the uncertainty band of the fit normalized by $\sigma_\text{data}$.}
\label{fig:ZJetsNormalization}
\end{figure}

The overall normalization of the V+jets background in the signal
region is determined from a fit to the \mJ~distribution in the lower
and upper sidebands of the observed data. The analytical form of the
fit function is chosen from simulation studies and the minor backgrounds are
taken from the simulation. Figures~\ref{fig:WJetsNormalization} and
\ref{fig:ZJetsNormalization} show the result of this fit procedure for
the \lnujet and \lljet analyses, respectively.
Tables~\ref{table:WWExpectedYields} and
\ref{table:ZZExpectedYields} show the predicted number of background events in the signal
region after the inclusion of the minor backgrounds and compare it
with the data.

\begin{table}[htbp]
\centering
\topcaption{
Observed and expected yields for the \lnujet analysis. The yields are quoted in the range $700 < \mWW <  3000\GeV$.  The expected background is quoted from the
sideband procedure.  The uncertainties in the background prediction from data are statistical in nature, as they depend on
the number of events in the sideband region. Statistical uncertainties in the signal expectations are negligible.}
\label{table:WWExpectedYields}
\renewcommand{\arraystretch}{1.2}
\begin{tabular}{cccccc}
\multicolumn{2}{c}{}  & $\mu\nu+\text{V-jet}$~HP & $\mu\nu+\text{V-jet}$~{LP} & ${\Pe\nu+\text{V-jet}}$~{HP} & ${\Pe\nu+\text{V-jet}}$~{LP}   \T\\
\hline \hline
\multicolumn{2}{c}{Observed yield}     & 1483  & 1546  & 892   & 988  \T\\
\multicolumn{2}{c}{Expected background}   & $1434 \pm 38$  & $1644 \pm 41$ & $878 \pm 30$  & $978 \pm 31$   \T\\
\hline \hline
\multicolumn{2}{c}{Bulk graviton (\ktilde = 0.5)}&\multicolumn{4}{c}{ Signal expectation (MC) }   \T\\
\hline
  \multicolumn{2}{c}{$\Mg = 800\GeV$}         &  12.8    &   5.1   & 10.1     & 3.9    \T\\
  \multicolumn{2}{c}{$\Mg = 1200\GeV$}        &   0.92   &  0.43   & 0.79     & 0.37    \T\\
\hline
\end{tabular}
\end{table}

\begin{table}[hbtp]
\centering
\topcaption{
Observed and expected yields for the \lljet analysis. The yields are quoted in the range 500 (650)
$< \mZZ <$ 2800\GeV for the HP (LP) category.
The expected background is quoted from the sideband procedure.  The uncertainties in the
 background predictions from data are statistical in nature, as they depend on
the number of events in the sideband region. Statistical uncertainties in the signal expectations are negligible.}
\label{table:ZZExpectedYields}
\renewcommand{\arraystretch}{1.2}
\begin{tabular}{cccccc}
\multicolumn{2}{c}{}  & ${\mu\mu+\text{V-jet}}$~{HP} & ${\mu\mu+\text{V-jet}}$~{LP} & ${\Pe\Pe+\text{V-jet}}$~{HP} & ${\Pe\Pe+\text{V-jet}}$~{LP} \T\\
\hline \hline
\multicolumn{2}{c}{Observed yield}     & 575 & 338 & 360 & 233  \T\\
\multicolumn{2}{c}{Expected background}   & $622 \pm 29$  & $338 \pm 22$ & $370 \pm 22$  & $207 \pm 17$   \T\\
\hline \hline
\multicolumn{2}{c}{Bulk graviton (\ktilde = 0.5)}&\multicolumn{4}{c}{ Signal expectation (MC) }   \T\\
\hline
  \multicolumn{2}{c}{$\Mg = 800\GeV$}         &   2.4   &   0.5    & 2.0     & 0.4    \T\\
  \multicolumn{2}{c}{$\Mg = 1200\GeV$}        &   0.16   &   0.04    & 0.14     & 0.035	\T\\
\hline
\end{tabular}
\end{table}

The shape of the \mVV distribution of the V+jets background in the
signal region is determined from the low \mJ~sideband only, through
an extrapolation function $\alpha_\mathrm{MC}(\mVV)$ derived from the
V+jets simulation, defined as:
\begin{equation}
\alpha_\mathrm{MC}(\mVV) = \frac{F_\mathrm{MC, SR}^{\mathrm{V+jets}}(\mVV)}{F_\mathrm{MC, SB}^{\mathrm{V+jets}}(\mVV)},
\end{equation}
where $F_\mathrm{MC, SR}^{\mathrm{V+jets}}(\mVV)$ and $F_\mathrm{MC,
SB}^{\mathrm{V+jets}}(\mVV)$ are the probability density functions used to
describe the \mVV spectrum in simulation for the signal region and
low \mJ~sideband region, respectively.
The high \mJ~sideband was not considered in order to exclude
possible contamination from beyond-SM resonances decaying into a V boson
and a SM Higgs boson, H, with mass of 125.6\GeV
\cite{CMS:HIG13002}, in addition to the VV final state considered here.
The partial compositeness model \cite{Contino:2006nn} is an example of
such a scenarios.  These signal events from HV resonances, in
which the Higgs boson is reconstructed as a jet in the CMS detector
and the V decays leptonically, would populate the high-mass sideband
region of both the
\lnujet ($\mJ \in [105, 130]\GeV$) and \lljet ($\mJ \in [110, 130]\GeV$) analyses.  This possibility cannot be ignored because this
search is not limited only to the bulk graviton model but includes
also a model-independent interpretation of the results
(Section~\ref{subsec:mod-indep-results}).

The \mVV distribution observed in the lower sideband region is
corrected for the presence of minor backgrounds in order to have an
estimation of the V+jets contribution in the control region of the
data, $F_{\text{DATA}, \mathrm{SB}}^{\text{V+jets}}(\mVV)$. The shape of the V+jets
background distribution in the signal region is obtained by rescaling $F_{\text{DATA},
\mathrm{SB}}^{\mathrm{V+jets}}(\mVV)$ for $\alpha_\mathrm{MC}(\mVV)$. The final
prediction of the background contribution in the signal region,
$N^\text{BKGD}_\mathrm{SR}(\mVV)$, is given by
\begin{equation}
N^\text{BKGD}_\mathrm{SR}(\mVV) = C_{\mathrm{SR}}^{\mathrm{V+jets}}\times F_{\text{DATA}, \mathrm{SB}}^{\mathrm{V+jets}}(\mVV)\times
\alpha_\mathrm{MC}(\mVV) + \sum_{k} C_{\mathrm{SR}}^{k}~F_\mathrm{MC, SR}^{k}(\mVV),
\end{equation}
where the index $k$ runs over the list of minor backgrounds and
$C_{\mathrm{SR}}^{\mathrm{V+jets}}$ and $C_{\mathrm{SR}}^{k}$
represent the normalizations of the yields of the dominant V+jets background and of the
different minor background contributions.
The ratio $\alpha_\mathrm{MC}(\mVV)$
reflects small kinematic differences between the signal region and
sideband, which are mostly independent from the theoretical prediction
of cross sections. To test the validity and the robustness of the
method, a closure test with data has been performed, predicting
successfully the normalization and shape of the V+jets background in
an upper sideband using the lower sideband data. The \mVV distribution
of the background in the signal region is described analytically by a
function defined as $f(x)\propto
\re^{-x/(c_0+c_{1}x)}$. Alternative fit functions have been studied but
their usage does not change the final performance. The
\mVV range of the fit determines the region of masses probed by the
searches. The range has been chosen such that there is a smoothly falling
spectrum, in order to have a stable fit and robust control of the
background estimation. For the \lnujet analysis, the fits are carried
out in the \mVV range [700, 3000]\GeV.
In the \lljet analysis, the ranges for the HP and LP categories are
[500, 2800] and [650, 2800]\GeV, respectively. The fits are always
unbinned. In the \lljet analysis, the shapes of the background distributions for the muon and
electron channels are found to be statistically compatible.  The final shape estimation
for the \lljet analysis has been carried out integrating over the two lepton flavors
in order to reduce the statistical uncertainties in the fitted parameters.

\begin{figure}[htbp]
\centering
\includegraphics[width=0.45\linewidth]{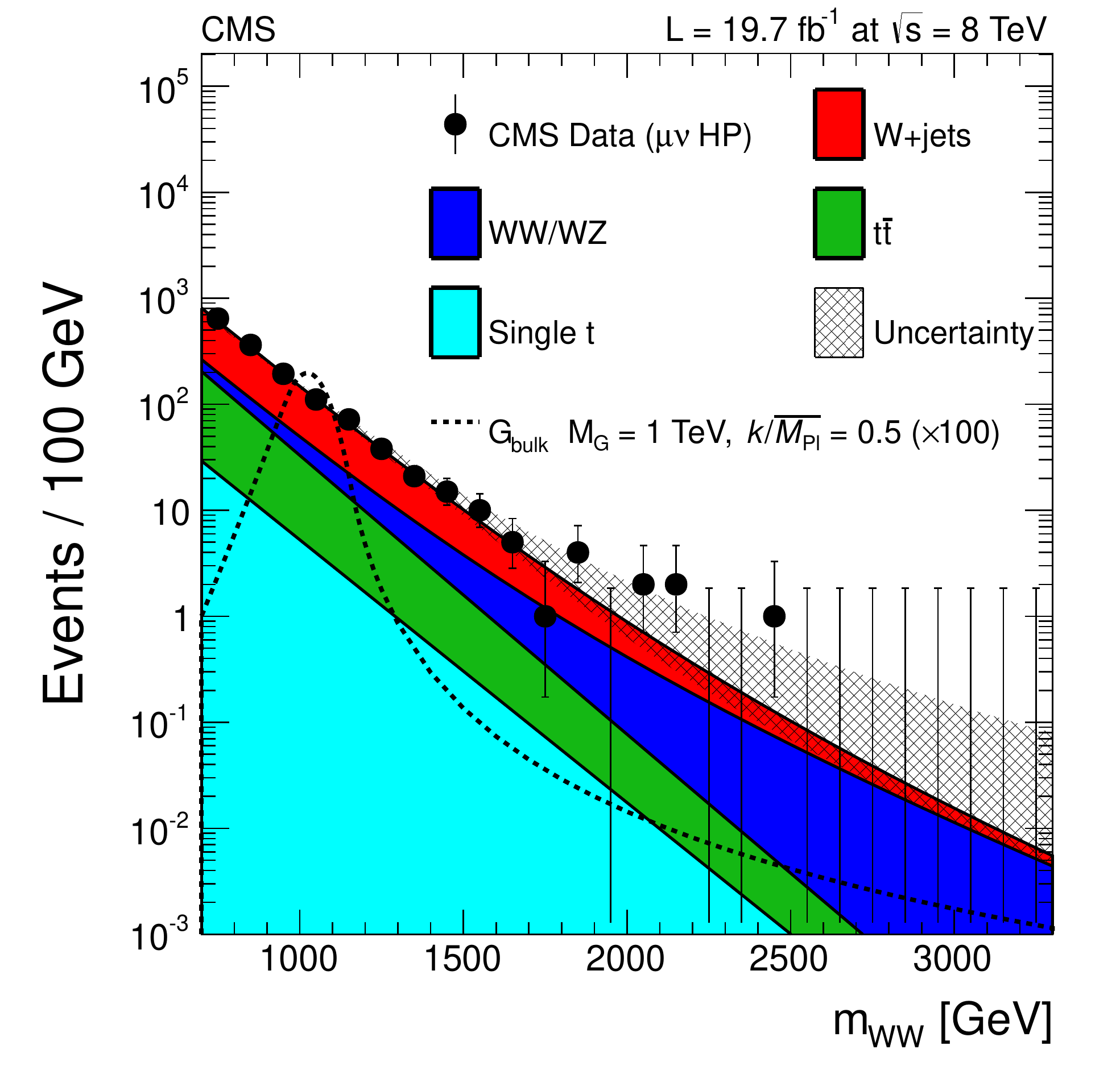}
\includegraphics[width=0.45\linewidth]{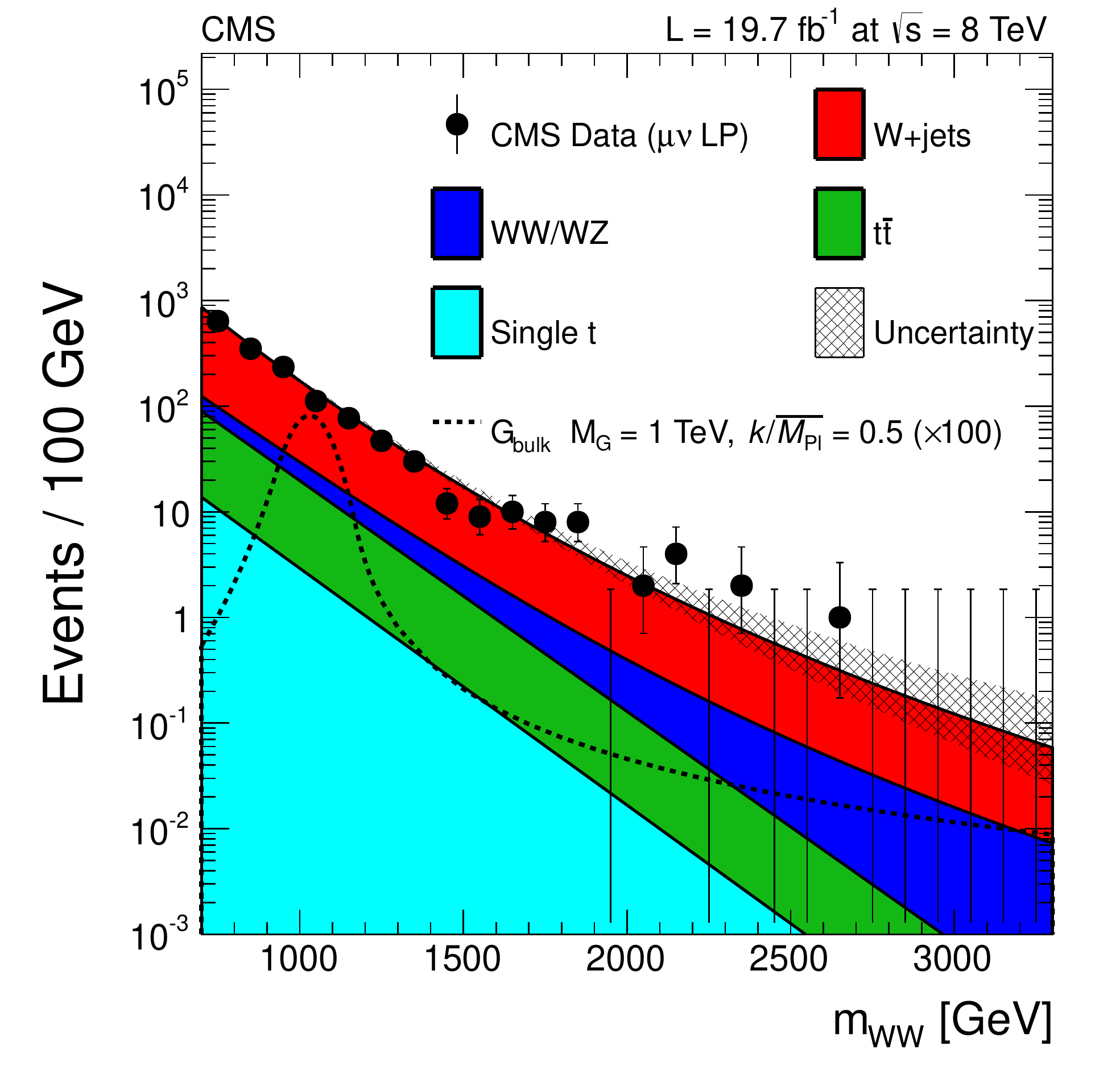}\\
\includegraphics[width=0.45\linewidth]{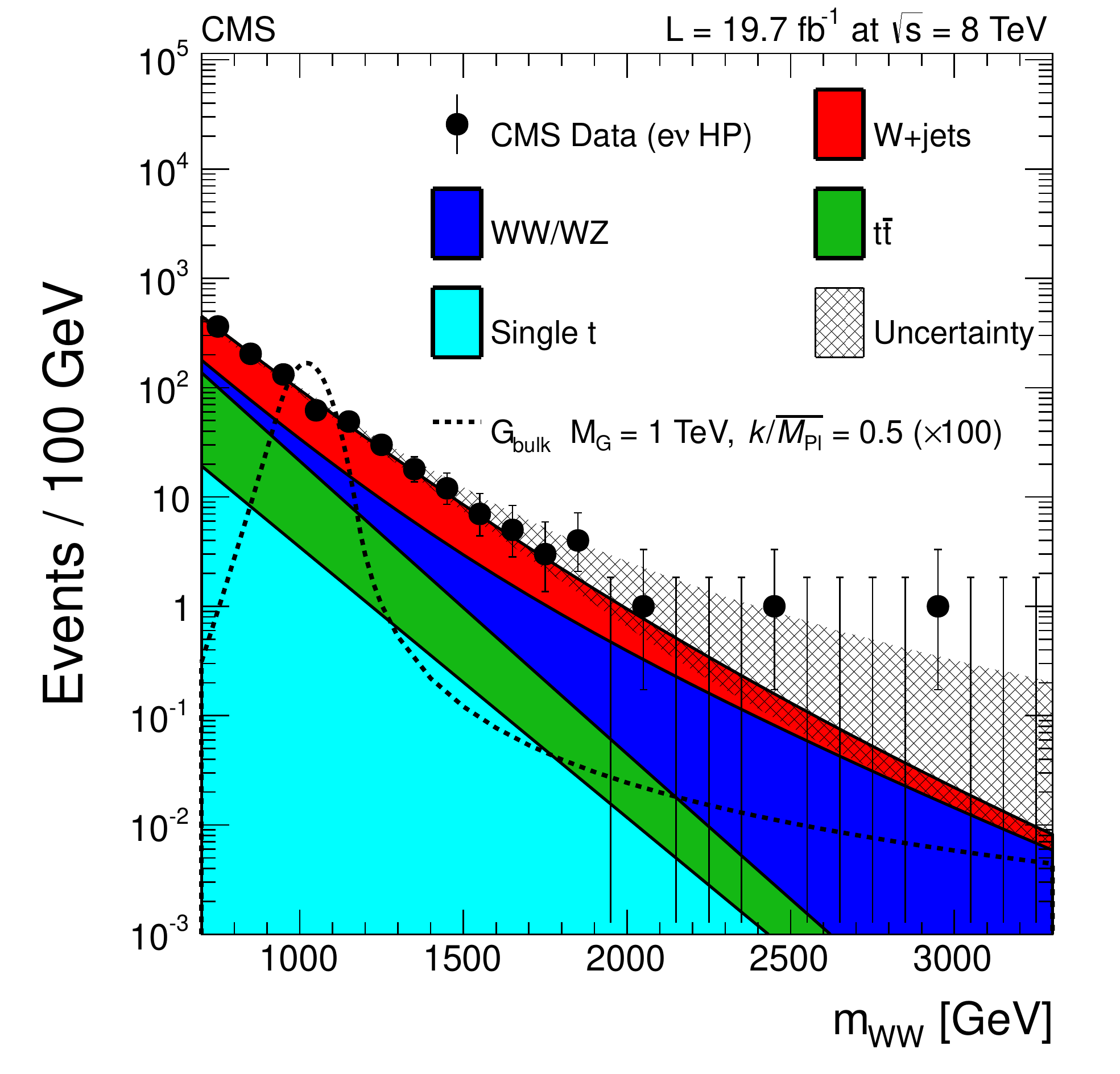}
\includegraphics[width=0.45\linewidth]{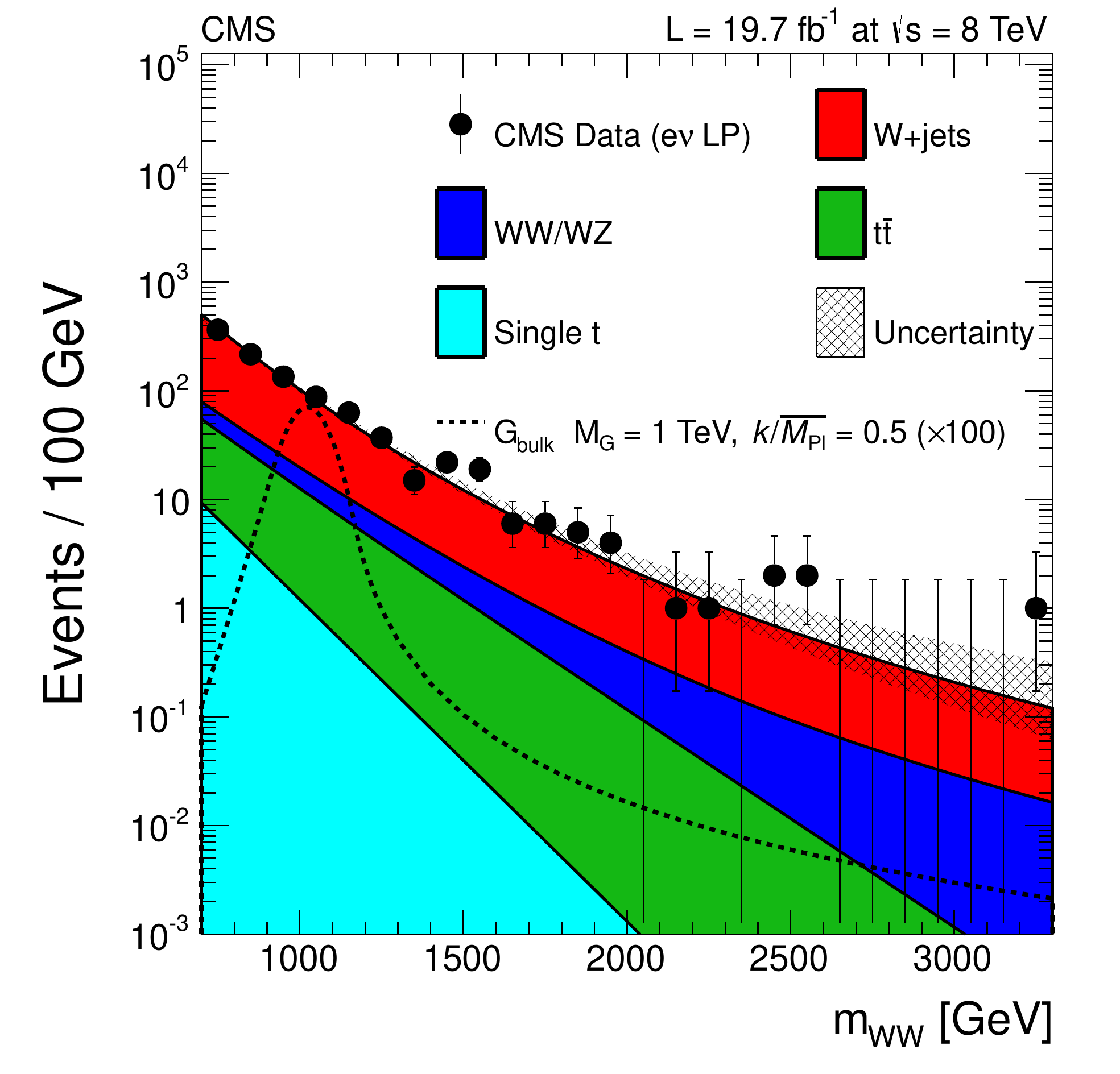}\\
\caption{
Final distributions in \mWW for data and expected backgrounds for both the muon
(top) and the electron (bottom) channels, high-purity (left) and low-purity (right)
categories.
The 68\% error bars for Poisson event counts are obtained from the Neyman construction as described in Ref.~\cite{Garwood}.
Also shown is a hypothetical bulk graviton signal with mass of 1000\GeV and $\ktilde=0.5$.
The normalization of the signal distribution is scaled up by a factor of 100 for a better visualization.
}
\label{fig:MZZwithBackgroundWW}
\end{figure}

\begin{figure}[htbp]
\centering
\includegraphics[width=0.45\linewidth]{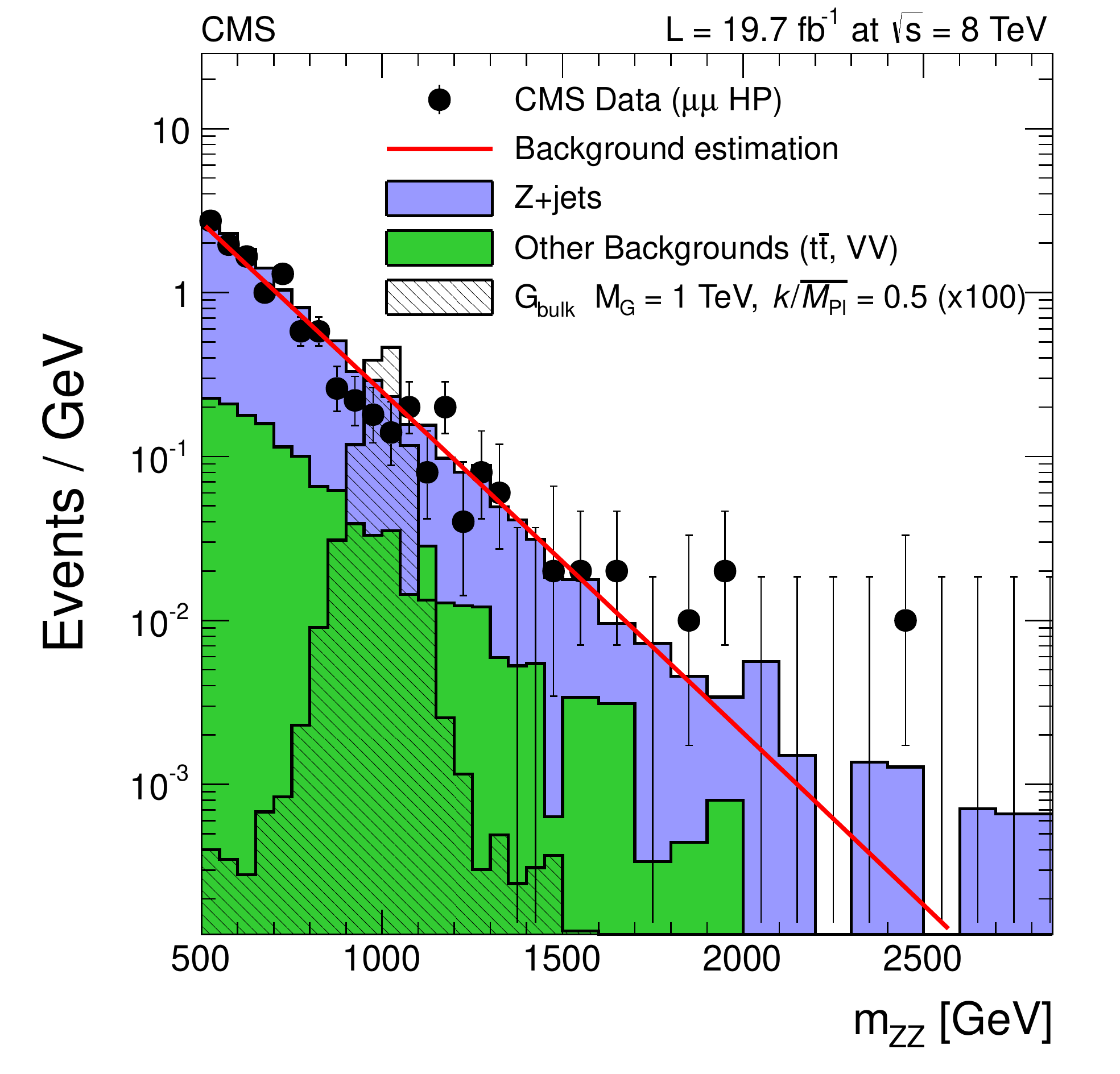}
\includegraphics[width=0.45\linewidth]{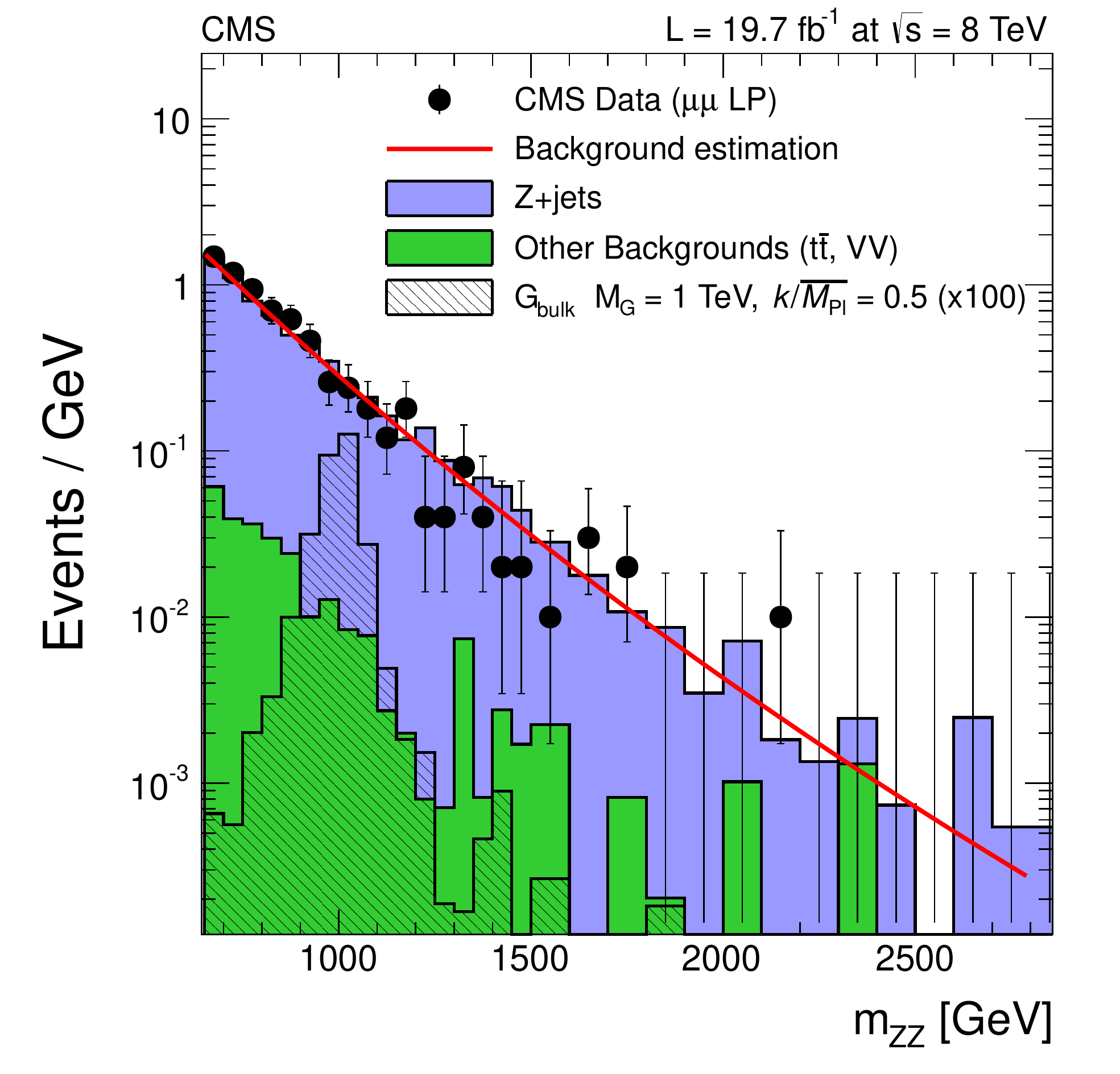}\\
\includegraphics[width=0.45\linewidth]{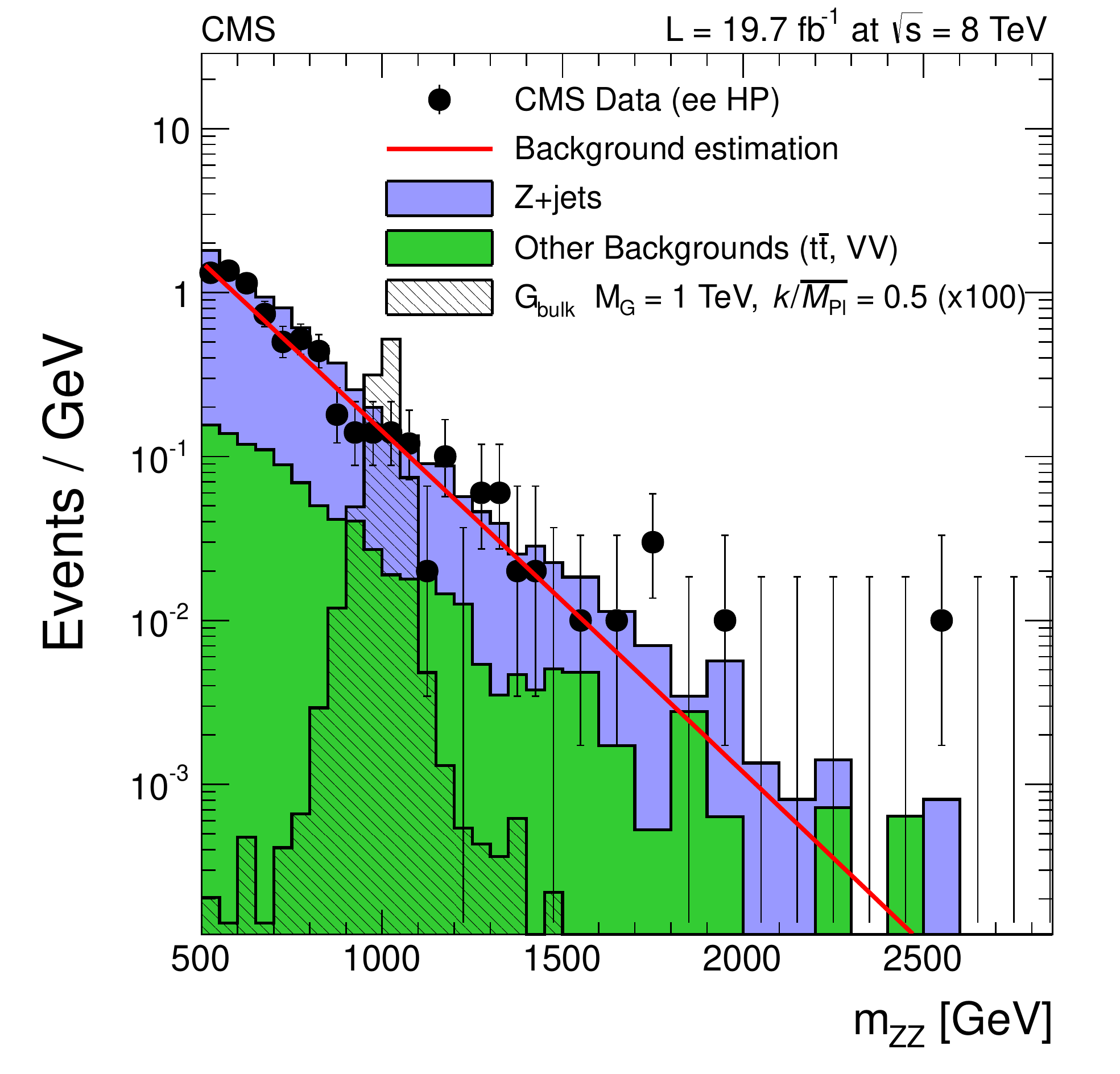}
\includegraphics[width=0.45\linewidth]{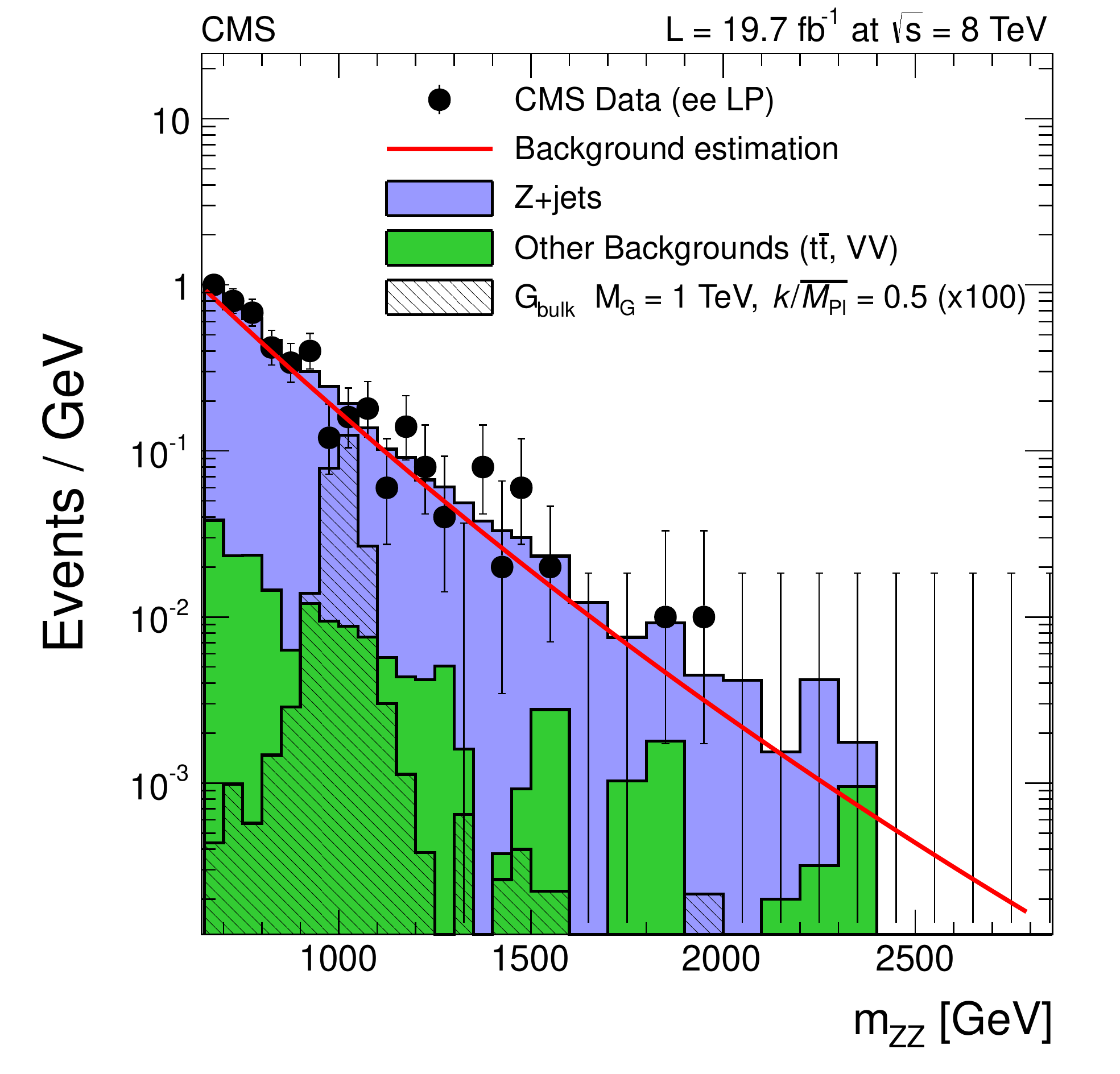}\\
\caption{
Final distributions in \mZZ for data and expected backgrounds for both
the muon (top) and the electron (bottom) channels, high-purity (left) and
low-purity (right) categories. Points with error bars show
distributions of data; solid histograms depict the different
components of the background expectation from simulated events.
The 68\% error bars for Poisson event counts are obtained from the Neyman construction as described in Ref.~\cite{Garwood}.
Also shown is a hypothetical bulk graviton signal with mass of 1000\GeV and
$\ktilde=0.5$. The normalization of the signal distribution is scaled
up by a factor of 100 for a better visualization. The solid line shows
the central value of the background predicted from the sideband
extrapolation procedure. }
\label{fig:MZZwithBackgroundZZ}
\end{figure}

Figure~\ref{fig:MZZwithBackgroundWW} shows the final observed spectrum
in \mWW of the selected events in the four categories of the \lnujet
analysis.
The observed data and the predicted background agree with each other.
The highest-mass event in the \lnujet channel is from
the electron LP category and it has $m_{\PW\PW}\approx
3200$\GeV.
This event is not included in the statistical
analysis of Section~\ref{sec:results}, which is performed up to
$m_{\PW\PW}$ of 3\TeV.  The impact of this event on the reported
results is negligible since we produce limits for a narrow bulk
graviton with a resonance mass up to 2.5\TeV. The observed event is
compatible at the 1$\sigma$ level with the background prediction
for \mWW above 2.5\TeV.

The \mZZ distribution of the selected events in the \lljet analysis is
presented in Fig.~\ref{fig:MZZwithBackgroundZZ}. Also in this case,
an overall good description in both normalization and shape of the
data by the background estimation is observed.  The highest mass event
in the \lljet channel is from the electron HP category and it has
$m_{\cPZ\cPZ}\approx 2600$\GeV.

\subsection{Modeling of the signal mass distribution}
\label{sec:signal}

The shape of the reconstructed signal mass distribution is extracted from the bulk
graviton MC samples generated with the coupling $\ktilde=0.2$, corresponding to an intrinsic
relative width of the resonance of about 0.2\%.
For models with $\ktilde \lesssim 0.5$, the natural width of the resonance is
sufficiently small to be neglected when compared to the detector resolution.
This makes our modeling of the detector effects on the signal shape independent
of the actual model used for generating the events.
In the final analysis of the
\mVV spectrum, the discovery potential and exclusion power both depend
on an accurate description of the signal shape. We adopt an analytical
description of the signal shape, choosing a double-sided Crystal-Ball
(CB) function (\ie, a Gaussian core with power law tails on both
sides)~\cite{CrystalBallRef} to describe the CMS detector resolution.
To take into account differences between muon and electron \pt resolutions at high \pt,
 the signal mass
distribution is parametrized separately for events with electrons and
muons.  No appreciable differences have been observed in the
\mVV signal shape between low- and high-purity categories. The
typical width of the Gaussian core is about 3\%--5\% of the nominal
mass in the \lljet channel or 4\%--6\% for the \lnujet channel.

\section{Systematic uncertainties}
\label{sec:sysunc}

\subsection{Systematic uncertainties in the background estimation}

Uncertainties in the estimation of the background affect both the
normalization and shape of the \mVV distribution. Uncertainties in the
background normalization are mainly statistical in nature and scale
with the amount of data in the sideband regions and the number of
events in the simulated samples.
Tables \ref{table:WWExpectedYields} and
\ref{table:ZZExpectedYields} show the uncertainties in the background
expectations for the \lnujet and \lljet analyses, respectively.
The systematic uncertainties in the V+jets background normalization are
dominated by the statistical uncertainty associated with the number of
events in data in the $m_{\text{jet}}$ sideband regions
(\WJETNORMUNCERT). The systematic uncertainty in the
\ttbar~normalization comes from the uncertainties in the
data-to-simulation scale factors derived in the top-quark enriched control
sample (\TTBARNORMUNCERT).  The systematic uncertainty in the WW
inclusive cross section is assigned to be
\VVNORMUNCERT, taken from the relative difference in the mean value
between the published CMS cross section measurement at $\sqrt{s}=8$\TeV
and the SM expectation~\cite{Chatrchyan:2013oev}.  An additional
systematic uncertainty in the WW normalization comes from the
uncertainty in the V-tagging scale factors derived in
Section~\ref{sec:ttbarcontrol}.  The same uncertainties derived for WW
are also used for \PW\Zo and \Zo\Zo processes.

Systematic uncertainties in the background shape are estimated from the
covariance matrix of the fit to the extrapolated sidebands and from the
uncertainties in the modeling of $\alpha_\mathrm{MC}(\mVV)$. They are both
statistical in nature, as they are driven by the available data in the
sidebands and the number of events generated for the simulation of the
V+jets background.

\subsection{Systematic uncertainties in the signal prediction}

Systematic uncertainties affect both the signal efficiency and the
\mVV shape. Table~\ref{table:systematics} presents the
 primary uncertainties in the signal normalization.
Among the sources of systematic uncertainty in
the signal efficiency are the muon momentum scale and resolution,
the electron energy scale and resolution, the jet energy scale and
resolution, and the unclustered energy in the event.
The event selection is applied to signal samples after varying
the lepton four-momenta within one standard deviation
of the corresponding uncertainty in the muon momentum scale~\cite{CMS:MUO10004}
or electron energy scale~\cite{Chatrchyan:2013dga}, or applying an appropriate momentum/energy
smearing in case of resolution uncertainties. The same procedure
is also applied for the jet four-momenta using the corresponding
energy scale and resolution uncertainties~\cite{CMS:JetCalibration}.
In this process, variations in the lepton and jet four-momenta
are propagated consistently to the \ETmiss~vector.
The signal efficiency is then recalculated using
modified lepton and jet four-momenta separately for
each source of systematic uncertainties.
The largest relative change in the signal efficiency compared to the
default value is taken as the systematic uncertainty for that specific source.
The muon, electron, and jet uncertainties are assumed to be uncorrelated.

The systematic uncertainties in the lepton trigger, identification,
and isolation efficiencies are derived using a dedicated tag-and-probe analysis
in $\cPZ\to \ell^{+}\ell^{-}$ events. The uncertainties in trigger and
identification+isolation efficiencies for muons are 3\% and 4\%, respectively.
The total uncertainty in the electron trigger, identification, and isolation efficiency is 3\%.
These uncertainties are evaluated taking into account the limited number
of data events in the boosted regime.
We also include systematic uncertainties in signal efficiency due to
uncertainties in data-to-simulation scale factors for the V-tagging
identification (derived from the top-quark enriched control sample, see
Section~\ref{sec:ttbarcontrol}),
and b-jet identification efficiencies (derived with the methods
described in Ref.~\cite{CMS:BTAG7TeV} and updated with the 8\TeV data).
The systematic uncertainties from pileup are assigned by re-weighting
the signal simulation samples such that the distribution of the number of interactions
per bunch crossing is shifted up and down by one standard deviation compared
with that found in data~\cite{Chatrchyan:2012nj}. The impact of these changes on the
signal efficiency is used to assess the systematic effect.
The impact of the proton PDF uncertainties on the signal efficiency is evaluated
with the PDF4LHC prescription~\cite{Botje:2011sn} using
MSTW~\cite{MSTW}
and NNPDF~\cite{NNPDF} PDF sets. The uncertainty in
the integrated luminosity is~\LUMIUNCERT \cite{CMS:LUM13001}.

\begin{table*}[htb]
\centering
\topcaption{Summary of systematic uncertainties in signal yield, relative
to the expected number of observed signal events. All systematic
uncertainties in the list are treated as uncorrelated.}
\label{table:systematics}
\small
\begin{tabular}{lcc}
{Source}     & \multicolumn{2}{c}{{Analysis}}\\
 &   {\lnujet}   &   {\lljet} \T\\
\hline
\hline
Muons (trigger and ID)       &{2\%} &{5\%}      \T\\
Muon scale                   &{1\%} &{2\%}    \T\\
Muon resolution              &{$<$0.1\%} &{0.5\%}    \T\\
Electrons (trigger and ID)   &{3\%} &{3\%}  \T\\
Electron scale               &{$<$0.5\%} &{$<$0.5\%}    \T\\
Electron resolution          &{$<$0.1\%} &{$<$0.1\%}    \T\\
Jet scale                    &{1--3\%} &{1\%}  \T\\
Jet resolution               &{$<$0.5\%} &{$<$0.1\%}  \T\\
Unclustered energy scale     &{$<$0.5\%} & \NA  \T\\
Pileup                       & {0.5\%} & {0.5\%}    \T\\
V tagging                    &  \multicolumn{2}{c}{\VTAGUNCERTHP (HP)}   \T\\
                             &  \multicolumn{2}{c}{\VTAGUNCERTLP (LP)}  \T\\
PDF                         &  \multicolumn{2}{c}{$<$0.5\%}  \T\\
Luminosity                   &  \multicolumn{2}{c}{\LUMIUNCERT}     \T\\
\hline
\end{tabular}
\end{table*}

Uncertainties in the scale and resolution of the four-momenta
of the reconstructed objects can bias the peak
and smear the width of the signal profile. The
systematic uncertainties considered to affect the signal shape are
the scale and resolution uncertainties on muons, electrons, jets,
and the unclustered energy scale. For each of these sources of
experimental uncertainty, the lepton/jets four-momenta and \ETmiss vector
are varied (or smeared) by the relative uncertainty.  In general, only
small effects on the peak position and the width of the Gaussian core
of the signal shapes have been found. The jet energy scale and
resolution introduce a relative uncertainty of about 3\% (2\%) in the
signal width for the
\lnujet (\lljet) channel. In the \lnujet channel, the unclustered
energy scale introduces a 1--3\% uncertainty in the signal width,
larger at low resonance masses. In the \lljet channel, the muon
resolution causes an additional relative uncertainty of 2\% in the
signal width. The uncertainty in the peak position of the signal is estimated
to be less than 1\%.

\section{Statistical interpretation}
\label{sec:results}

The comparison between the \mVV\ distribution observed in data and the
 background prediction from data is used to test for the presence of a
resonance decaying to vector bosons. We set upper limits on the
production cross section of a new resonance decaying to the \Wo\Wo~final
state or the \Zo\Zo final state by combining the four event categories of
the \lnujet analysis or the \lljet analysis, respectively. We follow
the modified frequentist prescription described in Ref.~\cite{CLs1,Junk:1999kv} ($\mathrm{CL}_S$ method). The limits are
computed using an unbinned shape analysis. Systematic uncertainties
are treated as nuisance parameters and profiled in the statistical
interpretation using log-normal priors.

\subsection{Limits on a narrow-width bulk graviton model}\label{subsec:bulk_limits}

\begin{figure}[htb]
\centering
     \includegraphics[width=0.45\linewidth]{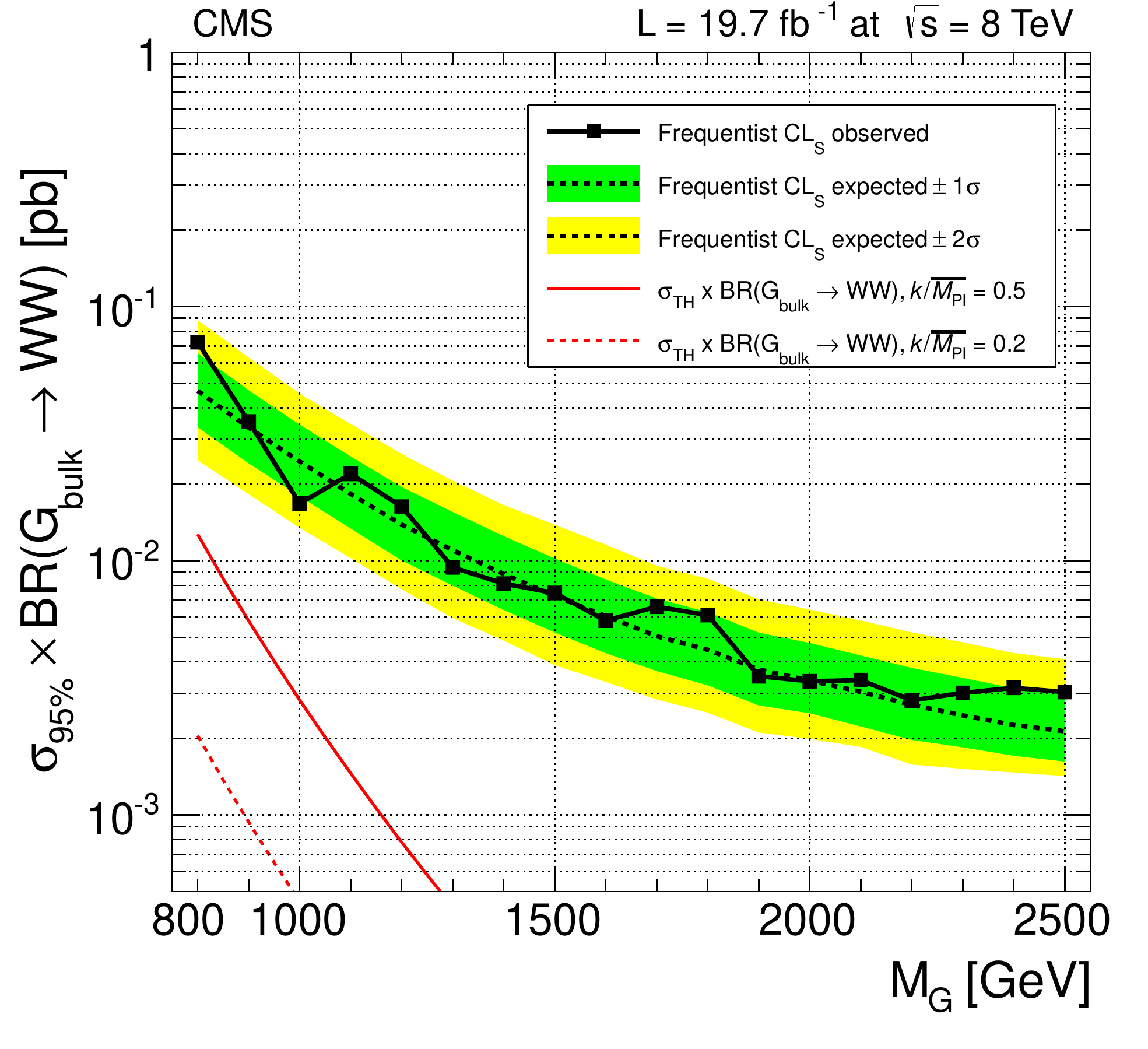}
     \includegraphics[width=0.45\linewidth]{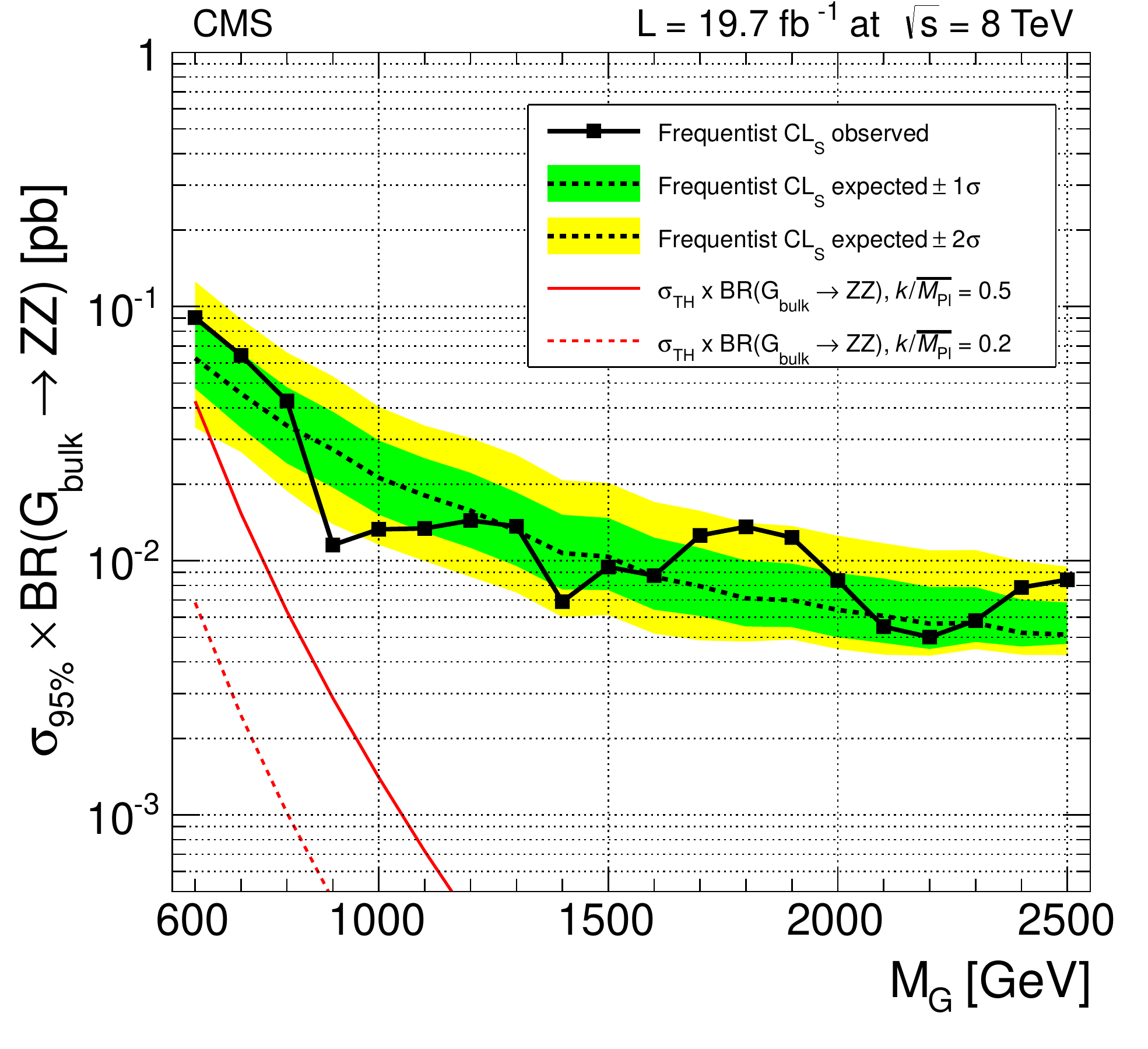}
\caption{
 Observed (solid) and expected (dashed) 95\% CL upper limits on the
  product of the graviton production cross section and the branching
  fraction of $\Grav\to \PW\PW$ (left) and $\Grav\to\Zo\Zo$ (right).
  The cross section for the production of a bulk graviton
  multiplied by its branching fraction for the relevant process is
  shown as a red solid (dashed) curve for $\ktilde =0.5$ (0.2), respectively.}
\label{fig:limitSingleChannel}
\end{figure}

Exclusion limits can be set in the context of the bulk graviton model,
under the assumption of a natural width negligible with respect to the
experimental resolution (narrow-width approximation). Figure~\ref{fig:limitSingleChannel} shows the 95\% confidence level (CL) expected and observed
exclusion limits as a function of \mG. The limits are compared with the
cross section times the branching fraction to $\Wo\Wo$ and $\Zo\Zo$  for
a bulk graviton with $\ktilde = 0.2$ and $\ktilde = 0.5$.
These results were cross-checked with an alternative background estimation
 from data, fitting the \mVV distributions for the same
selected events with a smoothly falling function. This approach,
common to previously released CMS results~\cite{CMS:EXO11095,
CMS:EXO12024,CMS:EXO12016}, provides results very close to the
baseline method described above, further strengthening our confidence
in the robustness of the background estimation method.

\begin{figure}[htb]
\centering
     \includegraphics[width=0.45\linewidth]{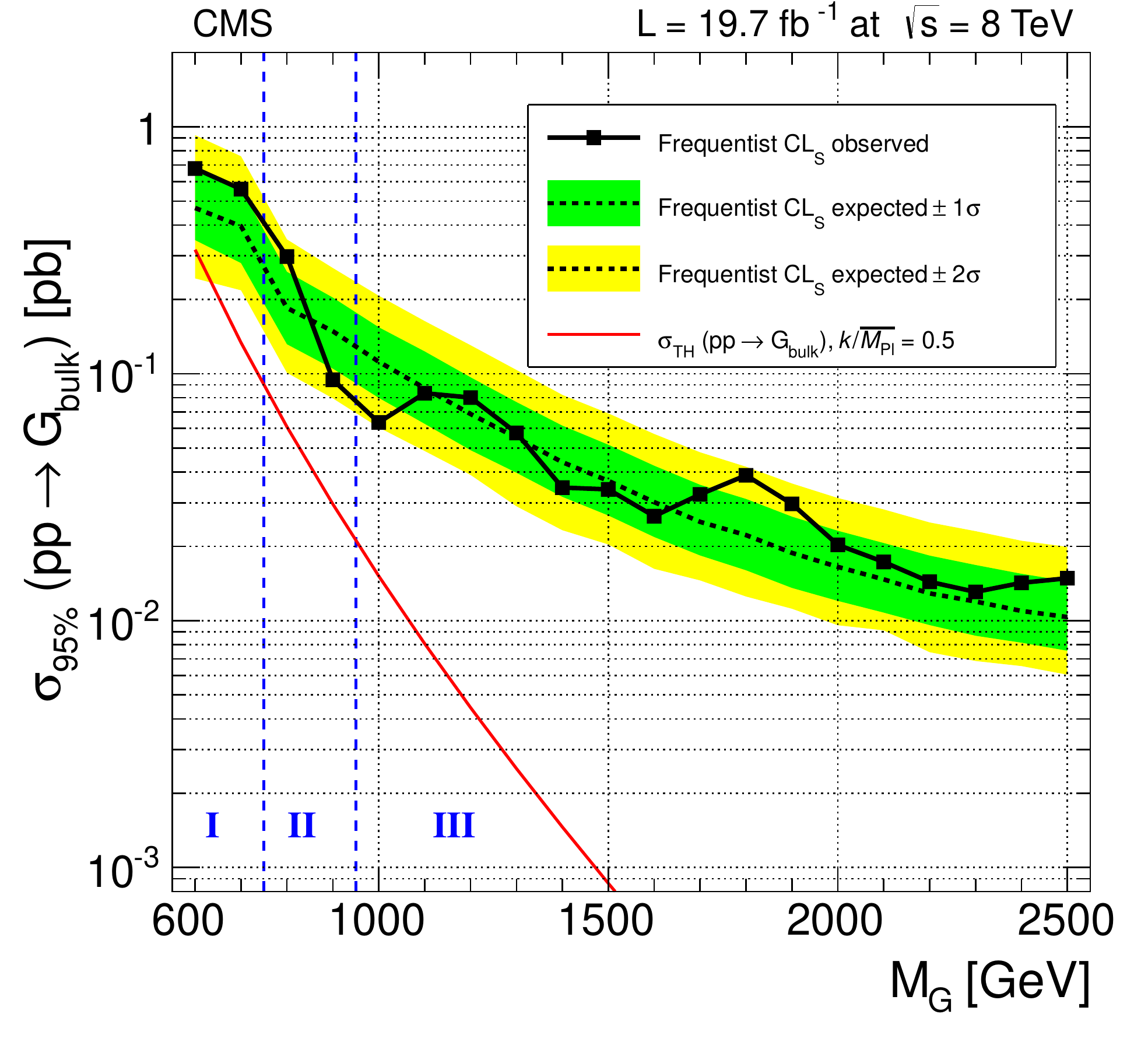}
     \includegraphics[width=0.45\linewidth]{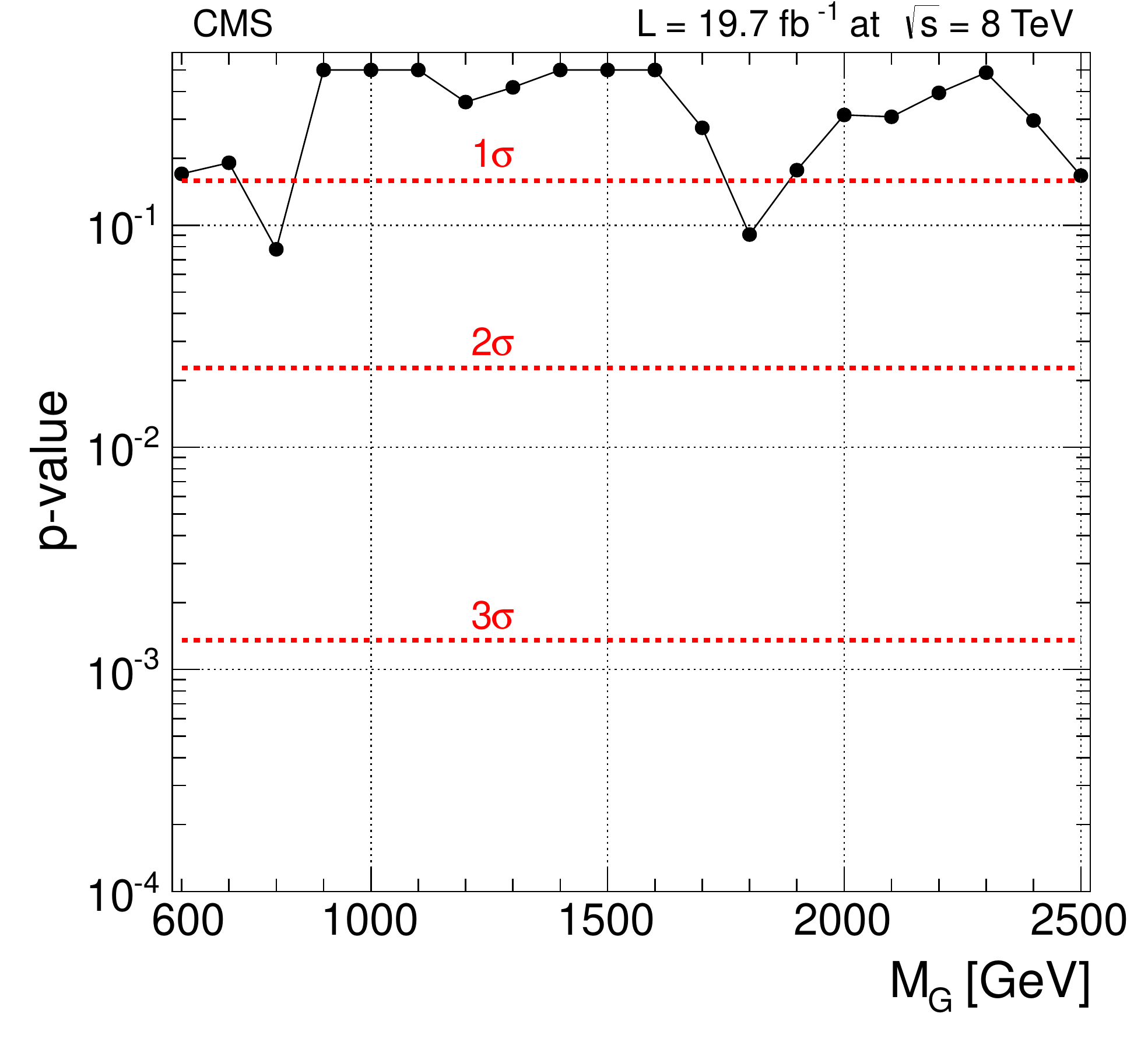}
\caption{Left: observed (solid) and expected (dashed) 95\% CL upper limit on
the graviton production cross section obtained with this analysis and
the analysis of the all-hadronic channel~\cite{CMS:EXO12024}. The cross section for
the production of a bulk graviton with $\ktilde =0.5$ is shown as a
red solid curve. In region I, only the \lljet channel contributes.  In
region II, both \lljet and \lnujet channels contribute. In region III,
both the semi-leptonic and all-hadronic channels contribute.  Right:
observed p-value as a function of the nominal signal mass.
Conversions of the p-value to the number of standard
deviations of a two-sided Gaussian distribution are drawn as dashed horizontal red
lines.}
\label{fig:limitCombined}
\end{figure}

The \lnujet and \lljet analyses are further combined together with a
complementary CMS search in the
$\Vo\Vo\to (\qqbarpr) (\qqbarpr) \to 2 \text{ V-jets}$
final state~\cite{CMS:EXO12024} (dijet channel),
in order to maximize the sensitivity of the search for this specific model.
The fully hadronic analysis uses the same techniques to identify V-jets discussed in Section~\ref{subsec:Vhadr}.
The systematic uncertainties in jet energy scale/resolution, V-tagging scale factors, and luminosity
are considered correlated at 100\% among the three analyses entering the statistical combination.
The systematic uncertainties in electrons and muons are considered correlated at 100\% between the \lljet and \lnujet channels.
The resulting 95\% CL upper limits on the signal cross section are shown in
Fig.~\ref{fig:limitCombined}.
The \lljet channel is the only one contributing to the limit for resonance masses below 800\GeV.
In the range 800--2500\GeV, the \lnujet channel dominates the sensitivity, although the
\lljet and dijet channels give  significant contributions to the combined limit in the region below
and above 1300\GeV, respectively. Because of the combination of the analyses,
the expected upper limits on cross section are made more stringent by about 15--20\%
compared to the individual \lnujet channel, depending on the resonance mass.
The integrated luminosity of the sample is not large enough to
allow us to set mass limits on the bulk graviton models with
$\ktilde = 0.2$ or 0.5.
Fig.~\ref{fig:limitCombined} (right) presents also
the local p-value of the significance of the excesses observed in the data.
No excesses with significances larger than two standard deviations are observed.

\subsection{Model-independent limits}
\label{subsec:mod-indep-results}

The analysis as presented above is specific to the case of a narrow
bulk graviton model,
but this is not the only extension of the SM predicting resonances decaying to vector
bosons.  Therefore it is useful to allow the reinterpretation of
these results in a generic model. In this section we present the
exclusion limits on the visible number of events after having
introduced some modifications to the analysis that greatly
simplify its structure, at a moderate price in terms of
performance. Together with the upper limits on the number of signal
events, we provide tables with the reconstruction and identification
efficiencies for vector bosons in the kinematic acceptance of the
analysis. Following the instructions detailed in Appendix
\ref{sec:mod-indep-instr}, it is possible to estimate the number of
events for a generic signal model that would be expected to be
detected in CMS with the collected integrated luminosity and to compare it with the
upper limit on the number of events.

To avoid the dependence on the assumptions in the construction of the
separate categories, we perform a simplified analysis, reducing
the event classification to one single category. We do this by adding the muon and
electron channels and dropping the low-purity category (whose sensitivity
is much smaller than the high-purity category). The loss in performance is very
small over a large range of masses.  The effect of dropping the LP category is visible
only at very high masses, where the upper limit on the cross section becomes 15\% less stringent.

A generic model cannot restrict itself to narrow signal widths, hence
we provide limits as a function of both mass (\mX) and natural width (\wX)
of the new resonance. The generated line shape is parametrized with a
Breit--Wigner function (BW) and its width is defined as the
$\Gamma$ parameter of the BW. The BW line shape is convoluted with the
double-sided CB introduced in Section~\ref{sec:signal} for describing the
detector resolution. While different values of \wX~are scanned, the
parameters of the double-CB function are kept fixed to the values determined under
the narrow-width approximation. It was checked that the
parametrization of the detector effects factorizes from the natural
width of the resonance and is stable as \wX~increases. The width scan
is done at regular steps of the relative width, $\wX / \mX$. The
range of values considered spans from the zero width approximation
(as in the nominal analysis), up to $\wX/\mX=0.40$, in regular steps of
0.05.

We provide the efficiency as a function of the vector boson kinematic variables,
as the efficiency can depend significantly on the production
and decay kinematic quantities of the new resonance. The efficiencies are extracted from
the bulk graviton samples generated for the baseline analysis. The
efficiencies are calculated by first preselecting simulated signal
events according to the acceptance requirements of the analysis. Thus the
usage of the tables is valid only within this kinematic region,
summarized in Tables~\ref{tab:eff_gensel_WW} and~\ref{tab:eff_gensel_ZZ}
of Appendix~\ref{sec:mod-indep-instr} for the \lnujet and \lljet analyses,
respectively. For preselected events, the reconstructed V candidates
are then independently checked to pass the full analysis selection.
The efficiency tables are presented as a function of the \pt and $\eta$
of the V boson from
the resonance decay prior to any simulation of detector effects.
Bins with fewer than 25 events generated therein are excluded
from the final tables. This choice
controls the statistical uncertainty of the parametrization and has a
very limited impact on the precision of the parametrized efficiencies
because they are located in extreme regions of phase space. All the reweighting and rescaling effects (including
lepton identification and trigger efficiencies, and V-tagging scale
factors) are included in the efficiencies.

The efficiencies of the second-lepton and b-jet vetoes in the
\lnujet analysis are found to be independent of the diboson event kinematic in signal events.
We use a constant efficiency of 91.5\% for the b-jet veto and 98.3\% for
the second-lepton veto, resulting in a total efficiency for the two
combined vetoes of $\varepsilon_{\text{vetoes}}=90\%$.

\begin{figure}[p!ht]
\centering
\includegraphics[width=\cmsFigWidth]{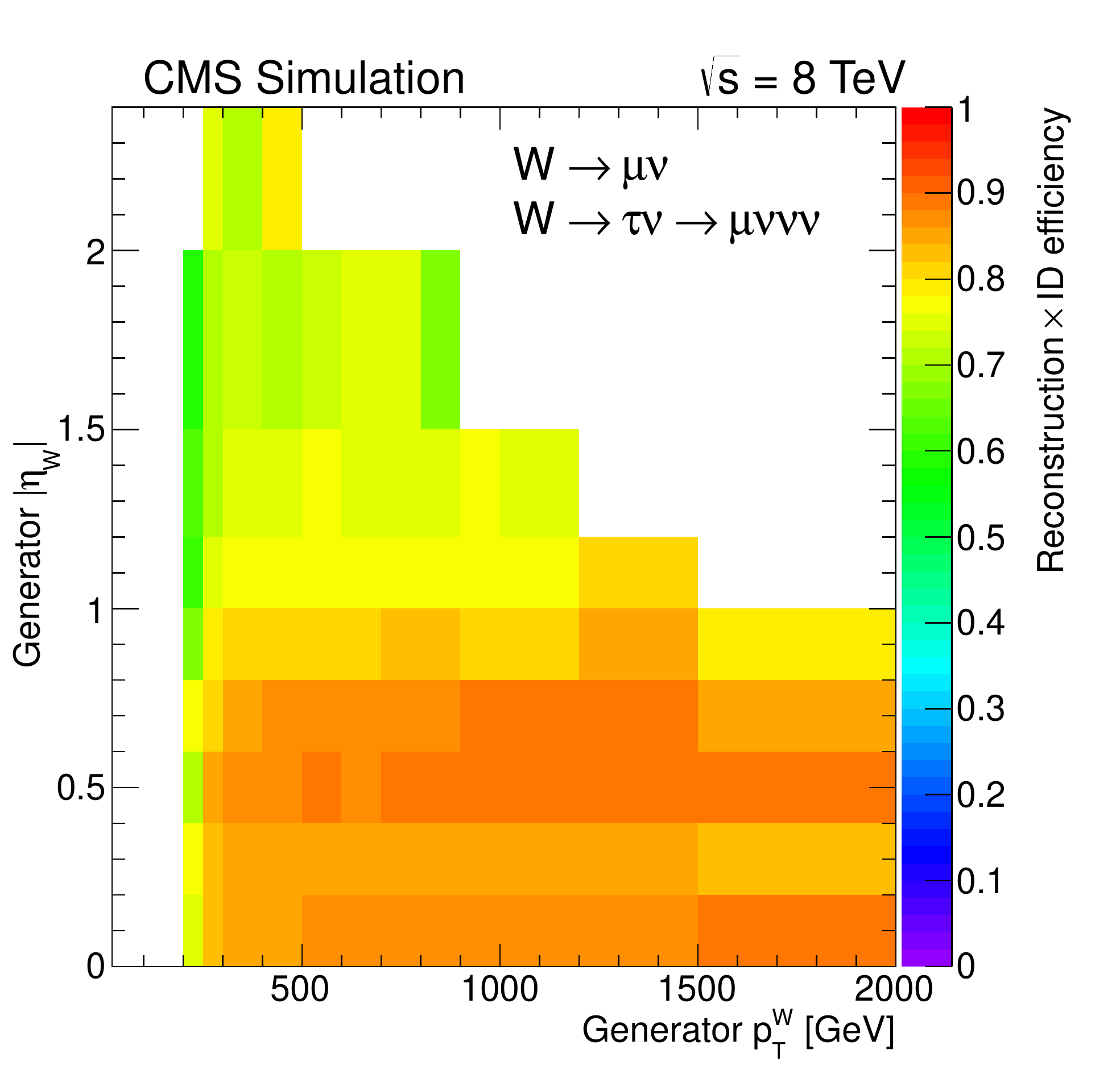}
\includegraphics[width=\cmsFigWidth]{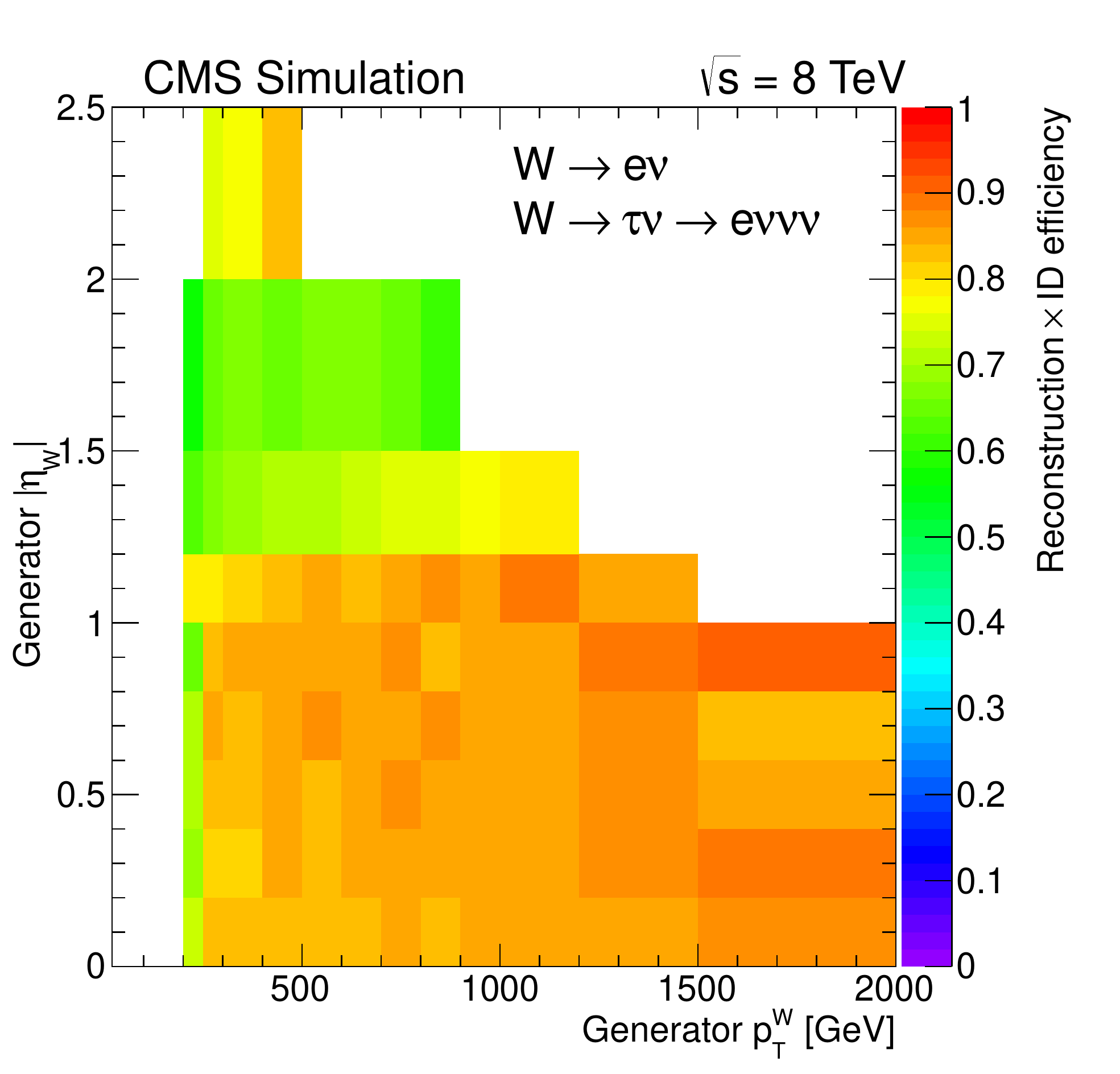}
\includegraphics[width=\cmsFigWidth]{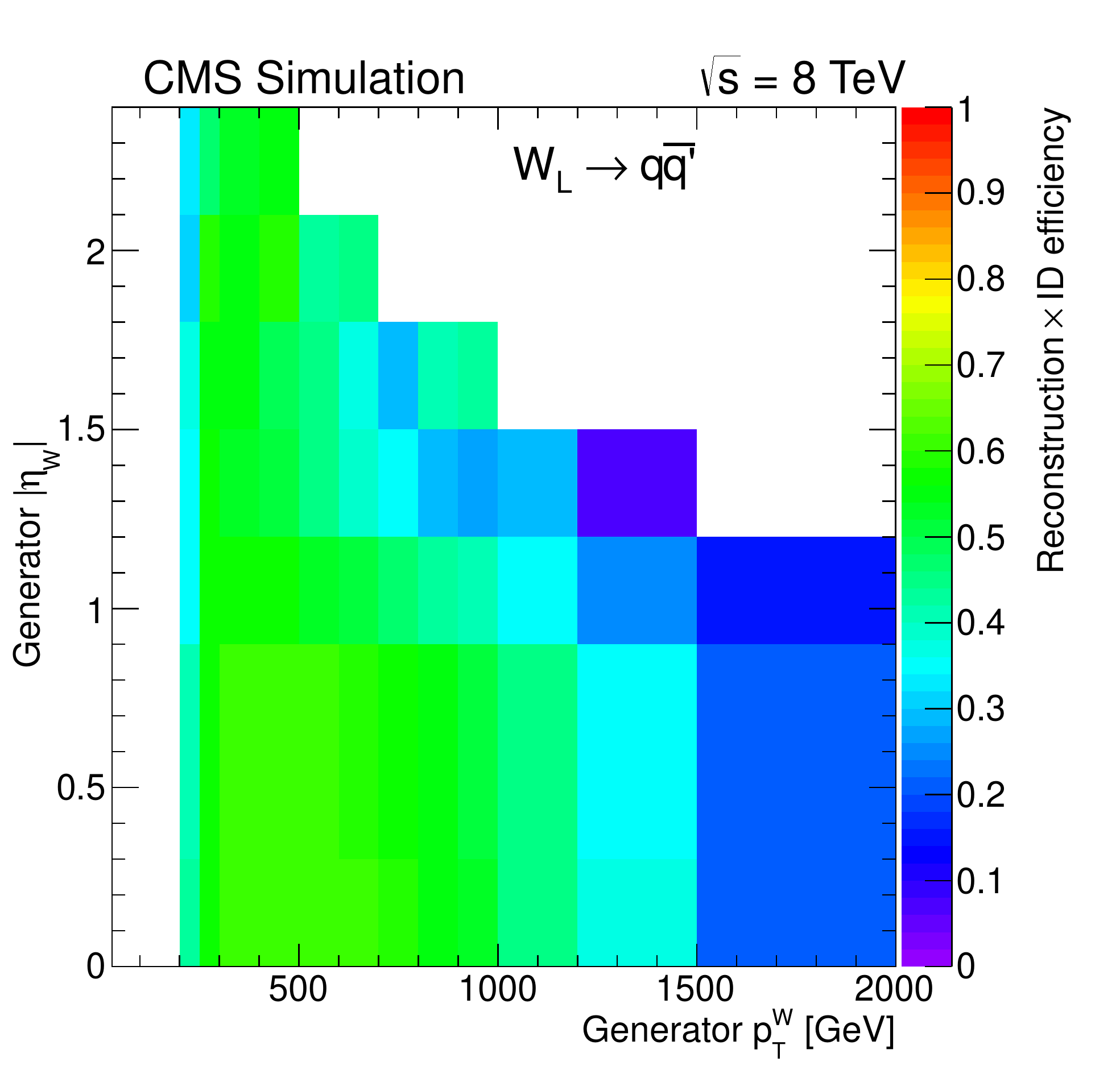}
\includegraphics[width=\cmsFigWidth]{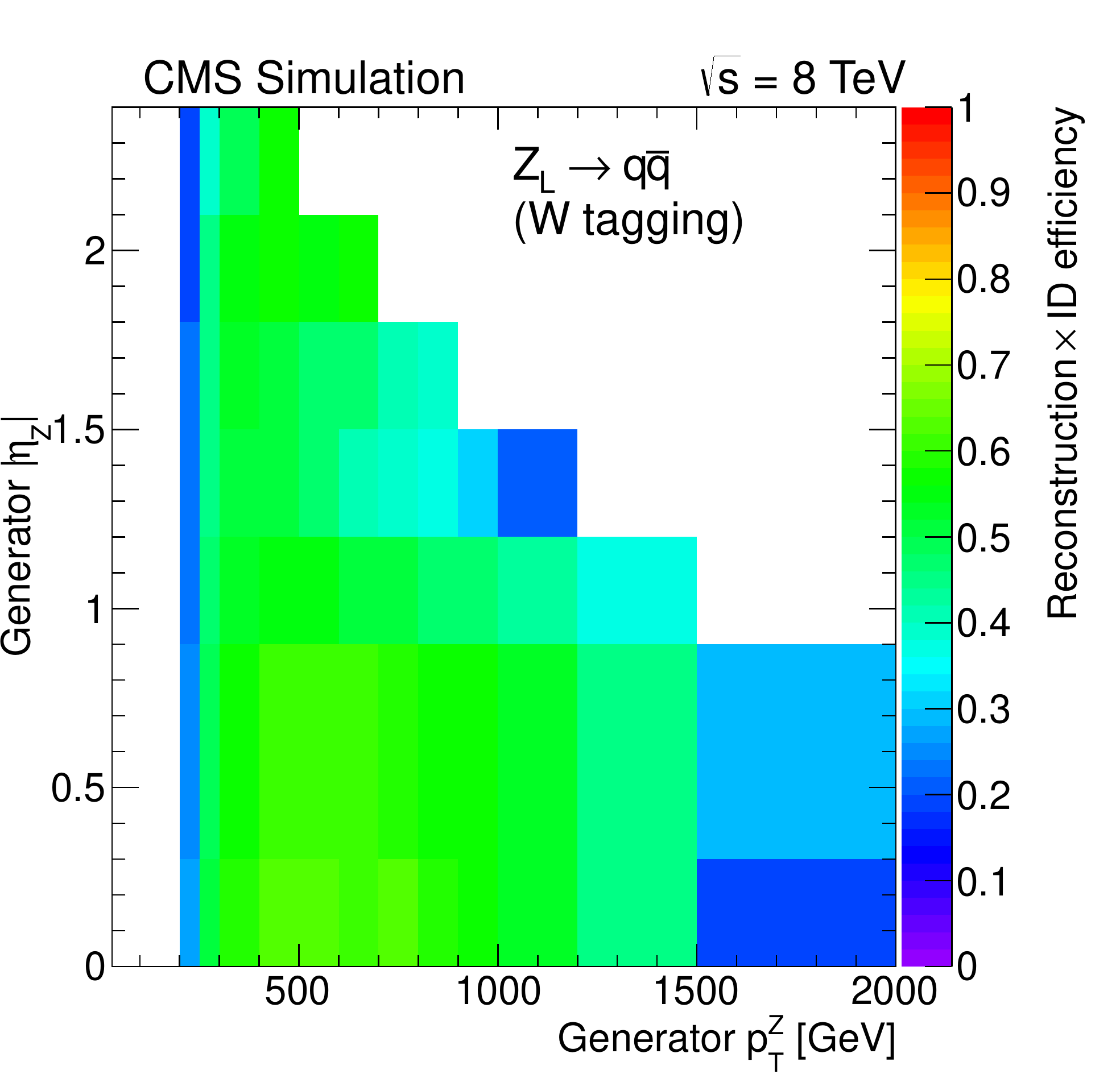}
\caption{Reconstruction and identification efficiencies for the
$\PW\to \mu\nu$ and $\PW\to \tau\nu \to \mu\nu\nu\nu$ (top left),
$\PW\to \Pe\nu$ and $\PW\to \tau\nu \to \Pe\nu\nu\nu$ (top right),
$\PW_\mathrm{L}\to \Pq\Paq'$ (bottom left), and $\cPZ_\mathrm{L}\to \Pq\Paq$ (bottom right) decays
as function of generated $\pt^\mathrm{V}$ and $\eta_\mathrm{V}$ using the W-tagging requirements for the hadronic V decays.}
\label{fig:eff_W}
\end{figure}

\begin{figure}[htp]
\centering
\includegraphics[width=\cmsFigWidth]{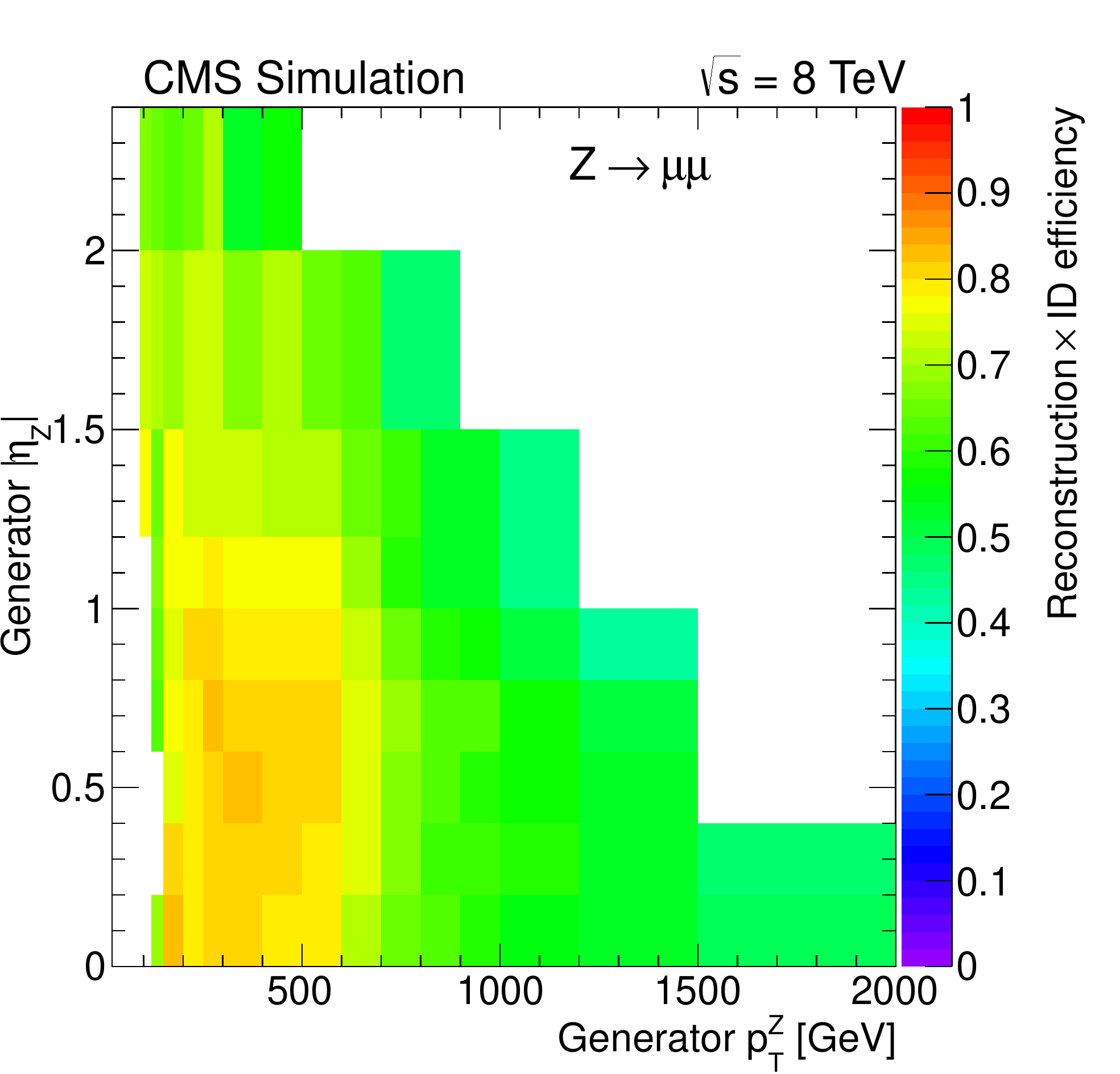}
\includegraphics[width=\cmsFigWidth]{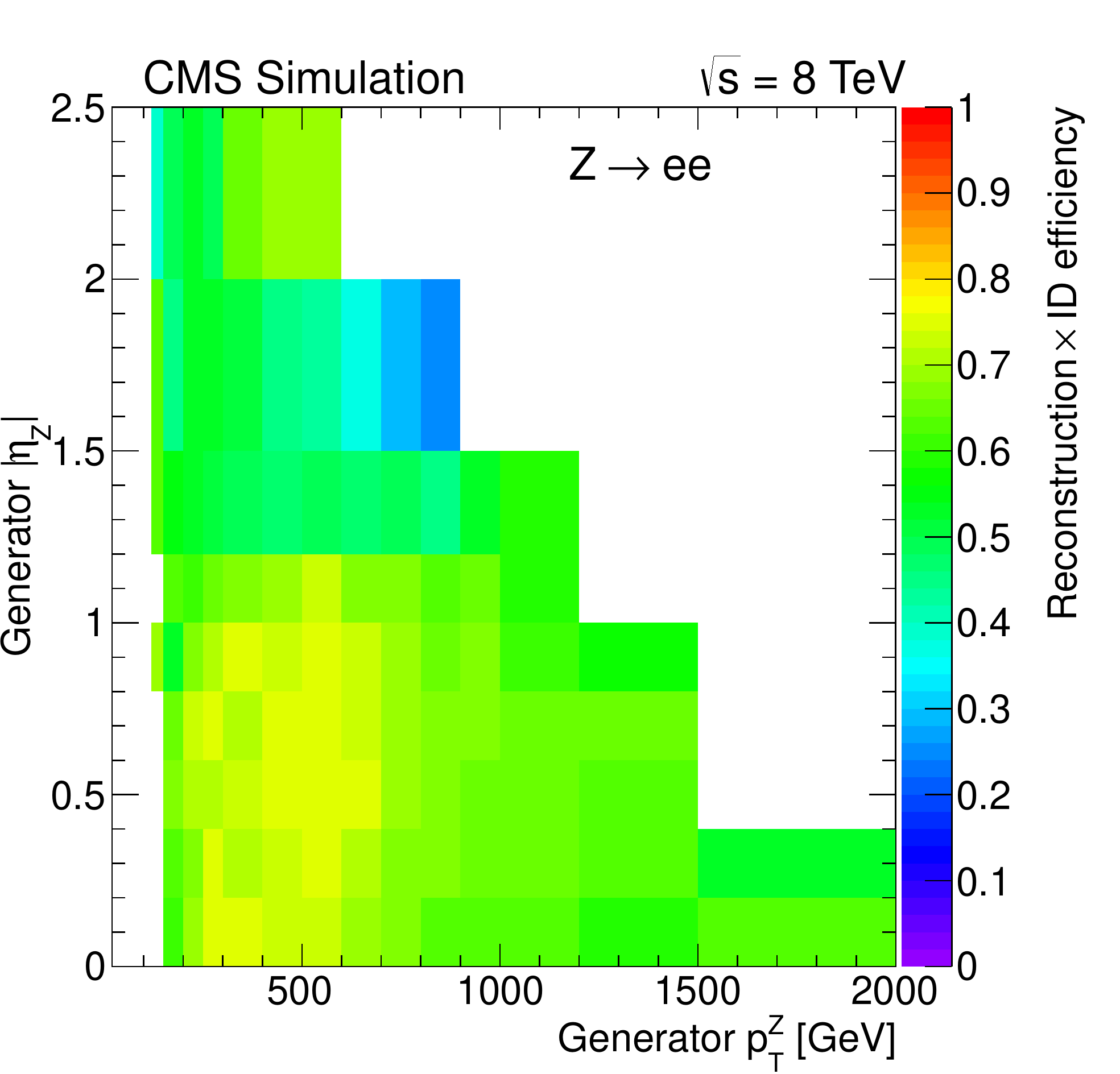}
\includegraphics[width=\cmsFigWidth]{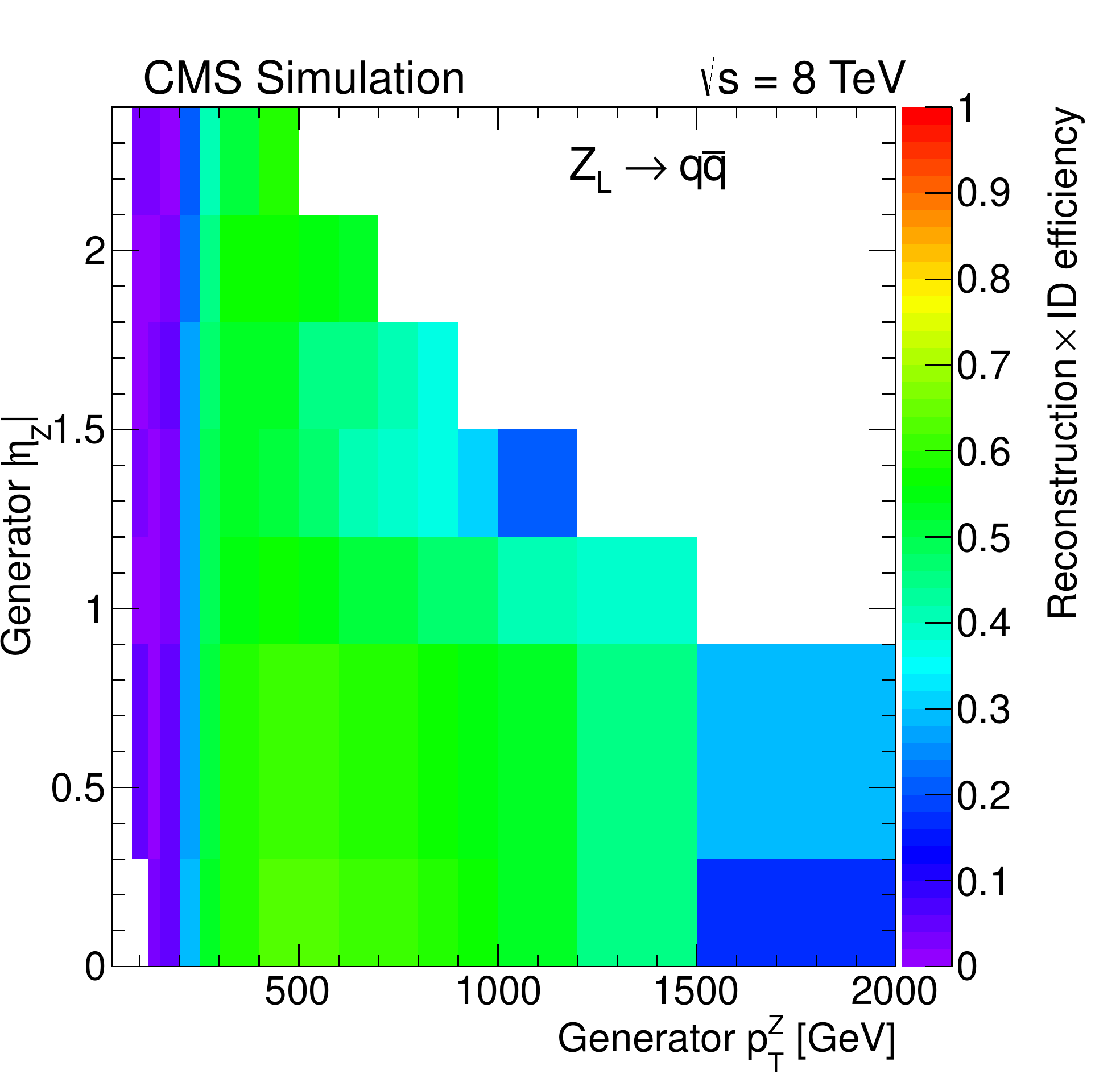}
\includegraphics[width=\cmsFigWidth]{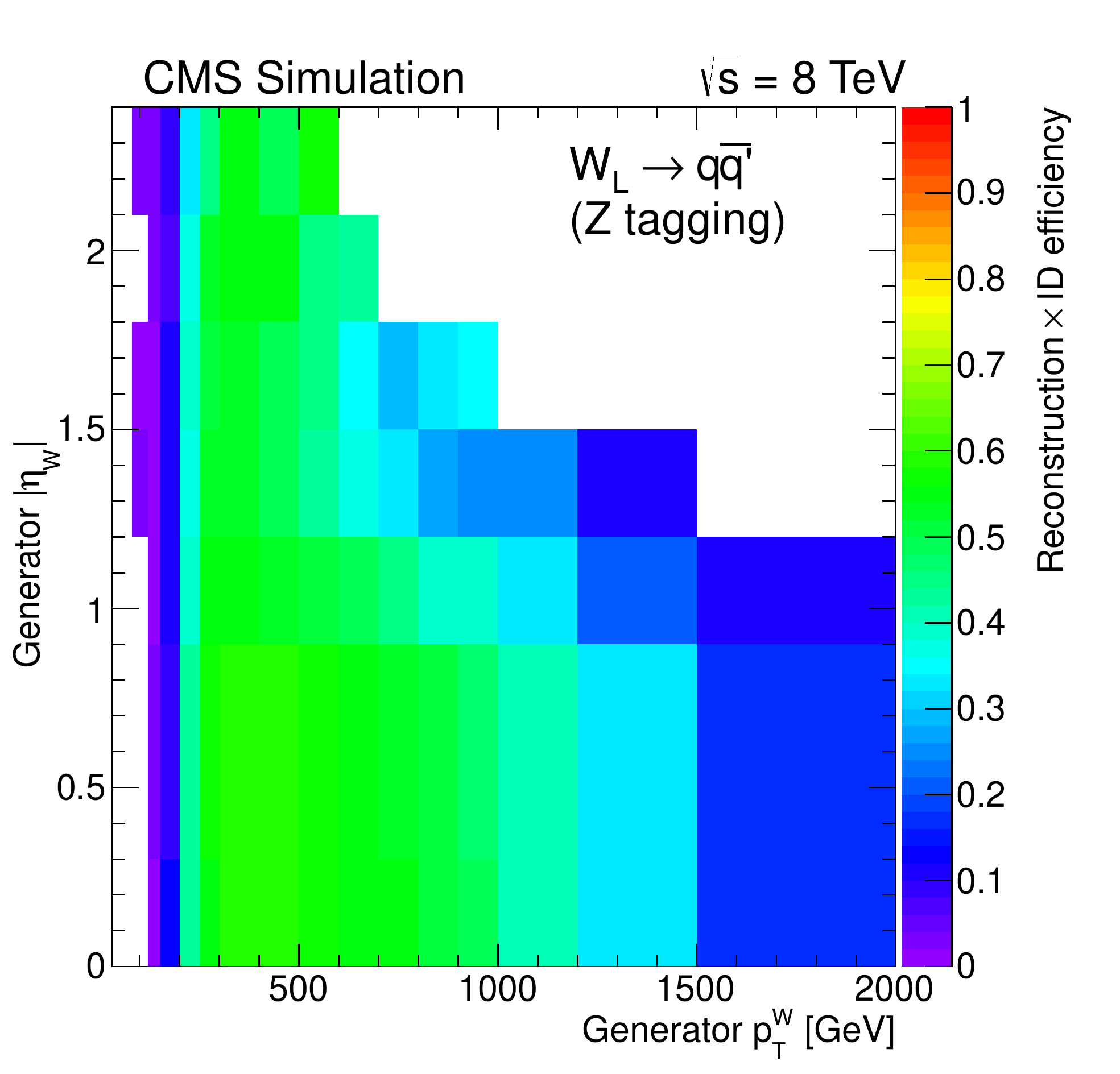}
\caption{Reconstruction and identification efficiencies for the
$\Zo\to \mu\mu$ (top left),  $\Zo\to\Pe\Pe$ (top right),
$\Zo_{\text{L}}\to \Pq\Paq$ (bottom left), and $\PW_\mathrm{L}\to \Pq\Paq'$ (bottom right)
decays as a function of generated $\pt^\mathrm{V}$ and $\eta_\mathrm{V}$ using the Z-tagging requirements for the hadronic V decays.}
\label{fig:eff_Z}
\end{figure}

It was checked that the dependence of the
total signal efficiency and acceptance on the width of
the generated sample is very mild. We
include this effect in the systematic uncertainties of the procedure as discussed later. The resulting
efficiencies are presented in Figs.~\ref{fig:eff_W} and~\ref{fig:eff_Z} for W and
\Zo bosons with longitudinal polarization, respectively.
The contribution from $\Zo\to \tau\tau$ decays with $\tau \to \ell \nu \nu$
is not reported since it is suppressed by the dilepton-mass requirement
of the analysis described in Section~\ref{subsec:Vlept}.
The same values are presented in tabulated form in Appendix \ref{sec:mod-indep-instr}.

Special care must be given to cases where the boson is transversely
polarized ($\text{V}_{\text{T}}$). The calculated efficiencies are based on
longitudinally polarized bosons, as
in the case of
the reference
bulk graviton model. The efficiency of the V-tagging selections depend
significantly on the degree of polarization of the vector
boson~\cite{CMS:JME13006}.  This effect is investigated with samples of RS1
gravitons produced with the \MADGRAPH generator. The V bosons
originating from the decays of RS1 gravitons are transversely polarized in
about 90\% of the cases. In the cases of bosons decaying leptonically, the
tables provided are still valid because of the generator-level
selection on the individual leptons, which guarantees that
polarization effects for the leptonic boson are included in the
acceptance. As shown in Ref.~\cite{CMS:JME13006}, the efficiency of the jet
substructure selection is found to be lower for transversely
polarized V bosons.  Studies of simulated RS1 graviton samples show
that the loss of efficiency is largely independent of the V
kinematic variables, so that the effect of the transverse polarization can be
adequately modeled by a constant scale factor of
0.85, independent of the $\eta$ and \PT of the
$\Vo \to \Pq\Paq$.

To validate the procedure, the resulting parametrized efficiencies
(including the event-veto efficiencies) are used to predict
the total efficiency for reconstructing bulk and
RS1 gravitons, and the estimation is compared to the exact number obtained
from performing the baseline analysis directly on the simulated samples.
In all cases, the agreement between the nominal and parametrized
efficiencies is within 10\% of their value.
Various approximations and uncertainties contribute to the final
additional systematic uncertainty in the efficiency;
the main ones are unaccounted correlations
between the physics objects, statistical uncertainties due to
limited size of the simulated sample, and residual dependencies on the natural width.
We assign an additional systematic uncertainty of 15\% on the total
signal efficiency when calculating the model-independent limits. This
additional systematic uncertainty conservatively addresses the remaining
imperfections in the parametrization of the efficiencies.

Figure~\ref{fig:UL_scan2D} and
Tables~\ref{tab:simplim_WW} and~\ref{tab:simplim_ZZ}
of Appendix~\ref{sec:mod-indep-instr} show the observed limits on the number of
events extracted from the simplified analysis, for the \lnujet and \lljet
analyses independently. The two analyses are not combined in order to
avoid assumptions on the branching fractions of a hypothetic resonance decaying to both WW
and ZZ channels. The limits are calculated using an asymptotic
approximation of the $\mathrm{CL}_S$ method \cite{AsymptCLs}.
Under the narrow-width
approximation, it is explicitly checked that the central values for
the expected and observed limits returned by the full hybrid
frequentist method and the asymptotic approximation match extremely closely
over all the range of the search. All the systematic uncertainties considered in
the baseline analysis are included in the calculation of these
limits, together with the additional 15\% uncertainty related to the
approximations used for parametrizing the efficiencies. The main
features of the observed limits presented in Section~\ref{subsec:bulk_limits}
are still visible. With increasing width, statistical fluctuations in
the limit tend to be smoothed out and the overall performance
degrades. For relative widths greater than 0.25, the deterioration
of the limit is very mild, because the sensitivity coming from the
knowledge of the signal shape is diluted by the very broad signal shape.

\begin{figure}[h!]
\centering
\includegraphics[width=\cmsFigWidth]{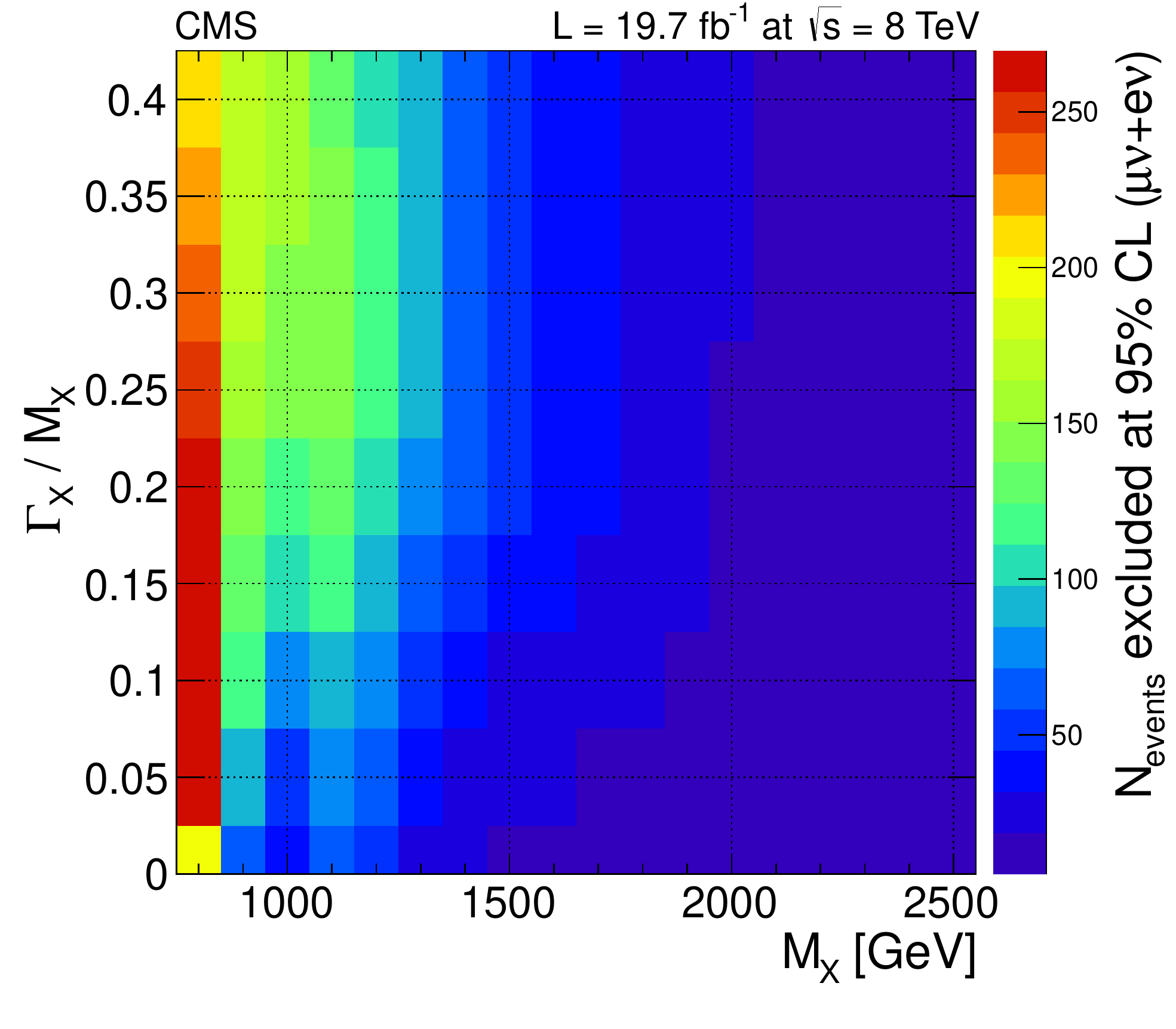}
\includegraphics[width=\cmsFigWidth]{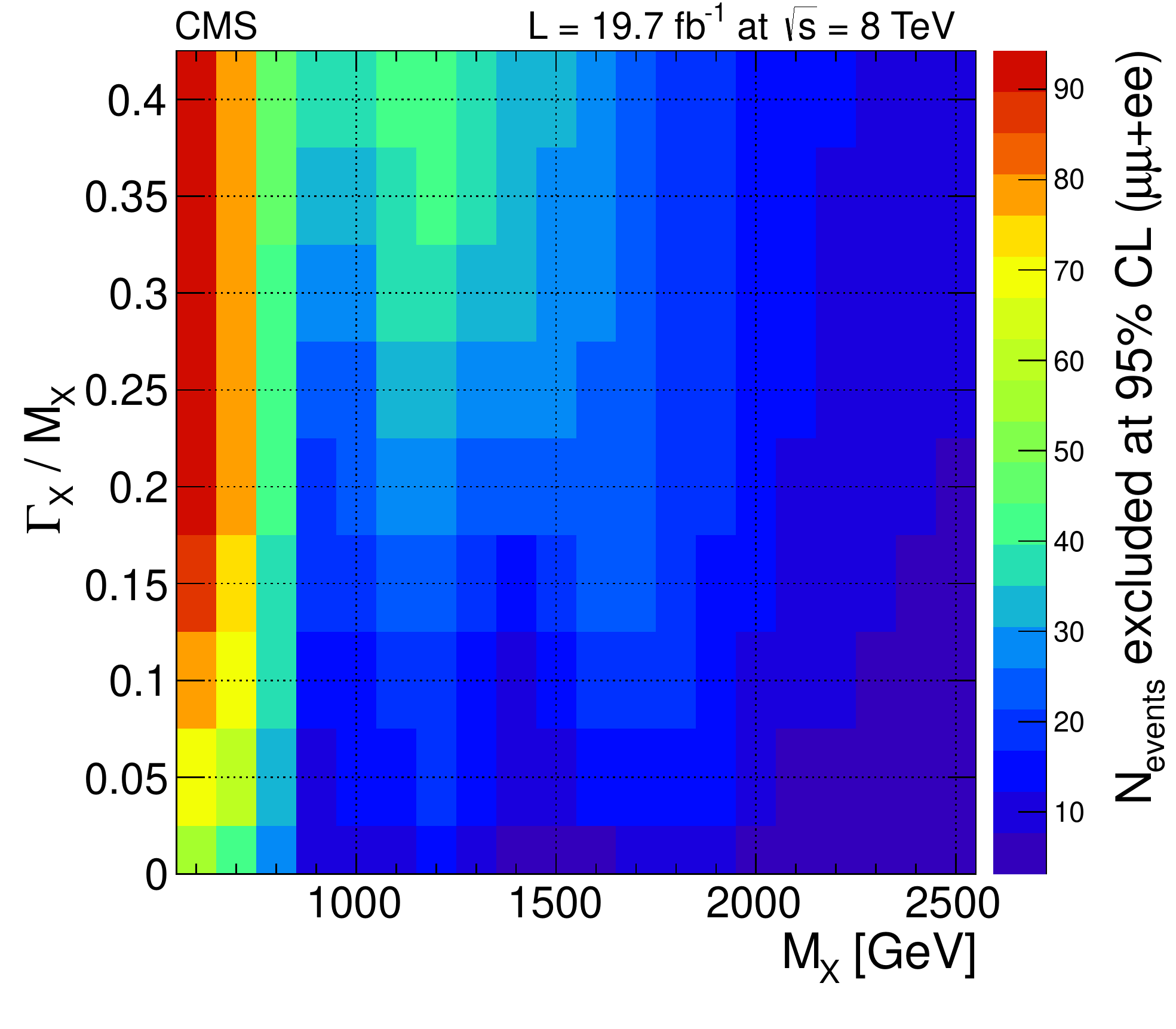}
\caption{Observed exclusion limits at 95\% CL on the number of events for
a $\PW\Vo\to \lnujet$~(left) and a $\cPZ\Vo\to \lljet$~(right) resonance, as a function
of its mass and normalized width.}
\label{fig:UL_scan2D}
\end{figure}

Although optimized for WW and \Zo\Zo resonances, the analysis is also sensitive
to charged resonances decaying to W\Zo, because of the loose requirement on
the V-jet mass. The efficiencies to identify a longitudinally polarized
\Zo (W) boson using W-jet (Z-jet) requirements are computed from
the bulk graviton samples using the same procedure described above,
and they are reported in Fig.~\ref{fig:eff_W} (Fig.~\ref{fig:eff_Z}).
The same values are presented in tabulated form in Appendix \ref{sec:mod-indep-instr}.
Also in this case, the effect of the transverse polarization of the bosons
is modeled by multiplying the aforementioned efficiencies by the constant scale
factor 0.85. In addition, in the \lnujet channel, the combined efficiency of the
second-lepton veto and b-jet veto becomes $\varepsilon_{\text{vetoes}}=81\%$, because of the
presence of $\Zo \to \bbbar$~decays, which can be rejected
by the requirement of the analysis that no jet is tagged as coming from a b quark.

\section{Summary}
\label{sec:summary}
We have presented a search for new resonances decaying to WW, ZZ, or WZ in which one
of the bosons decays leptonically and the other hadronically.
The final states considered are either $\ell \nu \qqbarpr$ or $\ell \ell \qqbarpr$ with
$\ell=\mu$ or \Pe.
The results include the case in which $\PW  \to \tau\nu$ or $\cPZ  \to \tau\tau$ where the tau decay is $\tau \to \ell\nu\nu$.
The events are reconstructed as a leptonic W or \Zo candidate
recoiling against a jet with mass compatible with the W or Z mass, respectively.
Additional information from jet substructure is used to reduce the
background from multijet processes. No evidence for a signal
is found, and the result is interpreted as an upper limit on the
production cross section as a function of the resonance mass in the
context of the bulk graviton model.
The final upper limits
are based on the statistical
combination of the two semi-leptonic channels considered here
with those of a complementary search in the fully-hadronic final state.
Upper limits at
95\% CL are set on the bulk graviton production cross section
in the range from 700 to 10\unit{fb} for resonance masses
between 600 and 2500\GeV, respectively. These limits are the
most stringent to date in these final states.
The two analyses in the semi-leptonic channels are repeated in a simplified scenario,
providing model-independent limits on the number of events.
The tabulated efficiency of reconstructing the vector bosons within
the kinematic acceptance of the analysis allows the reinterpretation
of the exclusion limits in a generic phenomenological model, including W\Zo resonances,
greatly extending the versatility of these results.

\section*{Acknowledgments}

\hyphenation{Bundes-ministerium Forschungs-gemeinschaft Forschungs-zentren}
We congratulate our colleagues in the CERN accelerator departments for
the excellent performance of the LHC and thank the technical and
administrative staffs at CERN and at other CMS institutes for their
contributions to the success of the CMS effort. In addition, we
gratefully acknowledge the computing centres and personnel of the
Worldwide LHC Computing Grid for delivering so effectively the
computing infrastructure essential to our analyses. Finally, we
acknowledge the enduring support for the construction and operation of
the LHC and the CMS detector provided by the following funding
agencies: the Austrian Federal Ministry of Science and Research and
the Austrian Science Fund; the Belgian Fonds de la Recherche
Scientifique, and Fonds voor Wetenschappelijk Onderzoek; the Brazilian
Funding Agencies (CNPq, CAPES, FAPERJ, and FAPESP); the Bulgarian
Ministry of Education and Science; CERN; the Chinese Academy of
Sciences, Ministry of Science and Technology, and National Natural
Science Foundation of China; the Colombian Funding Agency
(COLCIENCIAS); the Croatian Ministry of Science, Education and Sport;
the Research Promotion Foundation, Cyprus; the Ministry of Education
and Research, Recurrent financing contract SF0690030s09 and European
Regional Development Fund, Estonia; the Academy of Finland, Finnish
Ministry of Education and Culture, and Helsinki Institute of Physics;
the Institut National de Physique Nucl\'eaire et de Physique des
Particules~/~CNRS, and Commissariat \`a l'\'Energie Atomique et aux
\'Energies Alternatives~/~CEA, France; the Bundesministerium f\"ur
Bildung und Forschung, Deutsche Forschungsgemeinschaft, and
Helmholtz-Gemeinschaft Deutscher Forschungszentren, Germany; the
General Secretariat for Research and Technology, Greece; the National
Scientific Research Foundation, and National Office for Research and
Technology, Hungary; the Department of Atomic Energy and the
Department of Science and Technology, India; the Institute for Studies
in Theoretical Physics and Mathematics, Iran; the Science Foundation,
Ireland; the Istituto Nazionale di Fisica Nucleare, Italy; the Korean
Ministry of Education, Science and Technology and the World Class
University program of NRF, Republic of Korea; the Lithuanian Academy
of Sciences; the Mexican Funding Agencies (CINVESTAV, CONACYT, SEP,
and UASLP-FAI); the Ministry of Business, Innovation and Employment,
New Zealand; the Pakistan Atomic Energy Commission; the Ministry of
Science and Higher Education and the National Science Centre, Poland;
the Funda\c{c}\~ao para a Ci\^encia e a Tecnologia, Portugal; JINR,
Dubna; the Ministry of Education and Science of the Russian
Federation, the Federal Agency of Atomic Energy of the Russian
Federation, Russian Academy of Sciences, and the Russian Foundation
for Basic Research; the Ministry of Education, Science and
Technological Development of Serbia; the Secretar\'{\i}a de Estado de
Investigaci\'on, Desarrollo e Innovaci\'on and Programa
Consolider-Ingenio 2010, Spain; the Swiss Funding Agencies (ETH Board,
ETH Zurich, PSI, SNF, UniZH, Canton Zurich, and SER); the National
Science Council, Taipei; the Thailand Center of Excellence in Physics,
the Institute for the Promotion of Teaching Science and Technology of
Thailand, Special Task Force for Activating Research and the National
Science and Technology Development Agency of Thailand; the Scientific
and Technical Research Council of Turkey, and Turkish Atomic Energy
Authority; the Science and Technology Facilities Council, UK; the US
Department of Energy, and the US National Science Foundation.

Individuals have received support from the Marie-Curie programme and
the European Research Council and EPLANET (European Union); the
Leventis Foundation; the A. P. Sloan Foundation; the Alexander von
Humboldt Foundation; the Belgian Federal Science Policy Office; the
Fonds pour la Formation \`a la Recherche dans l'Industrie et dans
l'Agriculture (FRIA-Belgium); the Agentschap voor Innovatie door
Wetenschap en Technologie (IWT-Belgium); the Ministry of Education,
Youth and Sports (MEYS) of Czech Republic; the Council of Science and
Industrial Research, India; the Compagnia di San Paolo (Torino); the
HOMING PLUS programme of Foundation for Polish Science, cofinanced by
EU, Regional Development Fund; and the Thalis and Aristeia programmes
cofinanced by EU-ESF and the Greek NSRF.

\clearpage

\bibliography{auto_generated}   

\clearpage

\appendix

\section{Detailed instructions and additional material for generic interpretation of the results}
\label{sec:mod-indep-instr}

This section presents a technical description of the procedure for calculating
the signal yield expected to be observed in the CMS detector in a scenario with
a new resonance, X, decaying to two vector bosons (WW, ZZ, WZ) in the semi-leptonic
final states \lnujet and \lljet.
Tables \ref{tab:eff_W_mu}
to~\ref{tab:eff_W_jet} provide the efficiencies for the reconstruction
and identification of each of the two W vector bosons coming from the
$\mathrm{X}\to\PW\PW$ decay. Tables \ref{tab:eff_Z_mu} to~\ref{tab:eff_Z_jet} provide the
efficiencies for the reconstruction and identification of each of the
two \Zo bosons coming from the $\mathrm{X}\to\cPZ\cPZ$ decay.
In case of hadronic V decays from $\mathrm{X}\to\PW\Zo$ resonances the following tables should be:
Table~\ref{tab:eff_W_jet_forZqq} provides the efficiencies
for the reconstruction and identification of $\cPZ_\mathrm{L} \to \Pq\Paq$
using W-tagging requirements; Table~\ref{tab:eff_Z_jet_forWqq} provides the efficiencies
for the reconstruction and identification of $\PW_\mathrm{L}\to \Pq\Paq'$
using Z-tagging requirements.
The efficiencies are calculated using the reference bulk graviton samples, listed in
Section~\ref{sec:samples}.

These efficiencies can be applied to a generic model with the following procedure:
\begin{enumerate}
\item Generate a sample of events for a given mass and width of the X resonance;
the simulated process must include the decay of the X resonance
to leptons and quarks (including W$\to \tau\nu \to \ell\nu\nu\nu$ decays).
\item Filter the events according to the criteria listed in
Table~\ref{tab:eff_gensel_WW} (if the X resonance decays to WW) and
Table~\ref{tab:eff_gensel_ZZ} (if the X resonance decays to \Zo\Zo).
If the resonance decays to $\PW\cPZ\to \ell\nu \Pq\Paq$,
the criteria for hadronic W in Table~\ref{tab:eff_gensel_WW} should be applied on the
generated hadronic Z. If the resonance decays to $\cPZ\PW\to \ell\ell \Pq\Paq'$,
the criteria for hadronic \Zo in Table~\ref{tab:eff_gensel_ZZ} should be applied on the
generated hadronic W.
\item For each of the remaining events, calculate the efficiency
for reconstructing the $\Wo\to \mu\nu$ / $\Wo\to \tau\nu \to \mu \nu\nu \nu$,
$\Wo\to a \nu$ / $\Wo\to \tau \nu \to \Pe \nu\nu\nu$,
$\Zo\to \mu\mu$, and $\Zo\to \Pe\Pe$ using the Tables
\ref{tab:eff_W_mu}, \ref{tab:eff_W_ele}, \ref{tab:eff_Z_mu},
and \ref{tab:eff_Z_ele}, respectively. The tables provide the
efficiency parametrized as a function of \PT and $\eta$ of the W or \Zo boson.
\item In a similar way, for WW and ZZ resonances, calculate the efficiency of the hadronic W or Z
using the values in Table~\ref{tab:eff_W_jet} or~\ref{tab:eff_Z_jet}, respectively.
If the resonance decays to $\PW\cPZ\to \ell\nu \Pq\Paq$, calculate the efficiency of the hadronic Z
using Table~\ref{tab:eff_W_jet_forZqq}. If the resonance decays to $\cPZ\PW\to \ell\ell \Pq\Paq'$,
calculate the efficiency of the hadronic W using Table~\ref{tab:eff_Z_jet_forWqq}.
\item Weight each passing event with the product of the two efficiencies found
at steps 3 and 4. In case of a X resonance decaying to WW (\lnujet channel),
also multiply by the combined efficiency of the second-lepton
and b-jet vetoes, equal to $\varepsilon_{\text{vetoes}}=90\%$.
In case of a X resonance decaying to $\PW\cPZ \to \ell\nu \Pq\Paq$,
the combined efficiency of the two vetoes is equal to $\varepsilon_{\text{vetoes}}=81\%$.
\item The resulting sum of weights divided by the total number of
events provides an approximation to the total efficiency for the given model.
\end{enumerate}

The final number of events can be directly compared to the observed
limits in Fig.~\ref{fig:UL_scan2D} and
Tables~\ref{tab:simplim_WW} and~\ref{tab:simplim_ZZ},
 in order to assess the exclusion
power of the experiment with respect to the model considered.

The numbers provided refer to longitudinally polarized bosons. For
transversely polarized bosons that decay leptonically, the same
numbers are valid, as long as they are applied after the kinematic
acceptance requirements. If the boson decays to quarks and has a transverse
polarization, the efficiency must be scaled down by a factor of 0.85.

\begin{table}[ht]
\centering
\topcaption{Generator level requirements for the WW analysis, to be used for the
computation of the efficiency parametrization.
The vector sum of the transverse neutrino momenta $\sum{\vec{p}_{\mathrm{T},\nu}}$ is taken over all the neutrinos in the final state, coming either from
$\PW\to \ell\nu$ or $W\to \tau\nu \to \ell\nu\nu\nu$ decays with $\ell=\mu$ or \Pe.}
\label{tab:eff_gensel_WW}
\renewcommand{\arraystretch}{1.2}
\begin{tabular}{lr}
{Object} & {Requirement} \\
\hline
\hline
Muons & $\abs{\eta}<2.1$ \\
      & $\pt > 50\GeV$ \\
\hline
Electrons & $\abs{\eta}<2.5$ \\
          & $\pt > 90\GeV$ \\
\hline
$\sum{\vec{p}_{\mathrm{T},\nu}}$ (Muon ch.) & $\pt>40\GeV$\\
$\sum{\vec{p}_{\mathrm{T},\nu}}$ (Electron ch.) & $\pt>80\GeV$\\
\hline
W$\to \ell\nu$ or $W\to \tau\nu \to \ell\nu\nu\nu$ &  $\pt^{\Wo} >200\GeV$  \\
\hline
$\PW\to \Pq\Paq'$ & $\abs{\eta_{\Wo} }<2.4$ \\
                    & $\pt^{\Wo} >200\GeV$ \\
                    & $65 < m_{\Pq\Paq'} < 105\GeV$ \\
\hline
WW system      & $700 < \mWW < 3000\GeV$\\
               & $\Delta R(\Wo_{\Pq\Paq'},\ell)>\pi/2$\\
               & $\Delta\phi(\Wo_{\Pq\Paq'},\sum{\vec{p}_{\mathrm{T},\nu}})>2$\\
               & $\Delta\phi(\Wo_{\Pq\Paq'},\Wo_{\ell\nu})>2$\\
\hline
\end{tabular}
\end{table}

\begin{table}[ht]
\centering
\topcaption{Generator level requirements for the \Zo\Zo analysis, to be used for the
computation of the efficiency parametrization, with $\ell=\mu$ or \Pe. }
\label{tab:eff_gensel_ZZ}
\renewcommand{\arraystretch}{1.2}
\begin{tabular}{lr}
{Object} & {Requirement} \\
\hline
\hline
Muons & $\abs{\eta}<2.4$\\
      & $\pt > 20\GeV$\\
Highest-\pt muon & $\pt > 40\GeV$\\
\hline
Electrons & $\abs{\eta}<2.5$\\
          & $\pt > 40\GeV$\\
\hline
$\cPZ\to \ell\ell$ & $\pt^\Zo >80\GeV$ \\
                        & $70< m_{\ell\ell} < 110\GeV$ \\
\hline
$\cPZ\to \Pq\Paq$ & $\abs{\eta_{\Zo}}< 2.4$ \\
                        & $\pt^\Zo >80\GeV$ \\
                        & $70< m_{\Pq\Paq} < 110\GeV$ \\
\hline
ZZ system & $500 < m_{\cPZ\cPZ} < 2800\GeV$\\
\hline
\end{tabular}
\end{table}

\begin{table}[hbt]
\centering
\topcaption{Simplified limits on the number of visible events from generic WV resonances in the \lnujet channel as a function of resonance mass, $M_\mathrm{X}$, and normalized width, $\Gamma_\mathrm{X}/M_\mathrm{X}$. Shown are limits on the visible number of events at 95\% CL using the asymptotic $\mathrm{CL}_S$ approach. Results with $\Gamma_\mathrm{X}/M_\mathrm{X}=0$ are obtained using the resolution function only.}
\label{tab:simplim_WW}
\begin{tabular}{l c c c c c c c c c }
${M_\mathrm{X} [\GeVns]}$&\multicolumn{9}{c}{${\Gamma_\mathrm{X} / M_\mathrm{X}}$}\\
   & 0.00 & 0.05 & 0.10 & 0.15 & 0.20 & 0.25 & 0.30 & 0.35 & 0.40 \\
\hline
\hline
800& 200 & 260 & 270 & 270 & 260 & 250 & 240 & 230 & 210 \\
900& 70 & 93 & 113 & 132 & 150 & 160 & 170 & 170 & 170 \\
1000& 42 & 58 & 77 & 99 & 120 & 139 & 150 & 150 & 154 \\
1100& 62 & 78 & 95 & 110 & 130 & 139 & 140 & 140 & 137 \\
1200& 51 & 68 & 82 & 97 & 110 & 120 & 120 & 120 & 110 \\
1300& 30 & 42 & 54 & 69 & 82 & 89 & 91 & 89 & 85 \\
1400& 23 & 30 & 39 & 50 & 61 & 67 & 69 & 68 & 66 \\
1500& 18 & 24 & 32 & 41 & 48 & 52 & 53 & 53 & 51 \\
1600& 14 & 19 & 26 & 34 & 39 & 42 & 42 & 42 & 41 \\
1700& 13 & 18 & 24 & 29 & 32 & 33 & 34 & 34 & 33 \\
1800& 12 & 16 & 21 & 24 & 26 & 27 & 28 & 28 & 27 \\
1900& 8.7 & 13 & 16 & 19 & 21 & 22 & 22 & 22 & 22 \\
2000& 8.3 & 11 & 14 & 16 & 17 & 18 & 19 & 19 & 19 \\
2100& 7.5 & 9.5 & 12 & 13 & 14 & 15 & 16 & 16 & 16 \\
2200& 5.3 & 7.8 & 9.8 & 11 & 12 & 13 & 13 & 14 & 14 \\
2300& 5.4 & 7.1 & 8.6 & 9.7 & 11 & 11 & 12 & 12 & 13 \\
2400& 5.6 & 6.8 & 7.8 & 8.7 & 9.4 & 9.9 & 10 & 11 & 11 \\
2500& 5.3 & 6.5 & 7.3 & 8.0 & 8.6 & 9.1 & 9.6 & 10 & 10 \\
\hline
\end{tabular}
\end{table}
\begin{table}[hbt]
\centering
\topcaption{Simplified limits on the number of visible events from generic ZV resonances in the \lljet channel as a function of resonance mass, $M_\mathrm{X}$, and normalized width, $\Gamma_\mathrm{X}/M_\mathrm{X}$. Shown are limits on the visible number of events at 95\% CL using the asymptotic $\mathrm{CL}_S$ approach. Results with $\Gamma_\mathrm{X}/M_\mathrm{X}=0$ are obtained using the resolution function only.}
\label{tab:simplim_ZZ}
\small
\begin{tabular}{l c c c c c c c c c }
${M_\mathrm{X} [\GeVns]}$&\multicolumn{9}{c}{${\Gamma_\mathrm{X} / M_\mathrm{X}}$}\\
& 0.00 & 0.05 & 0.10 & 0.15 & 0.20 & 0.25 & 0.30 & 0.35 & 0.40 \\
\hline
\hline
600& 53 & 68 & 78 & 85 & 90 & 93 & 94 & 94 & 93 \\
700& 43 & 58 & 69 & 76 & 79 & 80 & 79 & 78 & 77 \\
800& 27 & 33 & 37 & 39 & 41 & 42 & 44 & 46 & 48 \\
900& 8.4 & 11 & 14 & 17 & 21 & 24 & 28 & 32 & 36 \\
1000& 11 & 14 & 16 & 19 & 22 & 25 & 29 & 33 & 37 \\
1100& 12 & 16 & 20 & 24 & 28 & 32 & 36 & 39 & 42 \\
1200& 14 & 18 & 21 & 25 & 29 & 33 & 37 & 41 & 43 \\
1300& 11 & 13 & 16 & 20 & 25 & 30 & 35 & 37 & 38 \\
1400& 5.2 & 7.9 & 11 & 16 & 22 & 28 & 31 & 33 & 33 \\
1500& 7.0 & 9.7 & 14 & 20 & 25 & 28 & 30 & 30 & 31 \\
1600& 7.5 & 12 & 18 & 22 & 25 & 26 & 26 & 27 & 27 \\
1700& 9.6 & 15 & 20 & 22 & 23 & 23 & 24 & 24 & 24 \\
1800& 10 & 15 & 18 & 19 & 20 & 20 & 21 & 21 & 21 \\
1900& 9.5 & 13 & 15 & 16 & 17 & 18 & 18 & 18 & 18 \\
2000& 6.3 & 9.5 & 12 & 14 & 14 & 15 & 15 & 16 & 16 \\
2100& 3.3 & 5.8 & 9.3 & 11 & 12 & 13 & 13 & 14 & 14 \\
2200& 3.1 & 5.4 & 7.9 & 9.2 & 10 & 11 & 12 & 12 & 13 \\
2300& 4.3 & 6.4 & 7.4 & 8.2 & 9.0 & 9.6 & 10 & 11 & 11 \\
2400& 5.9 & 6.5 & 7.0 & 7.6 & 8.2 & 8.8 & 9.3 & 9.9 & 10 \\
2500& 5.9 & 6.3 & 6.7 & 7.1 & 7.6 & 8.1 & 8.6 & 9.1 & 9.6 \\
\hline
\end{tabular}
\end{table}

\begin{table}[hbt]
\centering
\topcaption{Reconstruction and identification efficiency for the $\Wo\to \mu \nu$ and $\Wo \to \tau \nu \to \mu \nu\nu\nu$ decays
as function of generated $\pt^\Wo$ and $\abs{\eta_\Wo}$. Uncertainties on the efficiencies are included in the generic limit calculation as discussed in the text.}
\label{tab:eff_W_mu}
\resizebox{\textwidth}{!}{
\begin{tabular}{l c c c c c c c c c c }
${\pt^\Wo}$ {range} [\GeVns]&\multicolumn{10}{c}{${\abs{\eta_\Wo}}$ {range}}\\
 & 0.0--0.2 & 0.2--0.4 & 0.4--0.6 & 0.6--0.8 & 0.8--1.0 & 1.0--1.2 & 1.2--1.5 & 1.5--2.0 & 2.0--2.4 & 2.4--3.0 \\
\hline
\hline
200--250& 0.75 & 0.78 & 0.70 & 0.76 & 0.66 & 0.60 & 0.63 & 0.59 & \NA & \NA \\
250--300& 0.84 & 0.83 & 0.85 & 0.82 & 0.79 & 0.76 & 0.71 & 0.71 & 0.75 & \NA \\
300--400& 0.85 & 0.86 & 0.86 & 0.86 & 0.81 & 0.77 & 0.74 & 0.74 & 0.71 & \NA \\
400--500& 0.86 & 0.86 & 0.87 & 0.86 & 0.81 & 0.77 & 0.75 & 0.71 & 0.79 & \NA \\
500--600& 0.86 & 0.85 & 0.88 & 0.86 & 0.82 & 0.77 & 0.76 & 0.73 & \NA & \NA \\
600--700& 0.87 & 0.85 & 0.88 & 0.87 & 0.82 & 0.77 & 0.76 & 0.74 & \NA & \NA \\
700--800& 0.86 & 0.85 & 0.88 & 0.86 & 0.82 & 0.77 & 0.75 & 0.74 & \NA & \NA \\
800--900& 0.86 & 0.84 & 0.88 & 0.88 & 0.82 & 0.78 & 0.75 & 0.67 & \NA & \NA \\
900--1000& 0.87 & 0.85 & 0.89 & 0.88 & 0.81 & 0.77 & 0.76 & \NA & \NA & \NA \\
1000--1200& 0.87 & 0.85 & 0.90 & 0.88 & 0.82 & 0.77 & 0.76 & \NA & \NA & \NA \\
1200--1500& 0.87 & 0.85 & 0.90 & 0.89 & 0.84 & 0.80 & \NA & \NA & \NA & \NA \\
1500--2000& 0.90 & 0.82 & 0.89 & 0.85 & 0.80 & \NA & \NA & \NA & \NA & \NA \\
\hline
\end{tabular}
}
\end{table}

\begin{table}[hbt]
\centering
\topcaption{Reconstruction and identification efficiency for the $\Wo\to \Pe \nu$ and $\Wo \to \tau \nu \to \Pe \nu\nu\nu$ decays as a function of generated $\pt^\Wo$ and $\abs{\eta_\Wo}$. Uncertainties on the efficiencies are included in the generic limit calculation as discussed in the text.}
\label{tab:eff_W_ele}
\resizebox{\textwidth}{!}{
\begin{tabular}{l c c c c c c c c c c }
${\pt^\Wo}$ {range} [\GeVns]&\multicolumn{10}{c}{${\abs{\eta_\Wo}}$ {range}}\\
 & 0.0--0.2 & 0.2--0.4 & 0.4--0.6 & 0.6--0.8 & 0.8--1.0 & 1.0--1.2 & 1.2--1.5 & 1.5--2.0 & 2.0--2.5 & 2.5--3.0 \\
\hline
\hline
200--250& 0.73 & 0.68 & 0.70 & 0.71 & 0.65 & 0.80 & 0.62 & 0.57 & \NA & \NA \\
250--300& 0.84 & 0.80 & 0.83 & 0.85 & 0.84 & 0.79 & 0.67 & 0.65 & 0.75 & \NA \\
300--400& 0.84 & 0.82 & 0.83 & 0.83 & 0.85 & 0.81 & 0.69 & 0.68 & 0.76 & \NA \\
400--500& 0.84 & 0.85 & 0.86 & 0.85 & 0.85 & 0.83 & 0.71 & 0.65 & 0.83 & \NA \\
500--600& 0.84 & 0.83 & 0.83 & 0.86 & 0.84 & 0.84 & 0.72 & 0.68 & \NA & \NA \\
600--700& 0.83 & 0.85 & 0.85 & 0.85 & 0.84 & 0.83 & 0.73 & 0.68 & \NA & \NA \\
700--800& 0.85 & 0.85 & 0.87 & 0.85 & 0.86 & 0.84 & 0.75 & 0.65 & \NA & \NA \\
800--900& 0.84 & 0.85 & 0.85 & 0.87 & 0.84 & 0.86 & 0.75 & 0.61 & \NA & \NA \\
900--1000& 0.84 & 0.85 & 0.86 & 0.85 & 0.84 & 0.85 & 0.76 & \NA & \NA & \NA \\
1000--1200& 0.84 & 0.86 & 0.86 & 0.86 & 0.85 & 0.89 & 0.78 & \NA & \NA & \NA \\
1200--1500& 0.86 & 0.86 & 0.88 & 0.86 & 0.89 & 0.84 & \NA & \NA & \NA & \NA \\
1500--2000& 0.87 & 0.90 & 0.85 & 0.84 & 0.91 & \NA & \NA & \NA & \NA & \NA \\
\hline
\end{tabular}
}
\end{table}

\begin{table}[hbt]
\centering
\topcaption{Reconstruction and identification efficiency for the $\Wo_{\text{L}} \to \Pq\Paq'$ decay as a function of generated $\pt^\Wo$ and $\abs{\eta_\Wo}$ using W-tagging requirements. Uncertainties on the efficiencies are included in the generic limit calculation as discussed in the text.}
\label{tab:eff_W_jet}
\vspace*{\medskipamount}
\small
\begin{tabular}{l c c c c c c c}
${\pt^\Wo}$ {range} [\GeVns]&\multicolumn{7}{c}{${\abs{\eta_\Wo}}$ {range}}\\
 & 0.0--0.3 & 0.3--0.9 & 0.9--1.2 & 1.2--1.5 & 1.5--1.8 & 1.8--2.1 & 2.1--2.4 \\
\hline
\hline
200--250& 0.43 & 0.40 & 0.35 & 0.35 & 0.37 & 0.32 & 0.33 \\
250--300& 0.58 & 0.57 & 0.56 & 0.56 & 0.56 & 0.59 & 0.46 \\
300--400& 0.62 & 0.61 & 0.57 & 0.54 & 0.55 & 0.55 & 0.53 \\
400--500& 0.61 & 0.61 & 0.56 & 0.50 & 0.50 & 0.58 & 0.54 \\
500--600& 0.61 & 0.60 & 0.52 & 0.45 & 0.44 & 0.43 & \NA \\
600--700& 0.60 & 0.59 & 0.52 & 0.39 & 0.38 & 0.45 & \NA \\
700--800& 0.59 & 0.57 & 0.47 & 0.35 & 0.29 & \NA & \NA \\
800--900& 0.55 & 0.55 & 0.42 & 0.30 & 0.41 & \NA & \NA \\
900--1000& 0.52 & 0.51 & 0.41 & 0.28 & 0.43 & \NA & \NA \\
1000--1200& 0.45 & 0.44 & 0.35 & 0.28 & \NA & \NA & \NA \\
1200--1500& 0.36 & 0.35 & 0.24 & 0.07 & \NA & \NA & \NA \\
1500--2000& 0.21 & 0.21 & 0.16 & \NA & \NA & \NA & \NA \\
\hline
\end{tabular}
\end{table}

\begin{table}[hbt]
\centering
\topcaption{Reconstruction and identification efficiency for the $\Zo\to \mu\mu$ decay as a function of generated $\pt^\Zo$ and $\abs{\eta_\Zo}$. Uncertainties on the efficiencies are included in the generic limit calculation as discussed in the text.}
\label{tab:eff_Z_mu}
\resizebox{\textwidth}{!}{
\begin{tabular}{l c c c c c c c c c c }
${\pt^\Zo}$ {range} [\GeVns]&\multicolumn{10}{c}{${\abs{\eta_\Zo}}$ {range}}\\
 & 0.0--0.2 & 0.2--0.4 & 0.4--0.6 & 0.6--0.8 & 0.8--1.0 & 1.0--1.2 & 1.2--1.5 & 1.5--2.0 & 2.0--2.4 & 2.4--3.0 \\
\hline
\hline
90--120& \NA & \NA & \NA & \NA & \NA & \NA & 0.78 & 0.73 & 0.67 & \NA \\
120--150& 0.68 & \NA & \NA & 0.63 & 0.66 & 0.67 & 0.65 & 0.70 & 0.65 & \NA \\
150--200& 0.82 & 0.80 & 0.74 & 0.78 & 0.74 & 0.78 & 0.77 & 0.70 & 0.64 & \NA \\
200--250& 0.79 & 0.79 & 0.78 & 0.79 & 0.80 & 0.77 & 0.74 & 0.73 & 0.64 & \NA \\
250--300& 0.81 & 0.80 & 0.81 & 0.84 & 0.80 & 0.79 & 0.74 & 0.72 & 0.70 & \NA \\
300--400& 0.80 & 0.81 & 0.82 & 0.82 & 0.79 & 0.78 & 0.73 & 0.68 & 0.53 & \NA \\
400--500& 0.79 & 0.81 & 0.82 & 0.81 & 0.79 & 0.76 & 0.72 & 0.71 & 0.58 & \NA \\
500--600& 0.78 & 0.79 & 0.81 & 0.81 & 0.78 & 0.77 & 0.71 & 0.64 & \NA & \NA \\
600--700& 0.72 & 0.75 & 0.76 & 0.74 & 0.72 & 0.69 & 0.65 & 0.62 & \NA & \NA \\
700--800& 0.65 & 0.67 & 0.68 & 0.68 & 0.64 & 0.60 & 0.62 & 0.46 & \NA & \NA \\
800--900& 0.62 & 0.62 & 0.63 & 0.63 & 0.58 & 0.53 & 0.54 & 0.47 & \NA & \NA \\
900--1000& 0.58 & 0.62 & 0.59 & 0.62 & 0.56 & 0.54 & 0.52 & \NA & \NA & \NA \\
1000--1200& 0.55 & 0.58 & 0.58 & 0.57 & 0.51 & 0.45 & 0.46 & \NA & \NA & \NA \\
1200--1500& 0.54 & 0.54 & 0.53 & 0.51 & 0.43 & \NA & \NA & \NA & \NA & \NA \\
1500--2000& 0.49 & 0.46 & \NA & \NA & \NA & \NA & \NA & \NA & \NA & \NA \\
\hline
\end{tabular}
}
\end{table}

\begin{table}[hbt]
\centering
\topcaption{Reconstruction and identification efficiency for the $\Zo\to \Pe\Pe$ decay as a function of generated $\pt^\Zo$ and $\abs{\eta_\Zo}$. Uncertainties on the efficiencies are included in the generic limit calculation as discussed in the text.}
\label{tab:eff_Z_ele}
\resizebox{\textwidth}{!}{
\begin{tabular}{l c c c c c c c c c c }
${\pt^\Zo}$ {range} [\GeVns]&\multicolumn{10}{c}{${\abs{\eta_\Zo}}$ {range}}\\
 & 0.0--0.2 & 0.2--0.4 & 0.4--0.6 & 0.6--0.8 & 0.8--1.0 & 1.0--1.2 & 1.2--1.5 & 1.5--2.0 & 2.0--2.5 & 2.5--3.0 \\
\hline
\hline
120--150& \NA & \NA & \NA & \NA & 0.70 & \NA & 0.62 & 0.63 & 0.39 & \NA \\
150--200& 0.61 & 0.64 & 0.66 & 0.65 & 0.53 & 0.62 & 0.54 & 0.46 & 0.48 & \NA \\
200--250& 0.70 & 0.67 & 0.71 & 0.72 & 0.68 & 0.60 & 0.53 & 0.53 & 0.53 & \NA \\
250--300& 0.74 & 0.74 & 0.71 & 0.76 & 0.72 & 0.65 & 0.51 & 0.54 & 0.48 & \NA \\
300--400& 0.74 & 0.72 & 0.73 & 0.71 & 0.74 & 0.67 & 0.49 & 0.51 & 0.65 & \NA \\
400--500& 0.73 & 0.73 & 0.74 & 0.75 & 0.73 & 0.70 & 0.47 & 0.44 & 0.68 & \NA \\
500--600& 0.74 & 0.75 & 0.75 & 0.75 & 0.75 & 0.72 & 0.50 & 0.43 & 0.68 & \NA \\
600--700& 0.69 & 0.71 & 0.74 & 0.73 & 0.73 & 0.68 & 0.47 & 0.37 & \NA & \NA \\
700--800& 0.67 & 0.68 & 0.68 & 0.70 & 0.68 & 0.67 & 0.48 & 0.29 & \NA & \NA \\
800--900& 0.64 & 0.67 & 0.67 & 0.67 & 0.65 & 0.64 & 0.46 & 0.25 & \NA & \NA \\
900--1000& 0.63 & 0.65 & 0.65 & 0.66 & 0.67 & 0.66 & 0.54 & \NA & \NA & \NA \\
1000--1200& 0.63 & 0.64 & 0.65 & 0.65 & 0.60 & 0.58 & 0.58 & \NA & \NA & \NA \\
1200--1500& 0.60 & 0.63 & 0.63 & 0.66 & 0.58 & \NA & \NA & \NA & \NA & \NA \\
1500--2000& 0.63 & 0.54 & \NA & \NA & \NA & \NA & \NA & \NA & \NA & \NA \\
\hline
\end{tabular}
}
\end{table}

\begin{table}[hbt]
\centering
\topcaption{Reconstruction and identification efficiency for the $\Zo_{\text{L}}\to \Pq\Paq$ decay as a function of generated $\pt^\Zo$ and $\abs{\eta_\Zo}$ using Z-tagging requirements. Uncertainties on the efficiencies are included in the generic limit calculation as discussed in the text.}
\label{tab:eff_Z_jet}
\vspace*{\medskipamount}
\small
\begin{tabular}{l c c c c c c c}
${\pt^\Zo}$ {range} [\GeVns]&\multicolumn{7}{c}{${\abs{\eta_\Zo}}$ {range}}\\
 & 0.0--0.3 & 0.3--0.9 & 0.9--1.2 & 1.2--1.5 & 1.5--1.8 & 1.8--2.1 & 2.1--2.4 \\
\hline
\hline
80--120& \NA & 0.04 & 0.02 & 0.02 & 0.01 & 0.02 & 0.04 \\
120--150& 0.03 & 0.02 & 0.01 & 0.01 & 0.02 & 0.01 & 0.03 \\
150--200& 0.05 & 0.05 & 0.04 & 0.03 & 0.04 & 0.04 & 0.02 \\
200--250& 0.29 & 0.28 & 0.27 & 0.27 & 0.26 & 0.22 & 0.20 \\
250--300& 0.53 & 0.52 & 0.49 & 0.49 & 0.47 & 0.44 & 0.40 \\
300--400& 0.60 & 0.59 & 0.55 & 0.52 & 0.53 & 0.56 & 0.51 \\
400--500& 0.62 & 0.61 & 0.57 & 0.52 & 0.52 & 0.56 & 0.59 \\
500--600& 0.62 & 0.62 & 0.55 & 0.47 & 0.46 & 0.55 & \NA \\
600--700& 0.61 & 0.60 & 0.51 & 0.41 & 0.45 & 0.54 & \NA \\
700--800& 0.62 & 0.59 & 0.51 & 0.40 & 0.40 & \NA & \NA \\
800--900& 0.59 & 0.57 & 0.48 & 0.37 & 0.38 & \NA & \NA \\
900--1000& 0.57 & 0.56 & 0.47 & 0.31 & \NA & \NA & \NA \\
1000--1200& 0.53 & 0.52 & 0.41 & 0.22 & \NA & \NA & \NA \\
1200--1500& 0.45 & 0.45 & 0.38 & \NA & \NA & \NA & \NA \\
1500--2000& 0.18 & 0.30 & \NA & \NA & \NA & \NA & \NA \\
\hline
\end{tabular}
\end{table}

\begin{table}[hbt]
\centering
\topcaption{Reconstruction and identification efficiency for the $\Zo_{\text{L}}\to \Pq\Paq$ decay as a function of generated $\pt^\Zo$ and $\abs{\eta_\Zo}$ using W-tagging requirements. Uncertainties on the efficiencies are included in the generic limit calculation as discussed in the text.}
\label{tab:eff_W_jet_forZqq}
\vspace*{\medskipamount}
\small
\begin{tabular}{l c c c c c c c }
${\pt^\Zo}$ {range} [\GeVns]&\multicolumn{7}{c}{${\abs{\eta_\Zo}}$ {range}}\\
 & 0.0--0.3 & 0.3--0.9 & 0.9--1.2 & 1.2--1.5 & 1.5--1.8 & 1.8--2.1 & 2.1--2.4 \\
\hline
\hline
200--250& 0.27 & 0.25 & 0.23 & 0.23 & 0.24 & 0.20 & 0.19 \\
250--300& 0.51 & 0.50 & 0.46 & 0.45 & 0.46 & 0.45 & 0.39 \\
300--400& 0.59 & 0.58 & 0.53 & 0.51 & 0.52 & 0.55 & 0.50 \\
400--500& 0.62 & 0.61 & 0.55 & 0.51 & 0.51 & 0.57 & 0.58 \\
500--600& 0.63 & 0.62 & 0.55 & 0.47 & 0.48 & 0.56 & \NA \\
600--700& 0.61 & 0.60 & 0.51 & 0.41 & 0.47 & 0.57 & \NA \\
700--800& 0.62 & 0.60 & 0.51 & 0.38 & 0.41 & \NA & \NA \\
800--900& 0.60 & 0.58 & 0.48 & 0.37 & 0.38 & \NA & \NA \\
900--1000& 0.57 & 0.56 & 0.47 & 0.31 & \NA & \NA & \NA \\
1000--1200& 0.54 & 0.53 & 0.42 & 0.21 & \NA & \NA & \NA \\
1200--1500& 0.46 & 0.46 & 0.38 & \NA & \NA & \NA & \NA \\
1500--2000& 0.19 & 0.29 & \NA & \NA & \NA & \NA & \NA \\
\hline
\end{tabular}
\end{table}

\begin{table}[hbt]
\centering
\topcaption{Reconstruction and identification efficiency for the
$\Wo_{\text{L}} \to \Pq\Paq'$ decay as a function of generated $\pt^\Wo$ and $\abs{\eta_\Wo}$ using Z-tagging requirements. Uncertainties on the efficiencies are included in the generic limit calculation as discussed in the text.}
\label{tab:eff_Z_jet_forWqq}
\small
\begin{tabular}{l c c c c c c c }
${\pt^\Wo}$ {range} [\GeVns]&\multicolumn{7}{c}{${\abs{\eta_\Wo}}$ {range}}\\
 & 0.0--0.3 & 0.3--0.9 & 0.9--1.2 & 1.2--1.5 & 1.5--1.8 & 1.8--2.1 & 2.1--2.4 \\
\hline
\hline
80--120& \NA & \NA & \NA & 0.02 & 0.01 & \NA & 0.02 \\
120--150& 0.01 & 0.02 & 0.02 & 0.01 & 0.01 & 0.02 & 0.04 \\
150--200& 0.13 & 0.09 & 0.11 & 0.10 & 0.11 & 0.07 & 0.10 \\
200--250& 0.44 & 0.43 & 0.40 & 0.37 & 0.39 & 0.37 & 0.33 \\
250--300& 0.56 & 0.57 & 0.54 & 0.53 & 0.51 & 0.54 & 0.45 \\
300--400& 0.59 & 0.59 & 0.56 & 0.52 & 0.53 & 0.55 & 0.55 \\
400--500& 0.59 & 0.58 & 0.53 & 0.48 & 0.48 & 0.54 & 0.49 \\
500--600& 0.57 & 0.57 & 0.50 & 0.42 & 0.44 & 0.45 & 0.56 \\
600--700& 0.56 & 0.55 & 0.49 & 0.36 & 0.36 & 0.42 & \NA \\
700--800& 0.55 & 0.53 & 0.45 & 0.33 & 0.30 & \NA & \NA \\
800--900& 0.52 & 0.52 & 0.40 & 0.27 & 0.34 & \NA & \NA \\
900--1000& 0.48 & 0.47 & 0.38 & 0.26 & 0.35 & \NA & \NA \\
1000--1200& 0.42 & 0.41 & 0.33 & 0.26 & \NA & \NA & \NA \\
1200--1500& 0.32 & 0.33 & 0.22 & 0.10 & \NA & \NA & \NA \\
1500--2000& 0.18 & 0.18 & 0.11 & \NA & \NA & \NA & \NA \\
\hline
\end{tabular}
\end{table}

\cleardoublepage \section{The CMS Collaboration \label{app:collab}}\begin{sloppypar}\hyphenpenalty=5000\widowpenalty=500\clubpenalty=5000\input{EXO-13-009-authorlist.tex}\end{sloppypar}
\end{document}

%% file: EXO-13-009-authorlist.tex
\textbf{Yerevan Physics Institute,  Yerevan,  Armenia}\\*[0pt]
V.~Khachatryan, A.M.~Sirunyan, A.~Tumasyan
\vskip\cmsinstskip
\textbf{Institut f\"{u}r Hochenergiephysik der OeAW,  Wien,  Austria}\\*[0pt]
W.~Adam, T.~Bergauer, M.~Dragicevic, J.~Er\"{o}, C.~Fabjan\cmsAuthorMark{1}, M.~Friedl, R.~Fr\"{u}hwirth\cmsAuthorMark{1}, V.M.~Ghete, C.~Hartl, N.~H\"{o}rmann, J.~Hrubec, M.~Jeitler\cmsAuthorMark{1}, W.~Kiesenhofer, V.~Kn\"{u}nz, M.~Krammer\cmsAuthorMark{1}, I.~Kr\"{a}tschmer, D.~Liko, I.~Mikulec, D.~Rabady\cmsAuthorMark{2}, B.~Rahbaran, H.~Rohringer, R.~Sch\"{o}fbeck, J.~Strauss, A.~Taurok, W.~Treberer-Treberspurg, W.~Waltenberger, C.-E.~Wulz\cmsAuthorMark{1}
% ...
\vskip\cmsinstskip
% ...
1:~~Also at Vienna University of Technology, Vienna, Austria\\
2:~~Also at CERN, European Organization for Nuclear Research, Geneva, Switzerland\\
% ...
